%
%
%
%
%
%
%
%
%
%
%
%
%
%
\input phyzzx.tex
\input epsf
%
%
\catcode`\@=11 
\def\papersize{\hsize=40pc \vsize=53pc \hoffset=0pc \voffset=1pc
   \advance\hoffset by\HOFFSET \advance\voffset by\VOFFSET
   \pagebottomfiller=0pc
   \skip\footins=\bigskipamount \normalspace }
\catcode`\@=12 
\papers
\vsize=24.cm
\hsize=16.cm
\voffset=-1.cm
\newcount\figno
\figno=0
\def\fig#1#2#3{
\par\begingroup\parindent=0pt\leftskip=1cm\rightskip=1cm\parindent=0pt
\baselineskip=11pt
\global\advance\figno by 1
\midinsert
\epsfxsize=#3
\centerline{\epsfbox{#2}}
\vskip 12pt
{\bf Fig. \the\figno:} #1\par
\endinsert\endgroup\par
}
\def\figlabel#1{\xdef#1{\the\figno}}
\def\encadremath#1{\vbox{\hrule\hbox{\vrule\kern8pt\vbox{\kern8pt
\hbox{$\displaystyle #1$}\kern8pt}
\kern8pt\vrule}\hrule}}
%
\vsize=24.cm
\hsize=16.cm
\def\IM{\mathop{\Im m}\nolimits}
\def\RE{\mathop{\Re e}\nolimits}
\def\Z{{\bf Z}}
\def\R{{\bf R}}
\def\Q{{\bf Q}}

\def\C{${\cal C}$}
\def\M{${\cal M}$\ }

\def\nf{N_f}

\def\to{\rightarrow}

\def\ad{a_D}
\def\at{{\tilde a}}
\def\adt{{\tilde a_D}}
\def\net{{\tilde n_e}}
\def\nmt{{\tilde n_m}}
\def\st{{\tilde s}}
\def\e{\epsilon}
\def\l{\lambda}
\def\lu{\Lambda_1}
\def\ld{\Lambda_2}
\def\lt{\Lambda_3}
\def\lz{\Lambda_0}
\def\t{\tau}
\def\s{{\sigma}}
\def\d{{\rm d}}
\def\rd{\sqrt{2}}
\def\g{\gamma}

\def\a{\alpha}

\def\o{\omega}

\def\vac{\vert 0 \rangle}

\tolerance=500000
\overfullrule=0pt

\Pubnum={LPTENS-97/2 \cr
{\tt hep-th@xxx/9706145} \cr
June 1997}

\date={}
\pubtype={}
\titlepage
\title{{\bf The BPS Spectra and Superconformal Points
\break in Massive $N=2$ Supersymmetric QCD}}
\author{Adel~Bilal
}
\andauthor{Frank~Ferrari}
\vskip .5cm
\address{
CNRS - Laboratoire de Physique Th\'eorique de l'\'Ecole
Normale Sup\'erieure
\foot{{\rm Unit\'e Propre de Recherche 701
du CNRS, associ\'ee \`a l'\'Ecole Normale
Sup\'erieure et \`a l'Universit\'e Paris-Sud.}}    \break
24 rue Lhomond, 75231 Paris Cedex 05, France  \break
{\tt adel.bilal@physique.ens.fr,\
frank.ferrari@physique.ens.fr}
}

\vfill
\vskip 0.5cm
\singlespace
\abstract{\singlespace
We present a detailed study of the analytic structure, BPS spectra and
superconformal points of the $N=2$ susy ${\rm SU}(2)$ gauge theories 
with $N_f=1,2,3$ massive quark hypermultiplets. We compute the curves 
of marginal stability with the help of the explicit solutions for the low 
energy effective actions in terms of standard elliptic functions. We show 
that only a few of these curves are relevant.  As a generic example, the 
case of $N_f=2$ with two equal bare masses  is studied in depth. 
We determine the precise existence domains for each BPS state, and show 
how they are compatible with the RG flows. At the  superconformal point, 
where two singularities coincide, we prove that (for $N_f=2$) the massless spectrum 
consists of {\it four} distinct BPS states and is $S$-invariant. This is due 
to the monodromy around the superconformal point being $S$, providing 
strong evidence for exact $S$-duality of the SCFT. For all $N_f$,
we compute the slopes $\omega$ of the $\beta$-functions at the fixed point 
couplings and show that they are related to the anomalous dimensions $\alpha$
of $u=\langle {\rm tr}\, \phi^2\rangle$ by $\omega=2(\alpha -1)$.
}

%

\endpage
\voffset=1.cm

\singlespace
\pagenumber=1

 \def\PL #1 #2 #3 {Phys.~Lett.~{\bf #1} (#2) #3}
 \def\NP #1 #2 #3 {Nucl.~Phys.~{\bf #1} (#2) #3}
 \def\PR #1 #2 #3 {Phys.~Rev.~{\bf #1} (#2) #3}
 \def\PRL #1 #2 #3 {Phys.~Rev.~Lett.~{\bf #1} (#2) #3}
 \def\CMP #1 #2 #3 {Comm.~Math.~Phys.~{\bf #1} (#2) #3}
 \def\IJMP #1 #2 #3 {Int.~J.~Mod.~Phys.~{\bf #1} (#2) #3}
 \def\JETP #1 #2 #3 {Sov.~Phys.~JETP.~{\bf #1} (#2) #3}
 \def\PRS #1 #2 #3 {Proc.~Roy.~Soc.~{\bf #1} (#2) #3}
 \def\JFA #1 #2 #3 {J.~Funkt.~Anal.~{\bf #1} (#2) #3}
 \def\LMP #1 #2 #3 {Lett.~Math.~Phys.~{\bf #1} (#2) #3}
 \def\IJMP #1 #2 #3 {Int.~J.~Mod.~Phys.~{\bf #1} (#2) #3}
 \def\FAA #1 #2 #3 {Funct.~Anal.~Appl.~{\bf #1} (#2) #3}
 \def\AP #1 #2 #3 {Ann.~Phys.~{\bf #1} (#2) #3}
 \def\MPL #1 #2 #3 {Mod.~Phys.~Lett.~{\bf #1} (#2) #3}

\REF\SWI{N. Seiberg and E. Witten, {\it Electric-magnetic duality,
monopole
condensation, and confinement in $N=2$ supersymmetric Yang-Mills theory},
\NP B426 1994 19 , {\tt hep-th/9407087}.}

\REF\SWII{N. Seiberg and E. Witten, {\it Monopoles, duality and chiral
symmetry breaking in $N=2$ supersymmetric QCD}, \NP B431 1994 484 ,
{\tt hep-th/9408099}.}

\REF\FB{F. Ferrari and A. Bilal, {\it The strong-coupling spectrum of
Seiberg-Witten theory}, Nucl. Phys. {\bf B469} (1996) 387,
{\tt hep-th/9602082}.}

\REF\BF{A. Bilal and F. Ferrari, {\it Curves of marginal stability, and
weak and strong coupling BPS spectra in N=2 supersymmetric QCD},
Nucl. Phys. {\bf B480} (1996) 589,
{\tt hep-th/9605101}.}

\REF\ABDISC{A. Bilal,  {\it Discontinuous BPS spectra in $N=2$ susy QCD},
Nucl. Phys. B (Proc. Suppl.) {\bf 52A} (1997) 305-313, {\tt hep-th/9606192}.}

\REF\AD{P.C. Argyres and M.R. Douglas, {\it New phenomena in {\rm SU(3)}
supersymmetric gauge theory}, \NP B448 1995 93, {\tt hep-th/9505062}.}

\REF\APSW{P.C. Argyres, M.R. Plesser, N. Seiberg and E. Witten,
{\it New $N=2$ superconformal field theories in four dimensions},
\NP B461 1996 71 ,  {\tt hep-th/9511154}. }

\REF\FERI{F. Ferrari, {\it Charge fractionisation in $N=2$ supersymmetric
QCD}, \PRL 78 1997 795 , {\tt hep-th/9609101}.}

\REF\BRASTI{A. Brandhuber and S. Stieberger, {\it Self-dual strings and
stability of BPS states in $N=2$ ${\rm
SU}(2)$ gauge theories}, \NP B488 1997 199 , {\tt hep-th/9610053}.}

\REF\WAR{A. Klemm, W. Lerche, P. Mayr, C. Vafa and N. Warner, {\it Self-dual strings
and N=2 supersymmetric field theory}, \NP B477 1996 746 ,  {\tt
hep-th/9604034};\nextline
J. Schulze and N.P. Warner, {\it BPS geodesics in N=2 supersymmetric
Yang-Mills theory}, {\tt hep-th/9702012};\nextline
J. Rabin, {\it Geodesics and BPS states in N=2 supersymmetric QCD}, 
{\tt hep-th/9703145}.}

\REF\FERII{F. Ferrari, {\it The dyon spectra of finite gauge theories},
LPTENS
preprint 96/67, to appear in Nucl. Phys. B, {\tt hep-th/9702166}.}

\REF\LAG{L. Alvarez-Gaum\'e, M. Marino, F. Zamora,
{\it Softly broken N=2 QCD with massive quark hypermultiplets (I)},
{\tt hep-th/9703072}.}

\REF\INST{H. Aoyama, T. Harano, M. Sato and S. Wada, {\it Multi-instanton
calculus in $N=2$ supersymmetric QCD}, \PL B388 1996 331, {\tt
hep-th/9607076}\break
T. Harano and M. Sato, {\it Multi-instanton calculus versus exact results in
$N=2$ supersymmetric QCD}, {\tt hep-th/9608060}.}

\REF\ERD{A. Erdelyi et al, {\it Higher Transcendental Functions}, Vol 1,
McGraw-Hill,
New York, 1953.}

\REF\MACK{G. Mack, {\it All unitary ray representations of the conformal
group} SU(2,2) {\it with positive energy}, \CMP 55 1977 1 .}

\REF\ADEN{T. Eguchi, K. Hori, K. Ito and S.K. Yang, {\it Study of $N=2$
superconformal field theories in 4 dimensions}, \NP B471 1996 430 , {\tt
hep-th/9603002}.}

\REF\FGT{P.C. Argyres, {\it S-duality and global symmetries in $N=2$
supersymmetric field theory}, {\tt hep-th/9706095}.}

\REF\DS{M. Douglas and S. Shenker, {\it Dynamics of ${\rm SU}(n)$ supersymmetric gauge
theory}, \NP B447 1995 271 .}

\REF\WW{E.T. Whittaker and G.N. Watson, {\it A course of modern analysis},
Cambridge
University Press (1963).}

\REF\ELL{K. Chandrasekhara, {\it Elliptic functions}, Springer-Verlag
(Berlin,
1985).}

\REF\OHTA{Yuji Ohta, {\it Prepotentials of $N=2$ $SU(2)$ Yang-Mills
theories coupled with
massive matter multiplets}, J. Math. Phys. {\bf 38} (1997) 682, {\tt
hep-th/9604059}.}

\REF\CARDY{J.L. Cardy, \NP B170 1980 369  and \NP B205 1982 17 .}

{\bf \chapter{Introduction}}

Many new insights into the physics of strongly coupled gauge theories have
been obtained by the
study of $N=2$ supersymmetric Yang-Mills theories [\SWI,\SWII]. 
Two particularly interesting phenomena that occur in these four-dimensional theories
are the discontinuities of the BPS spectra [\FB,\BF,\ABDISC] on the Coulomb branch 
and the occurrence of
superconformal points  that are believed to lead to non-trivial
interacting 4D superconformal field theories (SCFT) [\AD,\APSW]. 

In the present paper we study the probably simplest case where both phenomena occur:
the ${\rm SU}(2)$ theories with massive quark hypermultiplets. The discontinuities of
the BPS spectra and the properties at the superconformal points are of course closely related
and our study of the first will shed important new light on the second, and vice versa.

The main ingredients in the study of the $N=2$ gauge theories are duality
and holomorphicity of the low-energy effective action. The properties of
the latter are encoded in
a (hyper) elliptic curve which also determines the abelian charges of the
theory.  The BPS states constitute a particularly
important sector of the Hilbert space:
all perturbative and known
solitonic states are BPS states. Their masses are determined by the
abelian charges and hence
by the (hyper) elliptic curve. Quite surprisingly, the latter also allows
us to
extract and determine the existence domains of the BPS states
on the Coulomb branch of the moduli space ${\cal M}$.
Indeed, BPS states generically are  stable, except on real codimension one
hypersurfaces
in ${\cal M}$ (i.e. real
dimension one curves for ${\rm SU}(2)$). These instability hypersurfaces
are
determined in terms of the abelian charges and the quantum numbers of the
BPS state.

For the
${\rm SU}(2)$ theories without or with $N_f$ {\it massless} quark
hypermultiplets ($N_f=1,2,3$) there
only is a {\it single} instability curve on  ${\cal M}\simeq {\bf C}$.
This curve is closed and goes through the singular
points on ${\cal M}$. The BPS spectra
are different inside and outside the curve [\FB,\BF,\ABDISC]. We found that in the
region
outside the curve, part of which is the semiclassical domain, all
semiclassically expected states
exist. These are: the dyons $(n_e,n_m)=(2n,1)$ for $N_f=0$, the quarks
$(1,0)$ and the dyons
$(n,1)$ for $N_f=1,2,3$, with in addition dyons $(2n+1,2)$ of magnetic
charge two for $N_f=3$.
There also is the W-boson $(2,0)$ for all $N_f$. In the region inside the
instability curve, which always
is a region of strong coupling, there only exist the states that are
responsible for the
singularities, i.e. the states that can become massless. For $N_f=0$ e.g.,
these are the dyons
$(0,1)$ and $(-2\e,1)$ while for $N_f=2$ these are $(0,1)$ and $(-\e,1)$.
Here $\e$ equals $+1$ in
the upper  and $-1$ in the lower half plane. Almost all BPS states
disappear when crossing the curve from outside to inside: they ``decay"
into the only two (or
three) existing states. In particular, there is no W-boson inside the
curve, although the gauge
symmetry is always broken ${\rm SU}(2)\to {\rm U}(1)$.

The determination of these BPS spectra was relatively easy since there was
only one instability curve. The
generic situation of ${\rm SU}(n)$  is much more complicated: each BPS
state has its own  family
of possible instability hypersurfaces. At first sight it seems hopeless to
study this general case by
the methods of [\FB,\BF]. However, a similar situation already occurs in
the
${\rm SU}(2)$ theories with {\it massive}  hypermultiplets. This is the case
we study in the present
paper. It will turn out that among the multitude of possible decay curves
only a relatively small
subset is  relevant, and one obtains a very clear picture of the existence
domains of
the various BPS states. We are quite confident that this structure can be
exploited to also determine
the exact BPS spectra at any point on the Coulomb branch of the ${\rm
SU}(n)$ moduli space.

Another feature, absent in the massless ${\rm SU}(2)$ theories, but present for the
higher ${\rm SU}(n)$ theories or the ${\rm SU}(2)$ theories with massive hypermultiplets,
is the existence of superconformal points. These can occur at special points on the
Coulomb branch in the ${\rm SU}(n)$ theories where singular lines intersect, or in the
massive ${\rm SU}(2)$ theories at special values of the masses when singular points
collide. In any case, at a superconformal point two or more mutually non-local BPS states
simultaneously become massless [\AD]. To study these points, it will turn out to be most
fruitful to view them as resulting from such a coincidence of individual singularities.

In this paper, we will discuss the BPS spectra of the ${\rm SU}(2)$
theories with $N_f=1,2,3$
massive quark hypermultiplets, and present
an in depth study of the $N_f=2$ theory with equal bare masses. This
latter case is sufficiently generic
to exhibit all interesting new phenomena. 
We determine the precise existence domains for each BPS state and confirm these results
by many additional consistency checks.
We will also see how the whole set
of decay curves and BPS spectra very consistently behaves under the RG flow
from one $N_f$ theory to
another. A subtlety present in the massive theories is related to the
abelian $s_i$-charges
($i=1,\ldots N_f$) [\FERI] which we need to determine for all BPS states.
Some indications on the
BPS spectra of the massive theories were already obtained in [\BRASTI]
within the
geodesics approach from string theory [\WAR]. However, it only provided some
partial and pointwise information on
${\cal M}$.\foot{
Where comparable, both results agree. Although there is a slight discrepancy with the published
version of [\BRASTI], after a first circulation of the present paper, the authors 
of [\BRASTI] have informed us that this is only due to some error when writing up their paper but
that they actually agree with our results, see Section 4 below.}

The analytic structure and the corresponding monodromies are fundamental for our study.
As we follow the various RG flows, we reach various superconformal points where the
monodromies are given by the products of the individual monodromies of the coinciding
singularities. Such a composite monodromy $M_{\rm sc}$
is quite special, acting e.g. as $S$-duality
relating mutually non-local states as monopoles and quarks. We argue that it should be 
an exact
quantum symmetry of the massless BPS sector and thus of the superconformal field
theories. This monodromy (completely) characterises the SCFT allowing us to compute 
scaling dimensions. We show that the SCFT is determined in terms of a single integer
$k=1,2$ or $3$ characteristic of $M_{\rm sc}$. This is reminiscent to the study of the
relevant deformations of the singular curves $y^2=x^3$ in [\APSW]. 
As we follow the RG flow further, the singularities separate again but
the analytic structure  is changed, providing us with an explanation of how the nature of
certain singularities can change from being due to  a massless quark at weak coupling to
being due to a massless dyon at strong coupling.

Let us outline the organisation of the present paper and some of our 
results: first, in Section 2, we recall the
elliptic curves for the massive $N_f=1,2,3$ theories
and discuss the quantum numbers of the  BPS states associated with the
singularities and how they change under the various RG flows one can
study.
This brings us naturally to a discussion of the superconformal points and how they are
classified by an integer $k=1,2,3$ through their monodromies.
Then, we compute the basic
functions $\ad(u)$ and $a(u)$,
which are the period integrals of a certain meromorphic one-form, in terms
of standard elliptic
integrals in a form immediately suited for numerical computations. 
Technical details are postponed to the appendices.
Similar
results were also obtained
independently and published recently in [\FERII,\LAG].
In Section 3, we
discuss general features of the BPS spectra and  the decay curves for all
the massive $N_f=1,2,3$
theories. In particular, 
we discuss the maximal possible set of BPS states compatible with the RG flow.

Section 4 then is an in depth study of the $N_f=2$ theory with two
equal bare masses $m$. There exists a superconformal point when
$m={\ld\over 2}$, and we discuss
separately the cases $m<{\ld\over 2}$ and $m>{\ld\over 2}$. We find that
all curves and existence
domains of BPS states have perfectly smooth RG flows. 
In particular, the crossover from small mass
($m<{\ld\over 2}$) to large mass ($m>{\ld\over 2}$) is perfectly smooth, 
except, in a certain sense, at the
superconformal point $u_*$ itself. Also, as
$m$ becomes very large, in
a basis $\adt,\ \at$ that flows to the $\ad^{(N_f=0)},\ a^{(N_f=0)}$ only
the states with $\st=0$
survive under the RG flow $m\to \infty$ to the pure gauge theory $N_f=0$.
At a given fixed point
$u\in {\cal M}$, almost all $\st\ne 0$ dyons disappear, already at finite
$m$, because they are
``hit" by their corresponding decay curves that move outwards (to large
$\vert u\vert$) as $m$ is
increased. Among the $\st\ne 0$ states only the quarks and some special dyons  exist
for all finite $m$, but as $m\to\infty$, their BPS masses diverge and thus
they simply drop out of
the spectrum for this reason. On the other hand, the different $\st=0$
states  decay on one and the same
curve which flows to the curve of the $N_f=0$ theory. Thus we are able to
see the flow to the
$N_f=0$ spectra in full detail.
For each case, $m<{\ld\over 2}$ and $m>{\ld\over 2}$, we first discuss the general
picture, which is then established by considering each class of BPS states separately.

The BPS spectra for $N_f=2$, $m={\ld\over 2}$ are then simply obtained as
the limit of those for either $m>{\ld\over 2}$ or $m<{\ld\over 2}$.
Nothing changes dramatically
for $m={\ld\over 2}$,
except at the value of $u$ equal to the two coinciding singularities which
is the superconformal
point. 
In Section 5, we discuss the physics of this superconformal point in some detail.
There, quite remarkably, one has {\it four}  massless states, namely 
the quark $(n_e,n_m)_s=(1,0)_1$ and
the monopole $(0,1)_0$ which are flavour doublets, and the two dyons 
$(1,1)_1$ and $(-1,1)_{-1}$ which are flavour singlets.
This spectrum has an $S$-duality invariance which is realized by
the monodromy matrix $M_{\rm sc}$ around the superconformal point. 
Since this massless sector constitutes the superconformal field theory, the latter
should have a quantum $S$-duality invariance.
We then discuss in general
the Argyres-Douglas ansatz [\AD] for the $\beta$-function of a theory
based on the massless states at the superconformal point. These authors conjecture
that, although there is no local action at our disposal to simultaneously describe all 
massless states, the beta might nevertheless be obtained by simply computing the contribution
of each massless particle in the formulation of the theory which describes it locally and
then adding together the suitably duality transformed contributions.
We show that despite its appealing features, this ansatz is not correct, as already
suspected by these authors because it let to irrational values for the slope $\omega$
of the $\beta$-function at the fixed point. We argue that the slopes
$\omega$ can be computed from the low-energy effective actions alone and find that 
they are
a rational numbers related to the scaling dimensions $\alpha$ of $\langle {\rm tr}\,
\phi^2\rangle$ as $\omega=2(\alpha-1)$ . We show that this relation is in perfect
agreement with $N=2$ superconformal invariance. 

Then follow four appendices. In appendix A,
we discuss the positions of the singularities and their different RG flows for
$N_f=1,2$ and $3$. In appendix B, we give details on the elliptic integrals needed in
Section 2 - and heavily used for the numerical computations of Section 4. In appendix
C, we express the period integrals in terms of the three elliptic integrals also for
$N_f=1,3$ and for $N_f=2$ with $m_1=m,\ m_2=0$, and check the RG flows on these
expressions. Finally, in appendix D, we study the RG flow of the $N_f=2$ integrals
with equal bare masses, thus providing some additional consistency checks.

{\bf \chapter{RG flows, analytic structure, superconformal points
and period integrals}}
\sectionnumber=0

The structure of the Coulomb branch of the asymptotically free SU(2) 
theories under study was derived
in [\SWI , \SWII]. It is given in terms of an elliptic curve 
of the form
$$y^2 = x^2 (x-u) + P_{N_f}(x,u,m_j,\Lambda _{N_f})\eqn\di$$
where\foot{This relation is only
valid up to a constant shift in the general case [\INST].}
$u=\langle {\rm tr}\,\phi ^2\rangle$
is the gauge invariant moduli, $\phi$ the scalar component of the $N=2$
vector multiplet, $m_j$ ($1\leq j\leq N_f$) the bare masses of the
hypermultiplets and $\Lambda _{N_f}$ the dynamically generated scale of the
theory. The polynomials $P_{N_f}$ are given by
$$\eqalign{
P_0 =& {\lz ^4\over 4}\, x\ ,\cr
P_1 =& {\lu^3\over 4}\, m_1 x - {\lu^6\over 64}\ ,\cr
P_2 =& - {\ld^4\over 64} \bigl( x-u\bigr) +{\ld^2\over 4} m_1 m_2 x -
{\ld^4\over 64} (m_1^2 + m_2^2)\ ,\cr
P_3 =& -{\lt^2\over 64}\, \bigl(x-u\bigr)^2 -{\lt^2\over 64}\, 
\bigl(x-u\bigr) \bigl( m_1^2+m_2^2 + m_3^2\bigr) \cr
&\qquad + {\lt\over 4} m_1 m_2 m_3 x -{\lt^2\over 64} 
\bigl(m_1^2 m_2^2 + m_1^2 m_3^2+ m_2^2 m_3^2\bigr) \ . \cr }
\eqn\dii $$
The masses of the so called BPS states, which
come in short representations of the supersymmetry algebra, can be 
computed using the fundamental formula
$$M_{\rm BPS}(u) =\rd\,
\left\vert n_m \ad(u) -n_e a(u) + \sum_i s_i {m_i\over\rd} 
\right\vert \cdotp
\eqn\diii$$
In this expression, $n_{e}$ and $n_{m}$ are two integers representing 
the electric and magnetic charges of the state, and the $s_{i}$ are 
integers or
half-integers which correspond to constant parts of the physical 
baryonic charges [\SWII ,\FERI]. If $\l$ is a meromorphic differential 
on the curve \di\ such that
$$ {\partial\l\over\partial u}= {\sqrt{2}\over 8\pi }\, {dx\over y}
\eqn\div$$
then the variables $a$ and $a_{D}$ are given by the contour integrals
$$a=\oint_{\g_1} \l\quad , \quad \ad =\oint_{\g_2} \l
\eqn\dv $$
for a certain homology basis $(\g _{1},\g _{2})$.
The BPS mass formula \diii\ is compatible with duality transformations 
of the form
$$ \pmatrix{a_{D}\cr a\cr m/ \sqrt{2}\cr}\longrightarrow M
\pmatrix{a_{D}\cr a\cr m/ \sqrt{2}\cr}\ ,\quad
\pmatrix{n_{e}\cr n_{m}\cr s\cr}\longrightarrow M^{*}
\pmatrix{n_{e}\cr n_{m}\cr s\cr}\eqn\dvi$$
where
$$  M=\pmatrix{\alpha & \beta & f\cr \gamma & \delta & g\cr 
0&0&1\cr}\ ,\quad
M^{*}=\pmatrix{\alpha & \beta & 0\cr \gamma & \delta & 0\cr 
\alpha g - \gamma f & \beta g - \delta f &1\cr}\eqn\dvii$$
with $\alpha$, $\beta $, $\gamma $, $\delta $ integers such that 
${\rm det}\, M=1$ and $f$, $g$ integers or half-integers. Note the 
useful relations
$$\bigl( M_{1}M_{2} \bigr) ^{*} = M_{1}^{*}M_{2}^{*}\quad ,
\quad \bigl(M^{*}\bigr)^{-1}=\bigl(M^{-1}\bigr)^{*} \ . 
\eqn\dviibis $$
One 
particularly important class of duality transformations corresponds to 
the monodromy transformations $a$ and $a_{D}$ undergo when encircling 
a singularity in the $u$ plane. These singularities are due to 
dyons $(n_e,n_m)_s^{\times d}$ lying in a $d$-dimensional representation of
the flavour group becoming massless and occur when the curve \di\ has a 
vanishing cycle, i.e. when the discriminant of the corresponding
polynomial \dii\ 
vanishes. When all the {\it non-zero} bare masses are equal, which is the 
only case we will study in the following for the seek of simplicity,
the mass term in \diii\ is of the form
$ms/\sqrt{2}$, and the monodromy matrices then read
$$\eqalign{
M_{(n_e,n_m)_s^{\times d}} &= \pmatrix{ 
1-d n_e n_m&d n_e^2&-d n_e s\cr -d n_m^2&1+d
n_m n_e&-d n_m s
\cr 0& 0&1\cr}\ , \cr
M^*_{(n_e,n_m)_s^{\times d}} &= 
\pmatrix{ 1-d n_e n_m&d n_e^2&0\cr -d n_m^2&1+d n_m
n_e&0
\cr -d s n_m& d s n_e&1\cr}
\ . }
\eqn\dviii$$
When two singularities coincide, the monodromy is given by the 
product of two matrices of the type \dviii . Note that the fact that 
$a$ and $a_{D}$ pick up constants multiple of $m/2\sqrt{2}$ under 
monodromy transformations is possible because the differential $\l$ 
has poles with residues such that 
$$2\pi i\ {\rm res}\l = \sum_i t_i {m_i\over \rd}\ \raise 2pt\hbox{,}
\quad t_i \in {1\over 2} \Z \ . \eqn\dix$$
\section{\bf RG flows and superconformal points}
The theories for different $\nf$ are related by the renormalization 
group flow when some of the bare masses of the hypermultiplets are 
sent to infinity and the corresponding states can be integrated out
(see Appendices A and C). 
For instance, if we let $m_{3}\rightarrow\infty $ while keeping $\ld 
^{2}=m_{3}\lt $ fixed, the polynomial $P_{3}$ in \dii\ flows to $P_{2}$. 
Another flow which we will study in great details in the following is 
$m_{1}=m_{2}=m\rightarrow\infty $ with $m\ld =\lz ^{2}$ fixed which allows 
to obtain the $\nf =0$ theory directly from the $\nf =2$ theory.
A particularly important phenomenon that must occur during the RG flow 
is that the quantum numbers of some of the particles becoming 
massless change. Typically, when $m\rightarrow\infty$ some 
singularities are at weak coupling and
correspond to elementary quarks becoming massless 
while when $m\rightarrow 0$ these singularities move towards the 
strongly coupled region where only magnetically charged states can 
become massless. The transformation of the quantum numbers must be
implemented by a SL$(2,\Z)$ matrix $U$.
Since the electric and magnetic 
quantum numbers of the singularities at $m=0$ can in general 
be deduced on physical grounds, by using for instance the discrete 
symmetry acting on the Coulomb branch when it exists, the general 
form of $U$ can be determined a priori. Two cases can then arise:
either the matrix $U$
does or does not correspond to any monodromy matrix of the form \dviii .
If it does not, we will see that (for real $m$) the quantum numbers change when at 
least two singularities coincide. Such points were  discovered 
by inspection in [\APSW]. The low energy theory at these points is 
believed to correspond to an interacting $N=2$ superconformal theory 
[\AD ,\APSW]. We will discuss the physics of these theories in Section 5,
but let us right now display the monodromy matrix around 
such a superconformal point, discuss some general constraints that it 
must satisfy in order to be compatible with conformal invariance, and 
determine the way the quantum numbers of the singularities change.

\fig{After the collision, the quantum numbers at the singularities and thus
the associated monodromy matrices change according to equations (2.15) or
(2.16).}{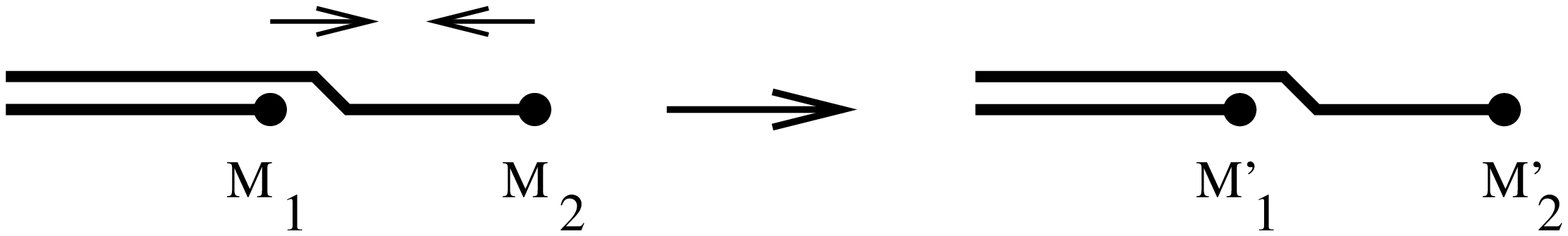}{12cm}\figlabel\figia

The generic case of two colliding singularities $\s _{1}$ and $\s _{2}$
is represented in Fig. 1. The singularity $\s _{j}$
whose monodromy is $M_{j}$ is due to $d_{j}$ hypermultiplets 
$(p_{j},q_{j})_{s_{j}}$ becoming massless. Note that we choose
$\s _{1}$ to be below the cut produced by $\s_{2}$. If the two cuts 
actually coincide on the left of $\s _{1}$, this will mean that the 
quantum numbers $(p_{1},q_{1})_{s_{1}}$ at $\s_{1}$ are
computed by looking at the solution $(a_{D},a)$ below the cut.
With this convention, the monodromy at the superconformal point is
$$ M_{\rm sc}=M_{2}M_{1}.\eqn\dx $$
The eigenvalues of this matrix depend on one symplectic invariant 
integer parameter
$$ k= d_{1}d_{2}(p_{1}q_{2}-p_{2}q_{1})^{2}\eqn\dxi $$
which is such that ${\rm tr}\, M_{\rm sc} = 2-k$. We must have
$k>0$ since otherwise the 
particles becoming massless are mutually local and the low energy 
theory is simply $N=2$ super QED with a dual photon. 
Superconformal invariance implies that the dimensions of $a$ and $a_D$ must
both be equal to one, and thus that near the superconformal point $u_*$ the
following expansion is valid
$$\eqalign{
a_{D}(u)= & a_{D}(u_{*})+c_{D}(u-u_{*})^{1/\alpha}+
o\bigl( (u-u_{*}) ^{1/\alpha}\bigr) \cr
a(u)=& a(u_{*}) + c (u-u_{*})^{1/\alpha}+o\bigl( (u-u_{*}) 
^{1/\alpha}\bigr),}
\eqn\dxii $$
where $c_{D}$ and $c$ are constants and $\alpha$ is the anomalous 
dimension of the operator $u$. This is possible only if the 
eigenvalues of $M_{\rm sc}$ are of modulus one, which rules out the 
cases $k>4$. Moreover, when $k=4$, $M_{\rm sc}$ is, up to a global 
sign, conjugate to a certain non-zero power of $T$. This would be the signal 
of logarithmic terms in the asymptotic expansion of $a$ and $a_{D}$, 
which again are ruled out by conformal invariance. Thus we conclude 
that $1\leq k\leq 3$. The three allowed values of $k$ correspond to
three inequivalent SCFT. We will bring some evidence in Section 4 that the
SCFT is actually (fully) characterized by this integer $k$, by computing critical
exponents as  functions of $k$. 

Denoting with a prime the quantities corresponding to the new quantum
numbers after the collision of the singularities, as indicated in Fig. 1, one must have
$$ M_2 M_1 = M_2' M_1'.\eqn\dxiii$$
The multiplicity $d_j$
of a given singularity $\s _j$, which can be directly read off the
discriminant of the relevant polynomial \di, \dii , cannot change at the
superconformal point. Thus one must either 
have $d_j = d_j'$, in which case \dxiii\ can be solved non trivially by 
$$ M'_j =  UM_jU^{-1}, \quad {\rm with}\quad U=M_{sc}^{-1},\eqn\dxiv $$
or $d'_1=d_2$, $d'_2=d_1$, in which case \dxiii\ can be solved by
$$ M'_1= UM_2U^{-1},\quad M'_2=UM_1U^{-1},\quad {\rm with}\quad
U=M_1^{-1}\quad {\rm or}\quad U=M_2.\eqn\dxv$$
The quantum numbers are then changed according to the matrix $U^*$, see 
\dvii . When $k=3$, there is another consistent solution to \dxiii ,
which corresponds to $U= I+ M_{\rm sc}^{-1}$ if $d_j=d_j'$ or to $U=M_1^{-1} +
M_2$ if $d'_1=d_2$. The matrix $U$ is then not of the form \dvii , but its
$2\times 2$ upper left block  part is still in SL$(2,\Z)$.\foot{
This would not be the case for $k=1$ or $k=2$.}
The corresponding matrix $U^*$ can then be straightforwardly deduced by
equating the central charges of the original and image states. This latter
case, with $d_1'=d_2$, is actually realized in the $N_f=3$ theory with three
equal masses which we study in the next subsection.
\section{\bf RG flows and analytic structure}
\fig{Shown are three different RG flows from $N_f$ to $N_f'<N_f\ $,
$m\rightarrow\infty$, $\Lambda _{N_f}\rightarrow 0$ with
$\Lambda _{N_f'}^{4-N_f'}=\Lambda _{N_f}^{4-N_f} m^{N_f - N_f'}$ fixed. 
The upper figure gives the flow from $N_f=2$ to $N_f'=1$, the middle figure the flow
$N_f=3$ to $N_f'=0$ and the lower one $N_f=2$ to $N_f'=0$. In this latter case 
which will be studied in detail in Section 4, we have indicated
the quantum numbers of the singularities on {\it both} sides of the branch
cuts.}{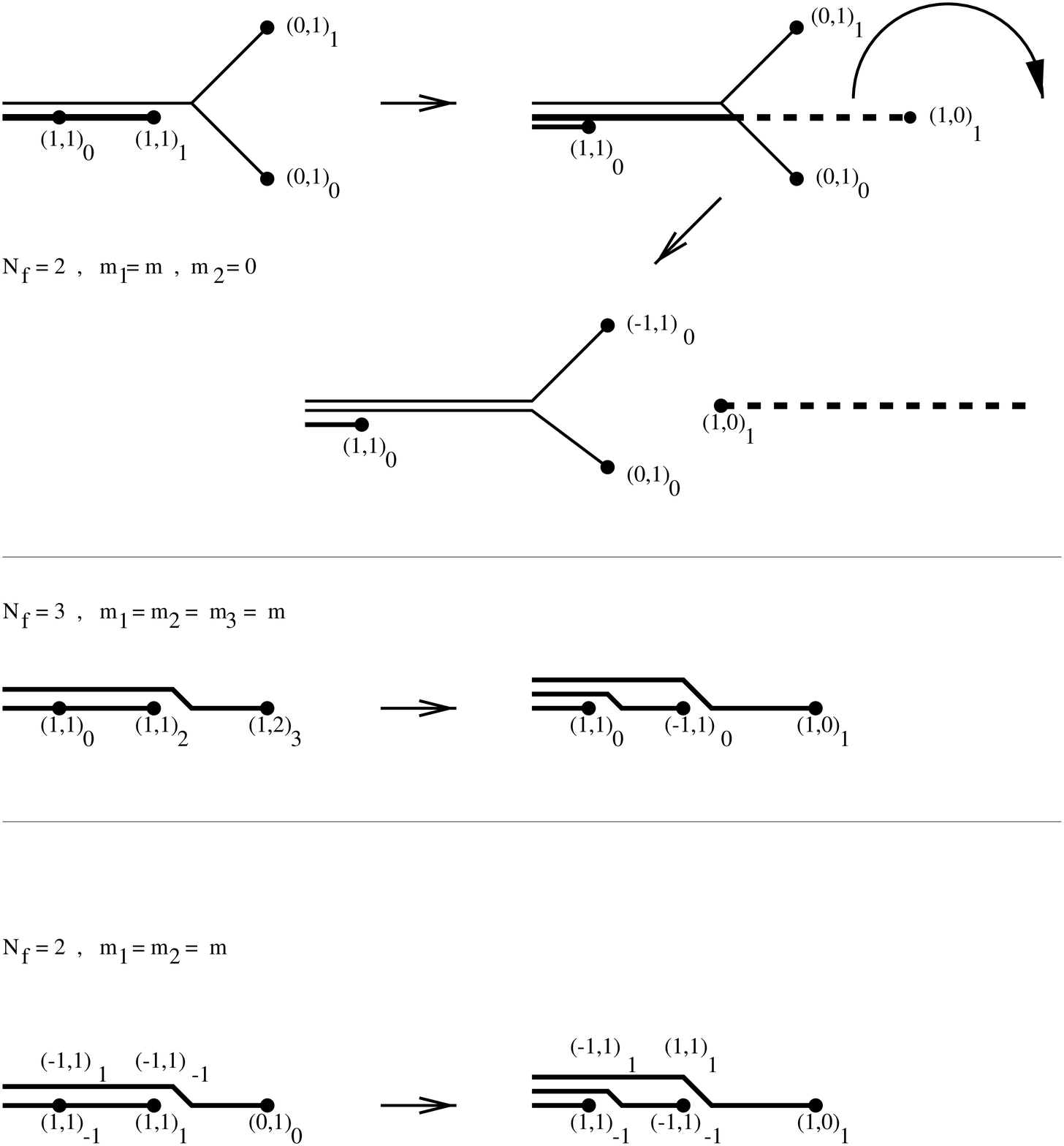}{15cm}\figlabel\figiia

In Fig. \figiia, we have reprensented three different RG flows which illustrate the
discussion of the previous subsection. 

The first case corresponds to the flow
from the $\nf =2$ theory to the $\nf =1$ theory obtained by sending $m_2$ to
infinity while keeping $m_2\ld^2=\lt^3$ fixed, and $m_1=0$. 
No superconformal point is needed in this case. The
dyon singularity $(1,1)_1$ changes to one component of
the elementary quark $(1,0)_1$
when crossing the cut produced by the $(0,1)_0$ singularity as $m_2$ is
increased. In order to recover a more conventional analytic structure in the
$m_2\rightarrow\infty$ limit, one can move the cut originating at $(1,0)_1$
to the right
as indicated in the Figure. Physically, this simply amounts to shifting the
$\theta$ angle by an integer multiple of $2\pi$ in the upper half $u$-plane,
which is an unphysical transformation. Mathematically, one keeps the solution
$(a_D,a)$ fixed for $\IM u<0$ and performs the transformation
$$ \pmatrix{a_D\cr a\cr m/\sqrt{2}\cr}\longrightarrow M_{(1,0)_1}^{-1}
\pmatrix{a_D\cr a\cr m/\sqrt{2}\cr}\eqn\dxvi$$
for $\IM u>0$. It follows that the quantum numbers of the massless state 
at the singularity which were $(0,1)_1$ before, now are 
$\bigl(M_{(1,0)_1}^{-1}\bigr)^* (0,1)_1 = (-1,1)_0$.

The second case corresponds to the flow from the $\nf =3$ theory with three
equal bare masses $m_1=m_2=m_3=m$ to the pure gauge theory $\nf =0$. Let us
discuss in this case the quantum numbers of the singularities in more detail.
When $m=0$, one has two singularities. One of them is due to a dyon $(1,1)$
lying in the spinorial representation {\bf 4} of the flavour group
Spin(6)=SU(4). A basis for this representation can be taken to be $(\vac
,\psi _i ^{\dagger}\psi _j ^{\dagger}\vac )$ where the $\psi _i ^{\dagger}$,
$1\leq i\leq\nf =3$, are the fermionic zero modes carrying flavour 
indices and one unit of $s_i$ charge.\foot{A 
semiclassical analysis is valid for studying the
flavour symmetry properties of the states becoming massless
even at strong coupling because these states can be continuously transported
to weak coupling by varying $u$.}
With this convention, all states with $n_m=1$ and odd electric charge will
lie in the same spinorial representation {\bf 4},
while states with $n_m=1$ and even
$n_e$ will lie in the other, complex conjugate, spinorial representation
${\bf 4}^*$. The second singularity is due to a $n_m=2$ flavour singlet
dyon. When $n_m=2$, the number of fermionic zero modes is doubled and one can
indeed construct SU(4) singlets from the ${\bf 4}\otimes {\bf 4}^*$ tensor
product. The ket associated with $(1,2)$ is then of the form
$$ |\psi\rangle = \Bigl( 
\psi ^{a \dagger}_1 \psi ^{a \dagger}_2 \psi ^{a \dagger}_3
+ \psi ^{a \dagger}_1 \psi ^{b \dagger}_2 \psi ^{b \dagger}_3
+ \psi ^{b \dagger}_1 \psi ^{a \dagger}_2 \psi ^{b \dagger}_3
+ \psi ^{b \dagger}_1 \psi ^{b \dagger}_2 \psi ^{a \dagger}_3\Bigr)  \vac
\eqn\dxvii $$
with $1\leq a,b\leq 2$ and $a\not =b$. Let $s_0$ be the $s=\sum s_i$ charge of
$\vac $. When $m>0$, the flavour symmetry group is broken from SU(4) to
SU(3). We see that the $(1,1)$ singularity will then split into one
SU(3) singlet of $s$ charge $s_0$, $(1,1)_{s_0}$, and one SU(3) vector of $s$
charge $s_0 + 2$, $(1,1)_{s_0 + 2}$. As for the $(1,2)$ singularity, is must
be a SU(3) singlet of $s$ charge $s_0 + 3$. At the expense of shifting the
variables $a_D$ and $a$ by  integer multiples of $m/\rd$, that is to say
choosing the cycles $(\g _1,\g _2)$ in \dv\ so that they encircle the poles
with non zero residues of the Seiberg-Witten differential $\lambda $ an
appropriate number of times, we can always choose $s_0 =
0$ as in Fig. 2, and the singularities are $(1,1)_0$, $(1,1)_2$ and
$(1,2)_3$. This is a natural choice because for example the $(1,1)_{0}$ state
must remain stable and  of finite mass for any $u$
when $m\rightarrow\infty$. This choice of $s$ charges will then insure that
the solution $(a_D,a)$ for the $\nf =3$ theory will flow smoothly towards the
solution for the pure gauge theory, without picking any (infinite) shift
proportional to $m$. When $m=\lt /8$, we reach a $k=3$ superconformal point
where the triplet $(1,1)_2^{\times 3}$ and the singlet $(1,2)_3$ cross.
Using the formula \dxv\ with $U=M_{(1,1)_2^{\times 3}}^{-1}+ M_{(1,2)_3}$,
we see that the singlet $(1,2)_3$ becomes the dyon $(-1,1)_0$ and the triplet
$(1,1)_2$ becomes the quark $(1,0)_1$. This is exactly what one would expect
on physical grounds. Note that we could have deduced that the $s$ charge of
$(1,2)$ must be $3$ independently of the previous semiclassical reasoning
by requiring that the elementary quark must have $s=1$.

Finally, the third case we have depicted in Fig. \figiia\ is
the flow of the $\nf=2$ theory
with equal bare masses towards the pure gauge theory. This case will be
extensively studied in Section 4. In particular, we will determine the
existence domains of all the BPS states along this flow. 
In the figure, we have also indicated
the quantum numbers of the singularities as viewed from the upper half
$u$-plane. We have a $k=2$
superconformal point when $m=\ld /2$. In the present case, the singlet
and doublet collide but do not cross each other and we are in the situation of eq.
\dxiv\ with $U=M_{\rm sc}^{-1}=\bigl( M_{(0,1)_0} M_{(1,1)_1} \bigr)^{-1}$.
Note that the upper left block of $U$, acting on $(n_e,n_m)$ is nothing but the matrix
$-S=\pmatrix{0&-1\cr 1&0\cr}$. Note also
that our choice of $s$ charges is not in this case the most
natural from the point of view of the RG flow. We prefered to make easier the
implementation of $CP$ invariance which we will use when determining the BPS
spectra, by choosing a solution $(a_D,a)$ satisfying
$$ a_D(\overline u) = - \overline a_D (u) ,\quad a(\overline u) =\overline a
(u).\eqn\dxviii$$
Then the transformation law of the quantum numbers under $CP$ is simply
$$ (n_e, n_m)_s \buildrel\hbox{CP}\over{\longrightarrow} 
(-n_{e},n_{m})_{-s}.\eqn\dxix$$
\section{\bf {The computation of the period integrals}}
In the following, we will need  explicit expressions for the periods 
$a_{D}$ and $a$, in order to compute the curves of marginal stability 
which determine the existence domains of the stable BPS states. A
general method, first introduced and used in [\FERI], and also 
independently in [\LAG], simply consists in expanding the period integrals 
$a_{D}$ and $a$
in terms of the three fundamental elliptic integrals, which can be 
expressed in terms of standard special functions well suited for a 
numerical computation. An efficient way of taking care of the precise 
definition of the cycles $(\g _{1},\g_{2})$ is to uniformize the 
cubics \di\ with the help of the Weirstra\ss\ $\wp$ function. 
For a cubic curve in Weierstra\ss\ normal form, i.e.
$$\eta^2=4\prod_{i=1}^3 (\xi-e_i) \ , \quad \sum_{i=1}^3 e_i=0
\eqn\dxx$$
the three fundamental elliptic integrals are
$$I_1^{(j)} = \oint_{\g_j} {\d\xi\over \eta} \ \raise 2pt\hbox{,} \quad
I_2^{(j)} = \oint_{\g_j} {\xi\d\xi\over \eta} \ \raise 2pt\hbox{,} \quad
I_3^{(j)}(c) = \oint_{\g_j} {\d\xi\over \eta (\xi-c)}\cdotp
\eqn\tvi$$
By convention, 
we choose the cycle $\g_1$ to encircle $e_3$ and $e_2$, and the
cycle
$\g_2$ to encircle $e_1$ and $e_2$. Of course, we still need to specify
how we
choose to number the three roots $e_i$ in each particular case.
Sometimes, if we
do not specify the contours, we simply write $I_1$ for $\int {\d\xi\over
\eta}$ etc.
\vskip 2.mm
\fig{The definition of the basic cycles $\g_1$ and $\g_2$
with respect to the roots $e_1$, $e_2$ and $e_3$.}{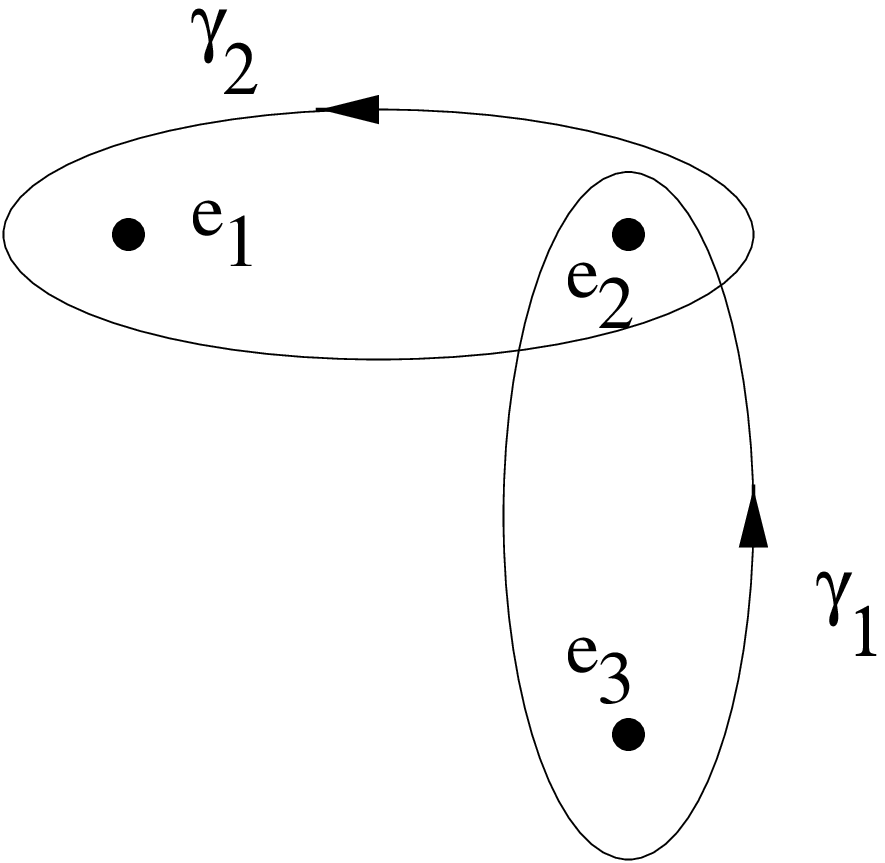}{5cm}
\figlabel\figiii
\vskip 2.mm
The necessity to introduce the elliptic integral of the third kind $I_3$ is
due to the
presence of poles with non-zero residues
in the massive theories. This also explains the failure of the 
standard methods of computation of the periods by means of the 
Picard-Fuchs equations. For the massless theories only
$I_1$ and $I_2$ occur which can be reexpressed in terms
of hypergeometric functions.

Let us focus on the case $\nf =2$ with $m_{1}=m_{2}=m$. 
The other cases are discussed in 
Appendix C. The Seiberg-Witten differential 
$\l \equiv \l^{\nf =2}_{m_{1}=m_{2}=m}$ is given by
$$\eqalign{
\l&=-{\rd\over 4\pi} {y\, \d x\over x^2-{\ld^4\over 64}}
=-{\rd\over 4\pi} {\d x\over y} {4 y^2\over \ld^2}
\left( {1\over x-{\ld^2\over 8}} -  {1\over x+{\ld^2\over 8}} \right) \cr
&=-{\rd\over 4\pi}  {\d x\over y}
\left[ x-u+{\ld^2\over 4}  {m^2\over x+{\ld^2\over 8}} \right] \ .}
\eqn\dxxi$$
Converting to Weierstra\ss\ normal form of the cubic by
$\eta=2y,\ \xi=x-{u\over 3}$ we arrive at
$$\oint_{\g_i}\l = {\rd\over 4\pi}
\left[ {4\over 3} u\,  I_1^{(i)} -2\, I_2^{(i)}
- {\ld^2\over 2} m^2\, I_3^{(i)}\left(-{\ld^2\over 8}-{u\over 3}\right)
\right] \ . \eqn\dxxii$$
One sees from \dxxi\ that $\l$ has two poles at $(x={-\ld^2\over 
8},y=\pm {i\over 4}m\ld ^{2})$
with residues $\mp {1\over 2\pi i} {m\over 2\rd}$ [\SWII].

In order to have an explicit formula for $a_{D}$ and $a$, we now have 
to choose the roots $e_{j}$ of the polynomial defining the cubic. One 
constraint comes from the asymptotic behaviour for large $|u|$, which 
is governed by asymptotic freedom:
$$a(u)\sim {1\over 2}\sqrt{2u} \ , \quad
\ad (u)\sim {i\over 2\pi} \sqrt{ 2u}\log {u\over \Lambda^2_{2}}\cdotp
\eqn\dxxiii$$
Since $a(u)$ is given by the integral over the cycle $\g_1$ surrounding
$e_2$ and $e_3$,
see Fig 3, the {\it set} $\{ e_2, e_3\}$ must be chosen such that the
large-$u$
asymptotics of $a$ does {\it not} contain a $ \sqrt{ 2u} \log u/
\Lambda^2$ term. The remaining
root necessarily is $e_1$. But which root in the set $\{ e_2, e_3\}$ is
called $e_2$, and
thus is encircled also by $\g_2$, is a matter of convention related 
to the possibility of shifting $a_{D}$ by an integer multiple of $a$.
A correct choice is the following
$$\eqalign{
e_1&={u\over 6} -{\ld^2\over 16}
+{1\over 2} \sqrt{u+{\ld^2\over 8} +\ld m} \sqrt{u+{\ld^2\over 8} -\ld 
m}\ ,\cr
e_2&=-{u\over 3}+{\ld^2\over 8} \ ,\cr
e_3&={u\over 6} -{\ld^2\over 16}
-{1\over 2} \sqrt{u+{\ld^2\over 8} +\ld m} \sqrt{u+{\ld^2\over 8} -\ld 
m}\ .}\eqn\dxxiv $$
One can then straightforwardly show (see Appendix D)
that in the limit $m\rightarrow 0$
the solution given by \dv\ and \dxxii\ converges toward the already 
known explicit solution of the massless $\nf =2$ theory [\BF] which also
corresponds to the electric and magnetic quantum numbers at the singularities
chosen in Fig. 2. It 
still remains to fix the positions of the cycles relatively to the 
poles with nonzero residues of $\l$, or, which is equivalent, to fix the
$s$ charges at the singularities. The solution, consistent with Fig. 2, is
given for $\IM u>0$ by
$$\eqalign{
a(u) & = {\rd\over 4\pi}
\left[ {4\over 3}\, u I_1^{(1)} -2\,  I_2^{(1)}
- {\ld^2\over 2} m^2\,  I_3^{(1)}\left(-{\ld^2\over 8}-{u\over 3}\right)
\right] + {m\over \rd} \cr
\ad(u) & = {\rd\over 4\pi}
\left[ {4\over 3}\,  u I_1^{(2)} -2\,  I_2^{(2)}
- {\ld^2\over 2} m^2\,  I_3^{(2)}\left(-{\ld^2\over 8}-{u\over 3}\right)
\right] }
\eqn\dxxv$$
with the $I_j^{(1)}$ precisely given by the formulae 
$$\eqalign{
I_1^{(1)} = 
& {2\over (e_1-e_3)^{1/2}} \, K(k)  \cr
I_2^{(1)} =
& {2\over (e_1-e_3)^{1/2}} \, \left[ e_1 K(k) +(e_3-e_1) E(k) \right] \cr
I_3^{(1)} =
& {2\over (e_1-e_3)^{3/2}} \, \Bigg[ {1\over 1-\tilde c +k'}\, K(k)\cr
& \phantom{XXX}+ {4 k'\over 1+k'}\, {1\over (1-\tilde c)^2- k'^2}\,
\Pi_1\left( \nu(c), {1-k'\over 1+k'}\right) \Bigg]  }
\eqn\txxiv$$
where
$$\eqalign{
k^2&={e_2-e_3\over e_1-e_3} \ \raise 2pt\hbox{,} \quad
k'^2=1-k^2 = {e_2-e_1\over e_3-e_1} \ \raise 2pt\hbox{,}\cr
\tilde c &= {c-e_3\over e_1-e_3} \ \raise 2pt\hbox{,} \quad
\nu(c) = - \left( {1-\tilde c+k'\over 1-\tilde c -k'} \right)^2
\left({1-k'\over 1+k'}\right)^2 \ \raise 2pt\hbox{,}}
\eqn\txxv$$
and the $I_j^{(2)}$ obtained form the $I_j^{(1)}$ by exchanging $e_1$ and
$e_3$. In \txxiv , $K$, $E$ and $\Pi _1$ are the three standard elliptic
integrals of [\ERD] whose integral representations are  given in Appendix B.
For $\IM u<0$ the solution then is obtained from \dxviii . In  Appendix B
we derive eqs. \txxiv\ as well as some useful relations between the
elliptic integrals. In Appendix D, we
discuss in some detail how one can understand the RG flow 
illustrated in Fig. \figiia\ directly from the explicit formulae \dxxv .
In particular, as $m\to\infty$, $m\ld=\lz^2$ fixed, we have
$$a(u) \to  a^{(0)}(u)\quad ,\quad  \adt(u) \to \ad^{(0)}(u)
\quad ,\quad {\rm where}\quad 
\adt(u) = \ad(u) - \e  a(u)  + \e  {m\over \rd} \ .
\eqn\basis$$
Here and in the following we always let $\e={\rm sign}(\IM u)$. 
The redefinition $\ad\to\adt$ 
corresponds to rotating the cut originating from
the massless quark singularity $\s_3$ to the right 
(and changing the contour $\g_2$). This is what one wants since this cut must 
disappear in the $m\to\infty$ limit in order to recover the standard
analytic structure of the pure gauge theory.
%
{\bf \chapter{The spectra of stable BPS states}}
\sectionnumber=0

\section{\bf  Decay curves}

In general, a given BPS state does not exist everywhere on the Coulomb
branch of the moduli space. For each of the {\it massless} theories with
$N_f\le 3$
[\BF,\FB] there is a single curve of marginal stability which goes through
the
singularities and separates the Coulomb branch into two regions. In the
region outside this
curve all semiclassically stable states exist, while inside the curve only
those BPS states
exist that are responsible for the singularities, in addition to the photon
vector multiplet.

The present cases of hypermultiplets with non-vanishing bare masses are very
different. Due
to the BPS mass formula $M_{\rm BPS}=\rd \vert Z\vert$ with
$$Z(u)=n_m \ad(u) - n_e a(u) +\sum_{i=1}^{N_f} s_i {m_i\over \rd}
\eqn\qi$$
a BPS state is stable against any decay of the type
$$
(n_e,n_m)_{s_i} \to k \times (n_e',n_m')_{s_i'} + l \times
(n_e'',n_m'')_{s_i''}
\eqn\qii$$
($k,l \in \Z$) unless  this satisfies at the same time the conservation of
charges and of
the total BPS mass:
$$n_e=k n_e'+l n_e '' \quad , \quad
n_m=k n_m'+l n_m '' \quad , \quad
s_i=k s_i'+l s_i'' \quad \Rightarrow \quad Z=k\, Z'+l\, Z''
\eqn\qiii$$
and
$$\vert Z \vert= \vert k\,  Z'\vert + \vert l\,  Z''\vert
\eqn\qiv$$
with obvious notations for $Z'$ and $Z''$. If all bare masses $m_i$ are
equal, due to the
${\rm SU}(N_f)$ flavour symmetry,
only the sum $s=\sum_i s_i$ is relevant and needs to be conserved. We
see
that a decay that satisfies the charge conservations \qiii\ is possible
only if
$${Z'\over Z} \equiv \zeta \in \R \ ,
\eqn\qv$$
and moreover if it is kinematically possible, i.e. if
$$0\le k \zeta \le 1 \ .
\eqn\qvi$$
For the case of vanishing bare masses, $m_i=0$ condition \qv\ reduces to
$\IM
{\ad(u)\over a(u)}=0$ which yields a single curve \C$^0$ on the Coulomb
branch
independent of the initial state $(n_e, n_m)_{s_i}$ considered. For
non-vanishing bare
masses however, we have a whole family of possible decay curves. Moreover,
a priori,
there is a different family of such curves for each BPS state. As an
example consider a
dyon with $n_m=1$. Then condition \qv\ reads
$$
\IM\ {n_m' \ad - n_e' a +\sum_i s_i' {m_i\over \rd} \over
\ad - n_e a +\sum_i s_i {m_i\over \rd} } = 0
\ \Leftrightarrow \
\IM\ {-(n_e'-n_m' n_e) a + \sum_i (s_i'-n_m' s_i) {m_i\over \rd} \over
\ad - n_e a +\sum_i s_i {m_i\over \rd} } = 0
\ .
\eqn\qvii$$
For fixed $n_e$ and $s_i$, this is an $N_f$-parameter family of curves
with rational parameters
$r_i=(n_e'-n_m' n_e)/(s_i'-n_m' s_i)$. Even though there are some
relations between the
possible quantum numbers $n_e'$ and $s_i',\, n_m'$ (see next subsection)
there are still
many possible values of $r_i$ and we expect a multitude of curves of
marginal stability
on the Coulomb branch of moduli space resulting in a rather chaotic
situation.
Fortunately not all of these curves satisfy the additional criterion
\qvi.
In particular, for
the case we will study in detail in the next section, namely $N_f=2$ with
equal bare
masses, where one expects a different one-parameter family of curves
labelled by
$r=(n_e'-n_m' n_e)/(s'-n_m' s)$, $s=s_1+s_2$, for {\it each} BPS
state, it turned
out that only one or two such curves in each family are relevant, i.e.
satisfy the
additional criterion \qvi. Hence the set of all relevant curves for {\it
all} BPS states
are nicely described by a single set of curves ${\cal
C}_{2n}^\pm,\ n\in\Z$, and rather
than having a chaotic situation one gets a very clear picture of which
states exist in
which region of the Coulomb branch.
It will become clear that this organizing scheme should be similarly at
work for all other massive theories with $N_f\le 3$.

One particularly simple case is the decay of states with $\sum s_i m_i =0$
into states
with $\sum s_i' m_i =0$. The corresponding decay curves all are
given by $\IM {\ad(u)\over a(u)}=0$, i.e. they all coincide with the curve
\C$^0$. We
will see that this is quite an important case, and that this curve \C$^0$
still plays a
priviledged r\^ole, even for non-zero bare masses.\foot{
This statement is of course meant for the choice of $\ad$ and $a$ that gives finite
limits under the RG flow as $m\to\infty$. For $N_f=2$ with equal masses, these are the
$\adt$ and $a$ of eq. \basis\ and $\sum s_i m_i=\sum s_i' m_i=0$ is meant to be
$\st=\st'=0$.}

Note that if we had considered decays into three independent BPS states,
$(n_e,n_m)_{s_i} \to k \times (n_e',n_m')_{s_i'} + l \times
(n_e'',n_m'')_{s_i''}
+ q \times (n_e''',n_m''')_{s_i'''}$, we would have {\it two }
conditions: eq. \qv\ would be
supplemented by ${Z''\over Z}\in \R$, so that such ``triple" decays can
only
occur at the intersection {\it points} of two curves. Below, when we
discuss how to
transport a BPS state along a path from one region to another, the path
can
always be chosen so as to avoid such
intersection points. Hence, triple decays  are irrelevant for establishing
the existence domains of the BPS states. Obviously, ``quadruple" and
higher decays, if possible at all, are just as irrelevant.

\section{\bf Our working hypothesis}

In order to determine the BPS spectra at
any point on the Coulomb branch, we will extensively use the
following claim:\hfill\break

{\bf P}: {\it At any point of the Coulomb branch of a theory having $N_f$ flavours with
bare masses $m_j$, $1\leq j\leq N_f$, the set of stable BPS states is included
into the set of stable BPS states of the $m_j=0$ theory at weak coupling.}

Note that the Coulomb branch of the $m_j=0$ theory is separated into
two regions, one containing all the BPS states stable at weak coupling, and
the other at strong coupling containing a finite subset of the BPS states
stable at weak coupling [\FB ,\BF]. One simple consequence of the 
claim (P)
is that the set of stable BPS states cannot enlarge when one goes from the
$N_f$ to the $N_f -1$ theory following the RG flow 
which is what one naturally expects. This is perfectly
consistent with the spectra determined for zero bare masses in [\FB ,\BF].
Another consequence, which plays a prominent r\^ole in the present work, is
that the possible decay reactions between BPS states are then extremely
constrained and thus the number of relevant
curves of marginal stability enormously decreased. This is explained in
detail for the $N_f=2$ theory with two equal bare masses $m_1=m_2=m$ in the
next section.

%
%

Let us give a strong argument motivating our fundamental claim (P). It is
inspired of ideas already discussed in [\FERII]. A crucial fact is that,
for any BPS state $p=(n_e,n_m)_s$ which does not belong to the weak coupling
spectrum of the $m_j=0$ theory, there always exist some values of the $m_j$
for which
$$ n_m a_D (u_0,m_j) -n_e a(u_0,m_j) + s {m\over\sqrt{2}} =0 \eqn\zeros$$
at some non-singular 
$u_0$ on the Coulomb branch. This may be proven by noting that for
sufficiently large $|m_j|$, the point $u_0$, if it exists, lies at large
$|u|\sim |m_j|$ where the formulas for $a_D$ and $a$ simplify and thus
where \zeros\ can be studied very explicitly. The set of 
curves of marginal stability $\cup _{r\in\Q} {\cal C}_p(r,m_j)$
a priori relevant for the decays of $p=(n_e,n_m)_s$, all cross at the point
$u_0$ and form a dense subset of the Coulomb branch.
The same properties are true for the complementary set $\cup _{r\in
(\R\setminus\Q)} {\cal C}_p(r,m_j)$. Now, suppose that $p$
exists in some open set of the Coulomb branch. It must then exist at some
points on a curve ${\cal C}_p(r,m_j)$ with $r$ an irrational number.
Since the state $p$ cannot decay on such a curve for $r$
irrational, it must also exist at the point $u_0$ where it is massless, which
would contradict the fact that $u_0$ is not a singular point. Thus $p$ cannot
exist as a stable state in any open region 
for the values of $m_j$ such that \zeros\ has a solution. Now, by 
varying the $m_{j}$, the curves ${\cal C}_p(r,m_j)$ loose their shape 
and will no longer cross at a single point , but we are 
still insured that $p$ cannot exist on any ${\cal C}_p(r,m_j)$ with 
$r$ irrational. We believe on physical grounds that the fact that $p$ 
cannot exist in any open region of the Coulomb branch means that $p$ 
simply cannot exist at all as a stable BPS state. Finally, note that 
instead of studying eq. \zeros , we could remark
that for arbitrarily small $|m_{j}|$, the spectra of BPS states should
be the same as the one for the $m_j=0$ theory, which proves that an 
undesirable state like $p$ cannot exist on ${\cal C}_p(r,m_j)$ for 
$r$ irrational and sufficiently small $|m_{j}|$, and thus for any 
$m_{j}$. However, this reasoning certainly is less rigorous than the 
one based on \zeros . To end this section, let us point out that 
the claim (P) is also strongly supported by the stringy approach used 
in [\WAR]. 
{\bf \chapter{The case of two flavours with equal masses}}
\sectionnumber=0
As an illustrative and representative example,  in this 
section we study 
the BPS spectra of the $N_{f}=2$ theory with any $m_{1}=m_{2}=m$ real 
positive bare mass in great detail. The analytic structures for small 
($0<m<\ld /2$) and large ($m>\ld /2$) bare masses is displayed in 
the lower Fig. 
2, while the explicit solution for the periods $a_{D}$ and $a$ is 
given by equations \dxxv\ . The three singularities on the 
Coulomb branch are at points $u=\sigma _{j}$, $\sigma _{1}\leq\sigma 
_{2}\leq\sigma _{3}$, such that
$$ \sigma _{1}=-{\ld ^{2}\over 8}-\ld m \ , \quad
\sigma _{2}=-{\ld ^{2}\over 8}+\ld m \ , \quad
\sigma _{3}=m^{2}+{\ld ^{2}\over 8}\cdotp \eqn\sing $$
When $m=\ld /2$, the singularities $\sigma _{2}$ and $\sigma _{3}$ coincide.
For small mass, $0<m<\ld /2$, the monodromies $M^{*}$ (see \dviii )
as viewed from the lower half $u$-plane are given by
$$M_1^*=\pmatrix{ \hfill 0&\hfill 1&0\cr -1&\hfill 2&0\cr\hfill
 1&-1&1\cr} \ , \quad
M_2^*=\pmatrix{\hfill 0&1&0\cr -1&2&0\cr -1&1&1\cr}\ ,\quad
M_3^*=\pmatrix{\hfill 1&0&0\cr -2&1&0\cr \hfill 0&0&1\cr}\ ,
\eqn\monodp$$
while for $m>\ld /2$,
$$M_1^*=\pmatrix{ \hfill 0&\hfill 1&0\cr -1&\hfill 2&0\cr\hfill
 1&-1&1\cr} \ , \quad
M_2^*=\pmatrix{\hfill 2&1&0\cr -1&0&0\cr \hfill 1&1&1\cr}\ ,\quad
M_3^*=\pmatrix{1&2&0\cr 0&1&0\cr 0&2&1\cr}\ .
\eqn\cix$$
The monodromy at the superconformal point, according to \dx,  is given by
$$ M_{\rm sc}^{*}=M_{2\cdot 3}^{*}=M_{3}^{*}M_{2}^{*}=
\pmatrix{\hfill 0&1&0\cr -1&0&0\cr -1&1&1\cr}\eqn\monodsc$$

Before embarking on the detailed derivation of the BPS spectra, let us first discuss
what one expects from the RG flow arguments. For $m=0$ there is a single decay curve, and
outside this curve all semiclassical states exist, namely all dyons $(n,1)$ and the
W-boson $(2,0)$ as well as the quarks $(1,0)$. The dyons are doublets in one or the other
spinor representation of the flavour ${\rm spin}(4)$ group, while the quarks are in the
vector representation and the W-boson is a singlet. Inside the curve, only the states that
can become massless and are responsible for the singularities exist, namely the monopole
$(0,1)$ and the dyon $(-\e,1)$. As soon as a non-zero bare mass $m$ is turned on, the
$s$-charge becomes relevant, and certain multiplets split. Semiclassically we still have
doublets of dyons $(2n,1)_0$ with even $n_e$ and $s=0$, while the doublets with odd $n_e$
split into two singlets $(2n+1,1)_{+1}$ and $(2n+1,1)_{-1}$. We have a quark doublet
$(1,0)_1$ and another quark doublet $(1,0)_{-1}$, as well as the W-boson $(2,0)_0$. According 
to our claim (P), this is the maximal set of stable BPS states:
$${\cal S}_{\rm max}=\left\{ (2n,1)_0^{\times 2},\ (2n+1)_{\pm 1},\ 
(1,0)_{\pm 1}^{\times 2},\ (2,0)_0 \right\} \ .
\eqn\smax$$
In the opposite limit, $m\to\infty$, $m\ld=\lz^2$ fixed, we expect to flow to the 
spectrum of the pure gauge theory. To describe this limit conveniently, we should move the
cut originating from the massless quark singularity $\s_3$ to the right so that it
disappears in the $m\to\infty$ limit. As explained at the end of Section 2, we can then
choose to work with $a$ and $\adt$ instead, which are the quantities that flow to
$a^{(0)}$ and $\ad^{(0)}$. We call the corresponding quantum numbers $\net$, $\nmt$ and
$\st$:
$$n_m=\nmt\quad , \quad n_e=\net + \e\, n_m \quad , \quad s=\st +\e\, n_m \ ,
\eqn\nqn$$
where we recall that $\e={\rm sign}(\IM u)$. Hence, in the $m\to\infty$ limit, we expect
to have semiclassically the W-boson $(\net,\nmt)_\st = (2,0)_0$ as well as the dyons 
$(\net,\nmt)_\st = (2k,1)_0$. Converting back to the $(n_e,n_m)_s$ we always use in this
paper, these are $(2,0)_0$ as well as the dyons $(2n+1,1)_\e$ : only dyons of {\it odd}
$n_e$ and $s=1$ in the upper or $s=-1$ in the lower half plane should survive the RG flow
to $N_f=0$ in the weak coupling region. All other states must disappear. There are two
possibilities. A state may simply drop out of the spectrum since its BPS mass diverges as
$m\to\infty$ as is the case of all states with $\st\ne 0$, i.e $s\ne \e\, n_m$. 
But a state can also disappear
at a point $u$ (kept fixed ), already at finite $m$, because it is ``hit" by its corresponding
decay curve which moves outwards as $m$ is increased. We will see below that this latter
possibility is realised for all dyons $(2n,1)_0,\ n\ne 0$, as well as for the the dyons
$(2n+1)_{-\e},\ n\ne -1,0$. The remaining undesired states, namely $(0,1)_0,\ (\pm
1,1)_{-\e}$ and $(1,0)_{\pm 1}$ simply disappear because their BPS masses diverge.

\section{\bf Possible decay reactions and decay curves}

In this section, we explain how the claim (P) stated in Section 3.2 
drastically restricts the number of possible decay reactions and 
decay curves on which they may occur. 
We will only use the basis of $a$ and $\ad$ and the
corresponding quantum numbers $(n_e,n_m)_s$. Recall that the BPS mass  is
$M_{\rm BPS}=\rd \vert n_m \ad -n_e a + s {m\over \rd}\vert$. As discussed above, the
maximal set of BPS states then consists of ${\cal S}_{\rm max}$ 
(as well as their antiparticles). These
states also constitute the semi-classical spectrum. Obviously, if at some point 
$u \in {\cal M}$ a
BPS state decays into two other states, the latter must be BPS states and must be
contained in this maximal spectrum ${\cal S}_{\rm max}$. 
As in [\BF] one may also check that matching of the flavour quantum numbers
does not  give rise to any new constraints: the flavour quantum numbers are related
to $n_e$ and $s$ in such a way that they match automatically if $n_e$ and $s$ do.
The following
results do not depend on whether the bare mass $m$ is larger or smaller than
${\ld\over 2}$. For reasons mentioned at the end of Section 3.1,
we will only need to consider 
decays into {\it two} types of BPS states:
$$(n_e, n_m)_s \ \to \ k\times (n_e', n_m')_{s'} + l\times (n_e'', n_m'')_{s''} \ .
\eqn\cxi$$
The quantum numbers $n_m',
n_m''$ always are either $0$ or $1$, since we can always choose $n_m',
n_m'' \ge 0$, i.e. a state $(n_e,-1)_s$ is written as $(-1)\times (-n_e,1)_{-s}$.
One has to establish the possible decays of each type of BPS state separately. 

\def\table#1#2{
\par\begingroup\parindent=0pt\leftskip=0.3cm\rightskip=1cm\parindent=0pt
\baselineskip=11pt
\midinsert
\epsfxsize=#2
\centerline{\epsfbox{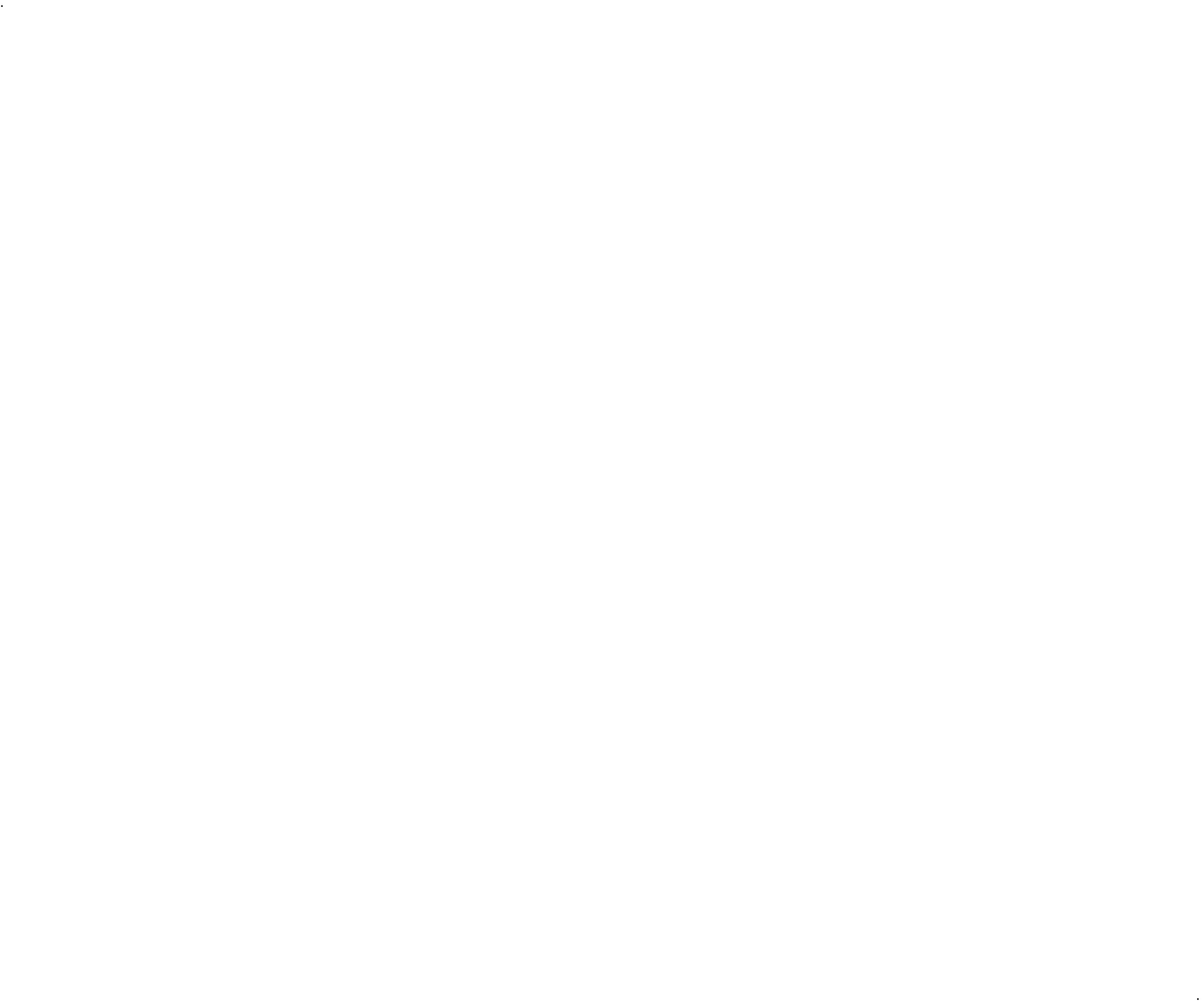}}
\vskip 12pt
\centerline{\bf Table 1:} \par
\vskip .5cm
#1\par
\vskip .5cm
\endinsert\endgroup\par
}
\def\encadremath#1{\vbox{\hrule\hbox{\vrule\kern8pt\vbox{\kern8pt
\hbox{$\displaystyle #1$}\kern8pt}
\kern8pt\vrule}\hrule}}

\table{
$\matrix{
{\rm initial\ state}    \hfill& {\rm decay\ products}                  		\hfill	&   			\hfill\cr
---			\hfill& ---						\hfill	& ---			\hfill\cr
{\rm W-boson} \ (2,0)_0 \hfill& (1,0)_1 + (1,0)_{-1}				\hfill	& r_{\rm w}=0		\hfill\cr
\IM {r_{\rm w}\ad+{m\over\rd}\over a}=0
			\hfill& (2p+2,1)_0 + (-1)\times (2p,1)_0		\hfill	& r_{\rm w}=\infty	\hfill\cr
r_{\rm w}=n_m'/s'	\hfill& (2p+1,1)_{\pm 1} + (-1)\times (2p-1,1)_{\pm 1}	\hfill	& r_{\rm w}=\pm 1	\hfill\cr
---			\hfill& ---						\hfill	& ---			\hfill\cr
{\rm quarks}\ (1,0)_{\pm 1}\hfill&(2,0)_0 + (-1)\times (1,0)_{\mp 1} 		\hfill	& r_{\rm q}=0		\hfill\cr
\IM {r_{\rm q}\ad+{m\over\rd}\over-a\pm{m\over\rd}}=0
			\hfill& (2p,1)_0 + (-1)\times (2p-1,1)_{\mp 1}		\hfill	& r_{\rm q}=\mp{1\over 2p}\hfill\cr
r_{\rm q}=n_m'/(s'\mp n_e')	
			\hfill& (2p+1,1)_{\pm 1} + (-1)\times (2p,1)_0 		\hfill	& r_{\rm q}=\mp{1\over 2p}\hfill\cr
---			\hfill& ---						\hfill	& ---			\hfill\cr
{\rm dyons}\ (2n,1)_0	\hfill& k\times (2,0)_0 + (2n-2k,1)_0			\hfill	& r_{\rm d}=\infty	\hfill\cr
			\hfill& (1,0)_{\pm 1} +(2n-1,1)_{\mp 1} 		\hfill	& r_{\rm d}=\pm 1	\hfill\cr
\IM {r_{\rm d} a-{m\over\rd}\over\ad-2n a}=0
			\hfill& (-1) \times (1,0)_{\pm 1} +(2n+1,1)_{\pm 1} 	\hfill	& r_{\rm d}=\pm 1	\hfill\cr
			\hfill& k\times (2p,1)_0 + (1-k)\times (2q,1)_0  	\hfill	&			\hfill\cr
r_{\rm d}=(n_e'-n_e n_m')/s'	\hfill& \quad {\rm with}\ n=kp+(1-k)q 		\hfill	& r_{\rm d}=\infty	\hfill\cr		
---			\hfill& ---						\hfill	& ---			\hfill\cr
{\rm dyons}\ (2n+1,1)_{\pm 1}
			\hfill& k\times (2,0)_0 +(2n-2k+1,1)_{\pm 1}		\hfill	& r_{\rm d}=\infty	\hfill\cr
			\hfill& (1,0)_{\pm 1} +(2n,1)_0			 	\hfill	& r_{\rm d}=\pm 1	\hfill\cr		
\IM {r_{\rm d} a-{m\over\rd}\over\ad-(2n+1)a\pm{m\over \rd}}=0
			\hfill& 2\times (1,0)_{\pm 1} +(2n-1,1)_{\mp 1}		\hfill	& r_{\rm d}=\pm 1	\hfill\cr
			\hfill& (-1)\times (1,0)_{\mp 1} +(2n+2,1)_0 		\hfill	& r_{\rm d}=\mp 1	\hfill\cr
r_{\rm d}={n_e'-n_e n_m'\over s'-s n_m'}
			\hfill& (-2)\times (1,0)_{\mp 1} +(2n+3,1)_{\mp 1} 	\hfill	& r_{\rm d}=\mp 1	\hfill\cr
			\hfill& (-1)\times (4p-2n-1,1)_{\mp 1} + 2\times (2p,1)_0\hfill & r_{\rm d}=\pm (2n-2p+1) \hfill\cr
			\hfill& k\times (2p+1,1)_{\pm 1} +(1-k)\times (2q+1,1)_{\pm 1} \hfill&			\hfill\cr
			\hfill& \quad {\rm with}\ kp+(1-k)q=n			\hfill	& r_{\rm d}=\infty 	\hfill\cr
}
$}{.5 cm}
As an axample, we present the discussion of all possible decays for the dyons $(2n,1)_0$.
The corresponding decay curves are
$$0=\IM {n_m'\ad - n_e' a +s'{m\over \rd} \over \ad -2n a  }
 \quad \Leftrightarrow \quad 
0=\IM {r_{\rm d} a  -{m\over \rd}\over \ad -2n a }\quad , 
\quad r_{\rm d}={n_e'-2n n_m'\over s'} \  .
\eqn\cxxi$$
Either $n_m'=0$ or $n_m'=1$. Consider first $n_m'=0$. Then this is the W-boson or a
quark. In the first case we simply have
$(2n,1)_0\ \to \ k\times (2,0)_0 + (2n-2k,1)_0$. The value of $r$ for this decay is 
$\infty$.
In the second case we can have $k$ quarks $(1,0)_{\pm 1}$ and the dyon
$(2n-k,1)_{s''}$ with $s''=\mp
k$. Hence $\vert k\vert =1$ and we have two possibilities:
$(2n,1)_0 \ \to \ (1,0)_{\pm 1} +(2n-1,1)_{\mp 1}$ and
$(2n,1)_0 \ \to \ (-1) \times (1,0)_{\pm 1} +(2n+1,1)_{\pm 1}$ with $r=\pm 1$. 
Consider now $n_m'=1$ (and $n_m''=1$, otherwise we are back to the previous case). Then
$(2n,1)_0=k\times (n_e',1)_{s'} +l\times (n_e'',1)_{s''}$ implies $l=1-k$ and $k
s'+(1-k) s''=0$. Since $\vert s'\vert, \vert s''\vert =0$ or $1$ this implies $s'=s''=0$
(since the cases $k=0$ or $k=1$, i.e. $l=0$ must be excluded because they do not give decays).
Hence $n_e'$ and $n_e''$ are even and $n=k n_e'+(1-k) n_e''$. Finally we get
$(2n,1)_0\ \to \ k\times (2p,1)_0 + (1-k)\times (2q,1)_0 \quad {\rm with}\
n=kp+(1-k)q$ with $r=\infty$ again.

We have determined  the possible decays for all the other states, namely 
the W-boson $(2,0)_0$, the dyons 
$(2n+1,1)_{\pm 1}$ and the quarks $(1,0)_{\pm 1}$ in exactly
the same way. The results are collected in Table 1. This Table also shows the
equations that determine the curves corresponding to the decays.

As already noted above and as will be discussed in detail 
below, not all curves are relevant.
It will turn out that the only relevant curves are
$${\cal C}^\infty\quad : \quad \IM {a\over \e\ad +{m\over \rd}} = 0 
\qquad, \qquad
{\cal C}^{\pm}_n\quad : \quad \IM {a\pm {m\over \rd}\over \ad - n a} = 0 \ .
\eqn\cxxib$$

\section{\bf  Decay curves  and BPS spectra for small mass ($m<{\ld\over 2}$)}

\subsection{The general picture}

We are now ready to establish the exact existence domains for every BPS state. In this
subsection, we will discuss the case of small mass, i.e. $m<{\ld\over 2}$. Although it
is not too much different, it is more convenient to discuss the case of $m>{\ld\over 2}$
separately in the next subsection.  We will consider each type of BPS state
separately. We have seen
that for the W-boson and the dyons $(2n,1)_0$ we only need to consider three
curves\foot{
The curve $r=0$ for the W-boson or for the quarks is simply $\IM a=0$ which is the
half-line $[\s_3,\infty)$. Since all states exist in the semiclassical region
$u\to\infty$ in the upper and in the lower half plane, every point in \M can be reached
from this region without crossing $[\s_3,\infty)$. Thus the $\IM a =0$ curve is
irrelevant.}:
$r=\pm 1,\ \infty$. For the quarks and the dyons $(2n+1,1)_{\pm 1}$ we have a priori
infinitely many possible decay curves ($p\in \Z$), but it is clear that only a few
values of $p$ will correspond to kinematically possible decays: although ${Z'\over Z}
\in \R$ for all $p$, only for a few $p$ we will have the additional condition \qvi\
that $0\le {k\, Z'\over Z}\le 1$. 
We have computed all curves numerically using
Mathematica. Their exact shape of course depends on the value of $m/\ld$, but 
is not of much importance. Only their positions
relative to each other and to the singularities actually matter and there is no qualitative change as
long as $m<{\ld\over 2}$.
To check whether a given curve is relevant, i.e. whether $0\le {k\, Z'\over Z}\le 1$ so
that the decay is kinematically possible and 
can really happen, it is enough to proceed as follows: if at {\it some} 
point
on the curve the decay is kinematically impossible (which can be easily checked by
computing $Z'$ and $Z$ numerically at this point) one can always transport the BPS state
through the curve at this point where it cannot decay, and thus the curve is
irrelevant.\foot{
Here we actually use the fact that the family of curves for a {\it given}
state is such that the curves do not cross each other, except possibly on the real axis.
Note that we always consider the parts of a curve in the lower and upper half plane separately.} 
On the other hand, to show that a curve is relevant we must make sure that the
decay is possible at any point of the curve. Since the real-valued function ${k\, Z'\over Z}$
varies smoothly along the curve, it is easy to see which real interval is its image and
whether it is entirely contained within $[0,1]$.

\vskip 2.mm
\fig{Shown are a sketch of the relative positions of the relevant decay curves 
for $m<{\ld\over 2}$ (for not too 
large $\vert n_e\vert$) as well as
the BPS states that decay across these curves. Three states do not decay
anywhere and still are present in the innermost region inside ${\cal
C}_0^-$. They are described as $(0,1)_0$ and $(-1,1)_{\pm 1}$ in the upper, 
and as $(0,1)_0$ and $(1,1)_{\pm 1}$ in the
lower half plane.
Note that, in reality, the angles at which the curves meet the real axis at the
points $x_k$ are slightly different from what they appear to be in
the Figure: indeed, the curves ${\cal C}^-_{-k-2}$, resp.  ${\cal C}^+_{k}$,
in the upper half plane are the smooth continuations of the curves 
${\cal C}^+_{-k}$, resp.  ${\cal C}^-_{k+2}$, in the lower half plane,
in agreement with the monodromy around infinity.}{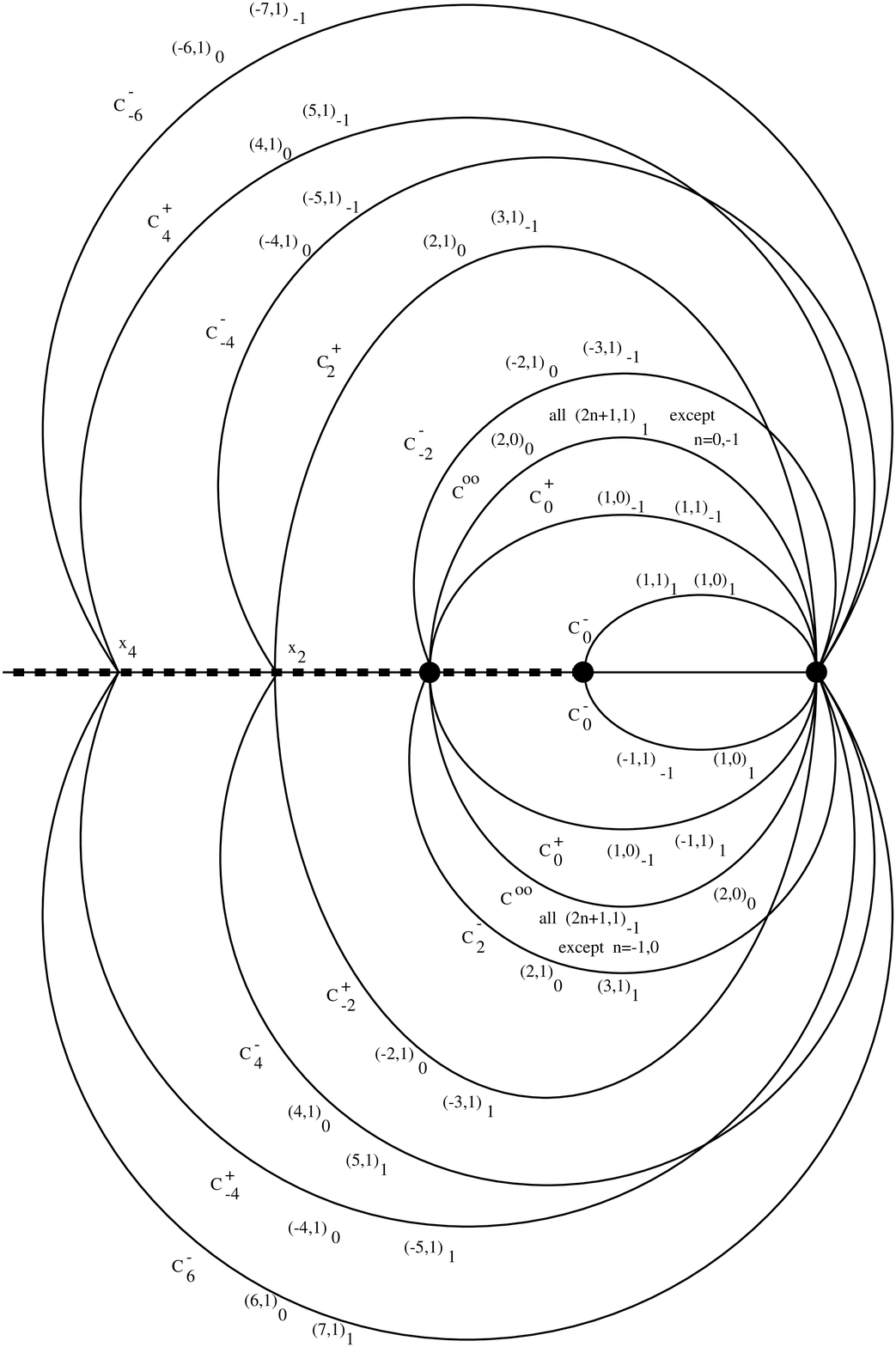}{13cm}
\figlabel\figxviii 
\vskip 2.mm
Let us already present the results of the analysis that will be given below.
We have assembled all the {\it relevant} decay curves into Fig. \figxviii\ that sketches their
relative positions and indicates the BPS states that decay across these curves.
All curves go through $\s_3$, while the other intersection point with the real axis
depends on the curve: $\s_2,\ \s_1$ and the points $x_{2n},\ n=1,2,\ldots$

There are several types of states: first, we have the states that become massless at the 
singularities. These are $(0,1)_0$ and, due to the cuts described differently 
in the two half planes, $(0,1)_0$ and $(-1,1)_{\pm 1}$ in the upper, 
and $(0,1)_0$ and $(1,1)_{\pm 1}$ in the
lower half plane. These states exist everywhere (throughout the corresponding half plane).

Second, we have the other dyons of $n_e=\pm 1$, the quarks and the W-boson. These states
decay on curves in the inner, strong coupling region of the Coulomb branch of moduli space:
The W-boson decays on ${\cal C}^\infty$, the quark $(1,0)_{-1}$ on ${\cal C}^+_0$ and the
quark $(1,0)_1$ on the innermost curve ${\cal C}^-_0$, while the dyons $(\e,1)_{-\e}$
decay on ${\cal C}^+_0$ and the dyons $(\e,1)_\e$ on ${\cal C}^-_0$.

Third, we have the dyons with $\vert n_e\vert \ge 2$. 
As discussed above, among these one must distinguish two
sorts: those that will survive the RG flow $m\to\infty$ 
to the pure gauge theory and those
that do not. The dyons that will survive this RG
flow are $(2n+1,1)_1$ in the upper half plane and  $(2n+1,1)_{-1}$ in the 
lower half plane. These dyons ($n\ne -1,0$) all decay on the curve 
${\cal C}^\infty$ which thus
plays a priviledged role. The other dyons, namely $(2n,1)_0$ ($n\ne 0$) and 
$(2n+1,1)_{-1}$ in the upper and $(2n+1,1)_1$ in the in the 
lower half plane ($n\ne -1,0$)
decay on curves ${\cal C}^\pm_{2k},\ k\ne 0$ (where $\vert 2k\vert$ equals 
$\vert n_e\vert$, $\vert n_e\vert+1$ or $\vert n_e\vert-1$).
There are only two states that decay on each of these curves
${\cal C}^\pm_{2k},\ k\ne 0$. These curves move more and more 
outwards as $m$ is increased.
Also, as $\vert k\vert$ gets bigger (i.e. the $\vert n_e\vert$ 
of the corresponding dyons increase) 
these curves more and more reach out towards the semiclassical region.
Conversely, as $m\to 0$, all curves flow towards a single curve, say 
${\cal C}^\infty$.

There are a couple of other points worth mentioning. First remark, that the whole
picture is compatible with the $CP$ transformation $(n_e,n_m)_s \to (-n_e,n_m)_{-s}$
under reflection by the real $u$-axis. Second, since all curves go through the
singularity $\s_3$, i.e. all existence domains touch $\s_3$, 
it follows that at this point all BPS
states exist. The same is true for the points $u$ that lie on the part of the real $u$ line to
the right of $\s_3$. Indeed, as $\vert n_e\vert$ is increased, the corresponding dyon
curves leaving $\s_3$ to the right with an ever smaller slope get closer and closer to any
given point on the real interval $(\s_3,\infty)$ but never touch it. 

Finally we note that the whole picture is perfectly consistent: 
if a BPS state decays across a given curve, the decay products are also
BPS states that must exist in the region considered, i.e. on both sides of the curve. 
Indeed, this is always the case. As
an example, consider  the dyons $(2n,1)_0$ ($n\ge 1$). In the upper half plane
they decay on the curves ${\cal C}^+_{2n}$ into the dyons $(2n-1,1)_1$ and the quark
$(1,0)_{-1}$. These dyons $(2n-1,1)_1$ exist everywhere in the upper half plane 
outside ${\cal C}^\infty$, while the quark $(1,0)_{-1}$ exists 
everywhere outside ${\cal C}^+_0$, and in
particular in the vicinity of the decay curves of $(2n,1)_0$ considered. 
Moreover, in many cases we perform several additional consistency checks to confirm
the existence domains we determined.

Now let us prove that our general picture we just described is indeed true.
We begin by studying the W-boson.

\subsection{The W-boson}

\fig{Sketch of the possible decay curves of the W-boson $(2,0)_0$ for 
$m<{\ld\over 2}$. The thick curves
$r_{\rm w}=1$ in the upper half plane and $r_{\rm w}=-1$ in the lower half plane are  the only ones 
that turn out to be relevant: they are ${\cal C}^\infty$.}{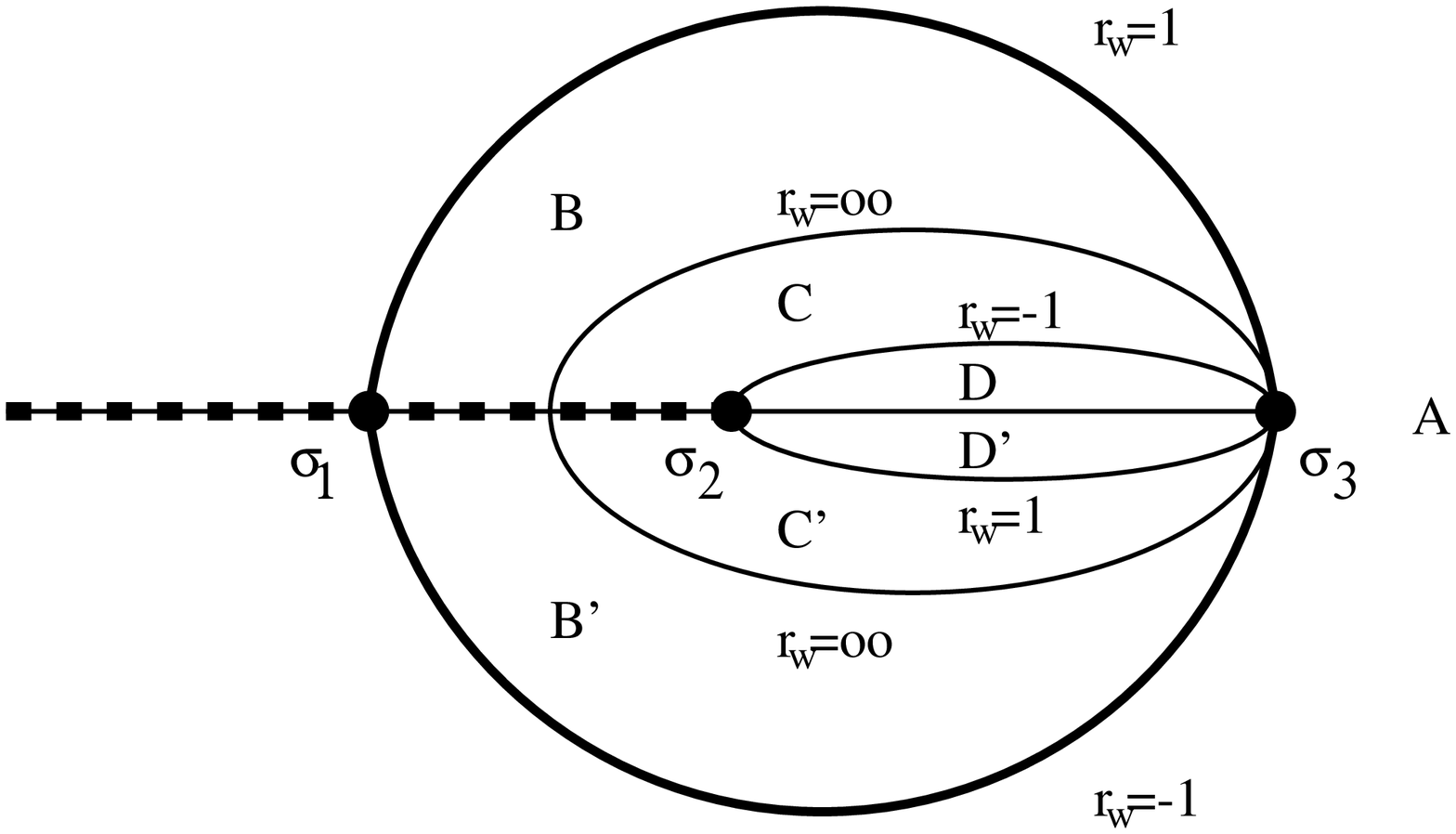}{10cm}
\figlabel\figix
From  Table 1 we see that the a priori possible decays are into a dyon and an anti-dyon.
The three curves $\IM {r_{\rm w} \ad +{m\over\rd} \over a}=0$ 
with $r_{\rm w}=\pm 1, \, \infty$ are shown in Fig. \figix.

First consider the curve $r_{\rm w}=\infty$. On this curve $\ad/a$ is real and varies from $-1$
to 0 in the upper half plane ($\e>0$) and from 0 to $+1$ in the lower half plane
($\e<0$). It is then easy to see that the only kinematically allowed decays  are
$(2,0)_0\to (0,1)_0 + (-1)\times (-2,1)_0$ for $\e>0$ and
$(2,0)_0\to (2,1)_0 + (-1)\times (0,1)_0$ for $\e<0$. On the $r_{\rm w}=+1$ curve we similarly
find that the kinematically possible decays are $(2,0)_0\to (1,1)_1 + (-1)\times
(-1,1)_1$ and on the $r_{\rm w}=-1$ curve  $(2,0)_0\to (1,1)_{-1} + (-1)\times
(-1,1)_{-1}$ for both $\e>0$ and $\e<0$.

Of course, as already mentioned above,
such decays can only take place if the final states do indeed exist on the
curve considered. This will  be checked below for the decay on ${\cal C}^\infty$,
but for the moment 
we invite the reader to consult the complete Figure \figxviii\ to convince
herself/himself that this is indeed the case.
We know that the W-boson exists in region
$A$ (connected to the semiclassical domain). Also we know that in the $m\to 0$ limit
$\s_2$ coincides with $\s_1$ and all curves become one and the same curve going through
$\s_1=\s_2$ and $\s_3$. From [\BF] we then know that the W-boson does not exist inside
this single $m=0$ curve. By the RG flow to finite $m$ it is then clear that it does not
exist in the innermost regions $D$ and $D'$. We will now show that the W-boson cannot exist in regions
$B, B', C$ or $C'$ either.

Suppose that $(2,0)_0$ exists in region $B$. Then we can transport it through the cut to region
$B'$ without crossing any decay curve. There it would be described\foot{
For typographical convenience we write $M^* (n_e, n_m)_s$ 
instead of $M^* \pmatrix{ n_e\cr n_m\cr s\cr}$.  
}
by $(M_{2\cdot 3}^*)^{-1} (2,0)_0 = (0,2)_{-2}$. 
Note  that the state $(0,2)_{-2}$ is different from 
$(2,0)_0$ and would have its own decay curves. We
do not say that $(0,2)_{-2}$ would exist in all of region $B'$, but at least 
it would exist in a region just below the
cut that separates $B$ and $B'$.
But such a state with $n_m=2$ should not
exist anywhere on \M, hence $(2,0)_0$ cannot exist in $B$. Suppose now $(2,0)_0$ exists
in $B'$. Then transport it through the cut into region $B$ where it is described as 
$M_{2\cdot 3}^*(2,0)_0 = (0,-2)_{-2}$ which again does not exist. Exactly the same
argument applies for regions $C$ and $C'$. Independently of the above RG-flow argument,
one can see similarly that $(2,0)_0$ cannot exist in $D$ or $D'$. If it would, there
would be a $(M_3^*)^{\mp 1} (2,0)_0 = (2,\pm 4)_0$ in $D'$ or $D$. We conclude that
$(2,0)_0$ cannot exist in any of the regions $B, B', C, C', D$ and $D'$. Hence it is the $r_{\rm w}=1$ curve for
$\e>0$ and the $r_{\rm w}=-1$ curve for $\e<0$ that border the existence domain of the W-boson. 
These are the two halves of the ${\cal C}^\infty$ curve as defined in  \cxxib. 
In this sense, ${\cal C}^\infty$ is the
only relevant curve for this BPS state.

Note that $M_{2 \cdot 3}$ acts on $(n_e,n_m)$  as $S\in {\rm SL}(2,\Z)$, i.e. it exchanges $n_e$
and $n_m$ (up to a sign). Thus whenever we have a state with $\vert n_e\vert \ge 2$ it
cannot exist in a region that is bounded by part of the cut $[\s_1, \s_2]$ (as are regions
$B, B', C$ and $C'$ for the W-boson) since this state would be described on the other
side of the cut by a $\vert n_m\vert \ge 2$. This is a very  useful fact which we will employ much in
the following.

\subsection{The quark $(1,0)_1$}

\fig{Sketch of the possible decay curves of the quark $(1,0)_1$ for 
$m<{\ld\over 2}$ labelled by the values of $r_{\rm q}$. The thick curve
$r_{\rm q}=\infty$ is ${\cal C}^-_0$ and 
is the only curve that turns out to be relevant.}{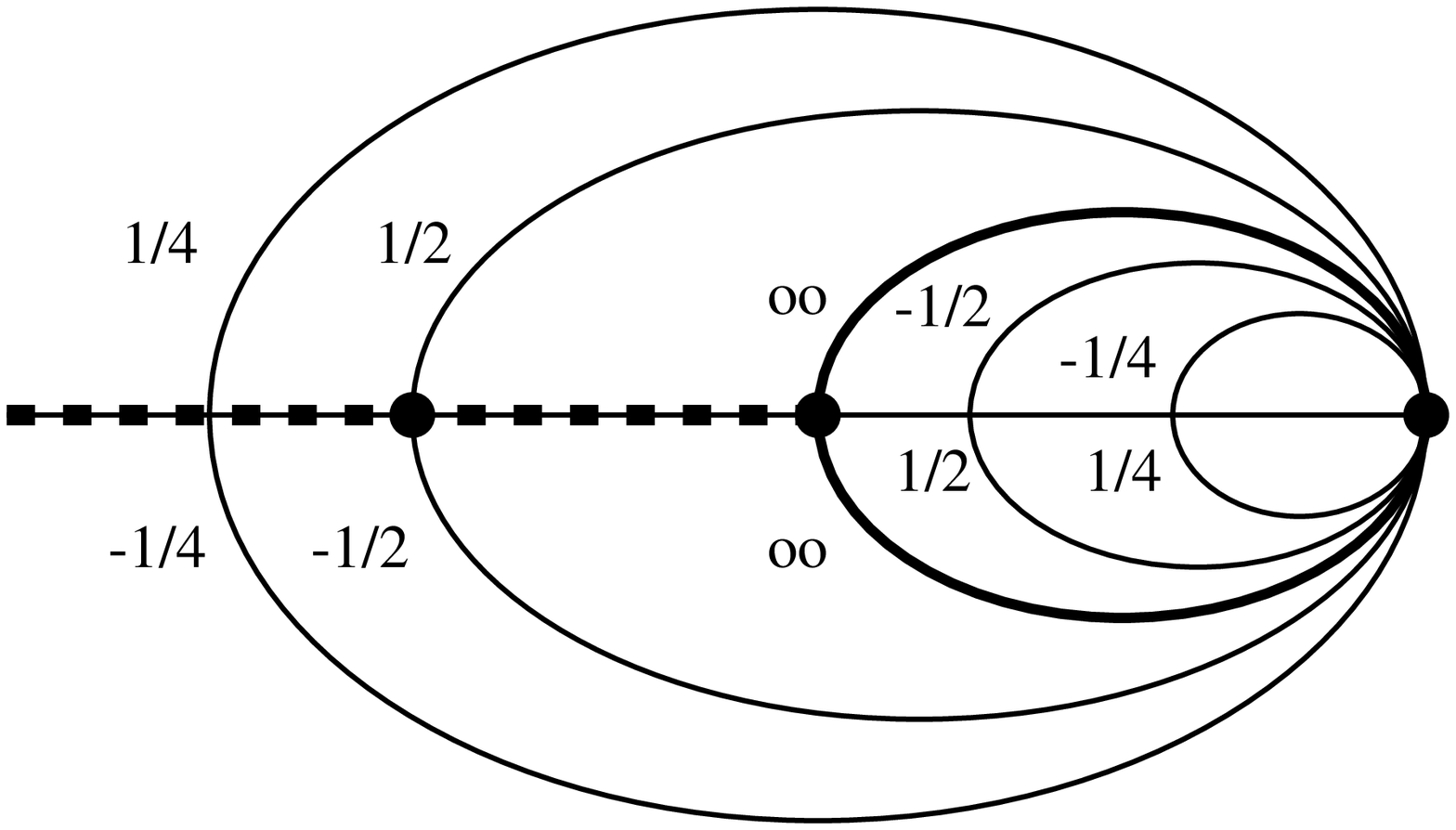}{8cm}
\figlabel\figx
%
From Table 1 we see that there is an infinity of a priori possible decays 
on the infinite set of curves labelled by
$r_{\rm q}=-{1\over 2p}$, including $r_{\rm q}=\infty$ for $p=0$, 
as indicated in Fig \figx.

One easily checks that for all curves with $p\ne 0$ the decays are kinematically
impossible. For $p=0$, however, $(1,0)_1\to (0,1)_0 + (-1)\times (-1,1)_{-1}$ is
kinematically possible for $\e>0$ while for $\e<0$ it is $(1,0)_1\to (-1)\times (0,1)_0
+(1,1)_1$. Note that in both cases the quark decays into the two BPS states that become
massless at the singularities $\s_2$ and $\s_3$, cf. Fig. \figiia,
and hence indeed exist throughout the corresponding half planes, as we will see below.
It is now clear that $(1,0)_1$ exists everywhere outside the $r_{\rm q}=\infty$ curve which
is nothing else than the curve ${\cal C}^-_0$ as defined in eq. \cxxib,
while it cannot exist inside, as one sees either from the RG-flow argument or otherwise
since one would get states $(M^*_3)^{\pm 1} (1,0)_1= (1,\mp 2)_1$ with $\vert
n_m\vert =2$  that do not exist.

\subsection{The quark $(1,0)_{-1}$}

%
\fig{Sketch of the possible decay curves of the quark $(1,0)_{-1}$ for 
$m<{\ld\over 2}$ labelled by the values of $r_{\rm q}$. The thick curve
$r_{\rm q}=\infty$ is ${\cal C}^+_0$ and is the only curve 
that turns out to be relevant.}{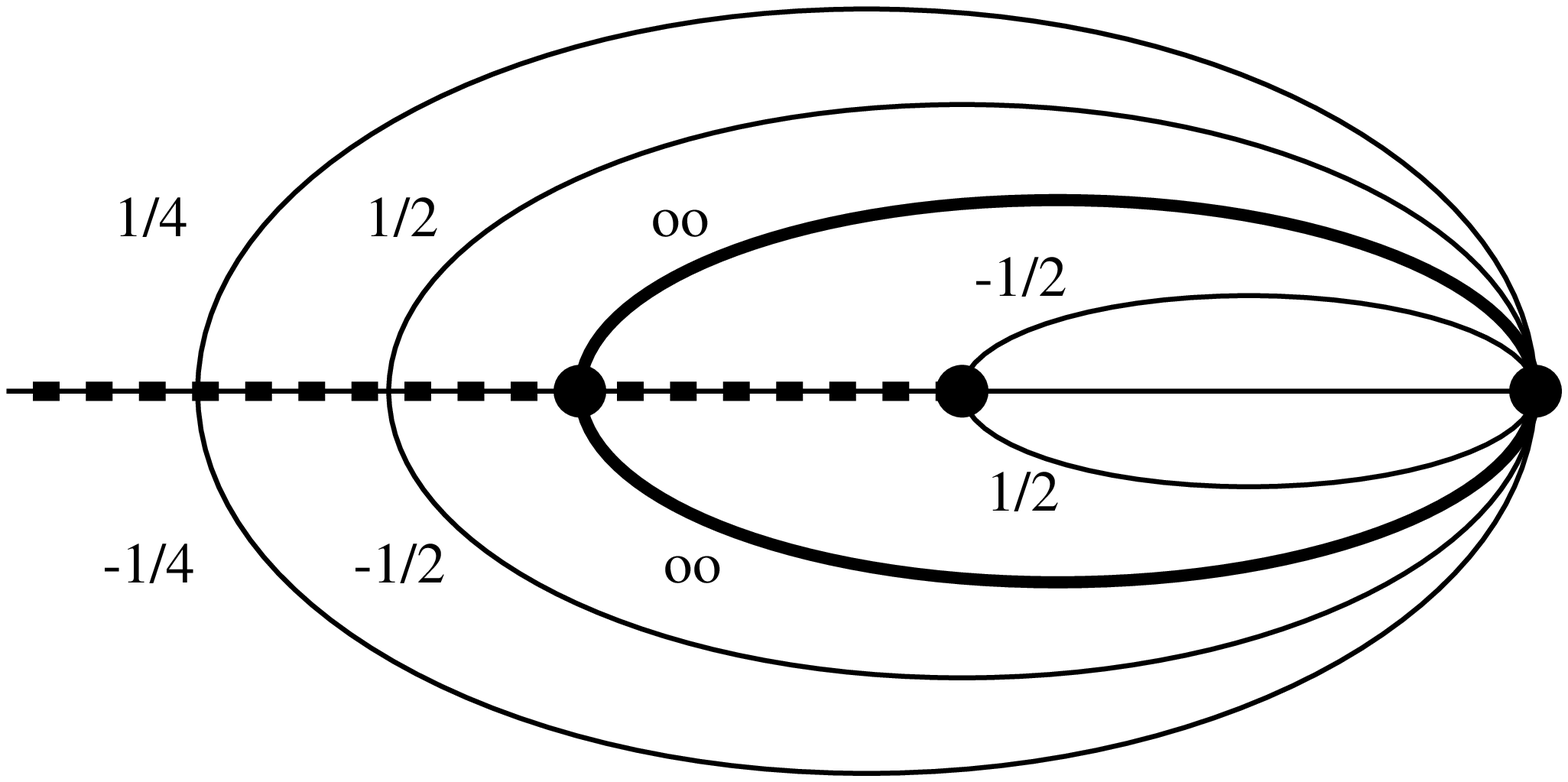}{8cm}
\figlabel\figxi
%
Again, there is an infinity of a priori possible decays on the infinite set of curves 
labelled by $r_{\rm q}={1\over 2p}$,
see Fig. \figxi.

We find again that all decays are kinematically impossible, except for $p=0$ on the
$r_{\rm q}=\infty$ curve:
$(1,0)_{-1} \to (0,1)_0 +(-1)\times (-1,1)_1$ for $\e>0$ and
$(1,0)_{-1} \to (-1)\times  (0,1)_0 + (1,1)_{-1}$ for $\e<0$. Note again that the relevant
curve goes through $\s_1$ and $\s_3$ and that the decay is precisely into the states that
are massless at $\s_1$ and $\s_3$, cf. Fig \figiia. Thus we see that the quark $(1,0)_{-1}$
exists everywhere outside the $r_{\rm q}=\infty$ curve and does not exist inside this curve. The
latter fact follows once more from the RG-flow from $m=0$ or using $(M_3^*)^{\pm 1}$ which
would generate a state with $\vert n_m\vert =2$ that does not exist. Note that here the
$r_{\rm q}=\infty$ curve is
${\cal C}^+_0$
which is different from the relevant decay curve ${\cal C}^\infty$ for the W-boson  (see
eq. \cxxib) although both curves go through $\s_1$ and $\s_3$: ${\cal C}^+_0$ lies inside 
${\cal C}^\infty$.

\subsection{The magnetic monopole $(0,1)_0$}

%
\fig{Sketch of the possible decay curves of the magnetic monopole $(0,1)_0$ for 
$m<{\ld\over 2}$ labelled by the values of $r_{\rm d}$. }{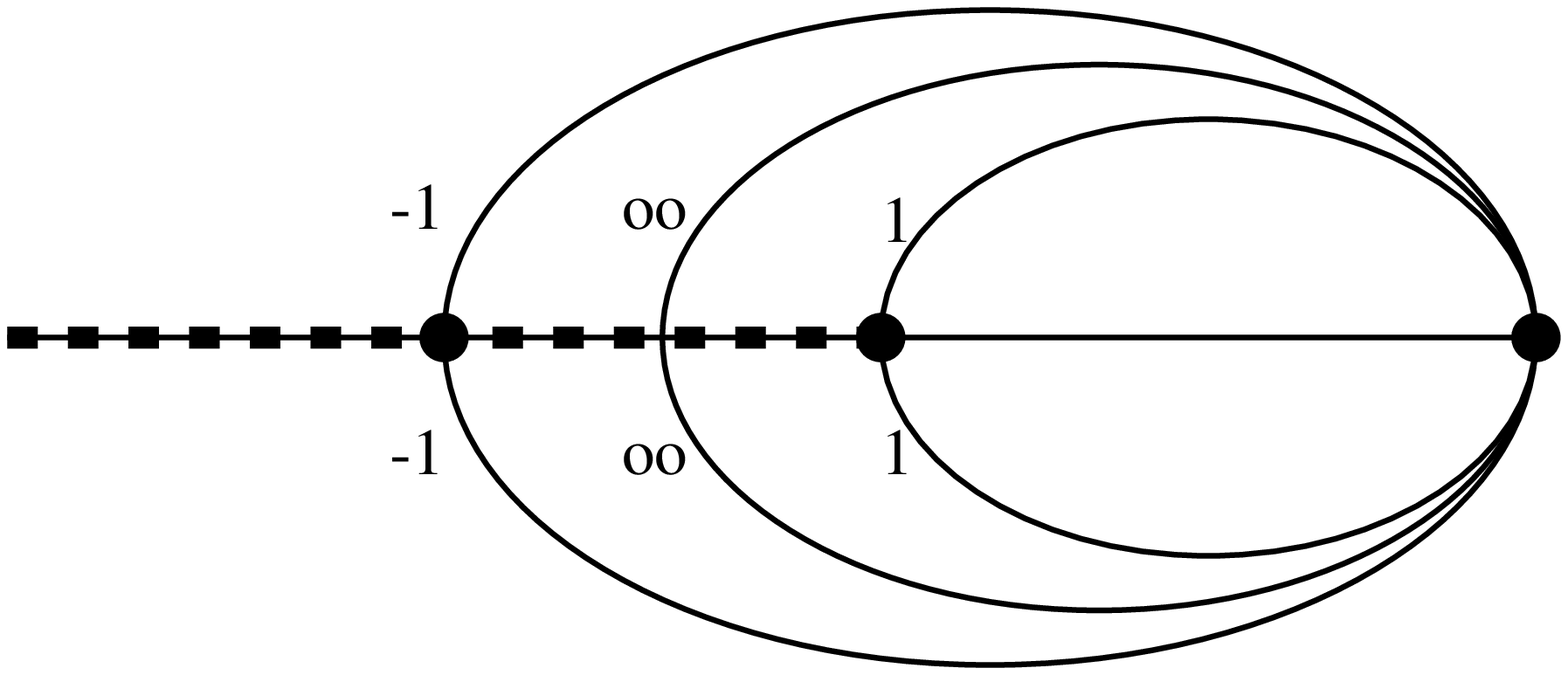}{8cm}
\figlabel\figxii
%
As for all dyons with even $n_e$, there are only three possible decay curves: $r=\pm 1$ and
$\infty$. For the magnetic monopole, which is massless at $\s_3$ these three curves are as
shown in Fig. \figxii.

All curves go through $\s_3$. Hence any point $u$ can be reached by a path that starts at
$\s_3$ and does not cross any curve. Since the monopole certainly exists at $\s_3$, we
conclude that it must exist everywhere. As a consistency check, one can verify that after
transporting it through any of the cuts it is decribed by quantum numbers that still
correspond to allowed BPS states that indeed do exist there.

\subsection{The dyons $(2n,1)_0$ with $n\ne 0$}

%
\fig{Sketch of the possible decay curves of the dyons $(2n,1)_0$, $2n\ge 4$, for 
$m<{\ld\over 2}$ labelled by the values of $r_{\rm d}$. The thick curves
are those that turn out to be relevant. They are ${\cal C}^+_{2n}$ in
the upper half plane and ${\cal C}_{2n}^-$ in the lower half plane.}{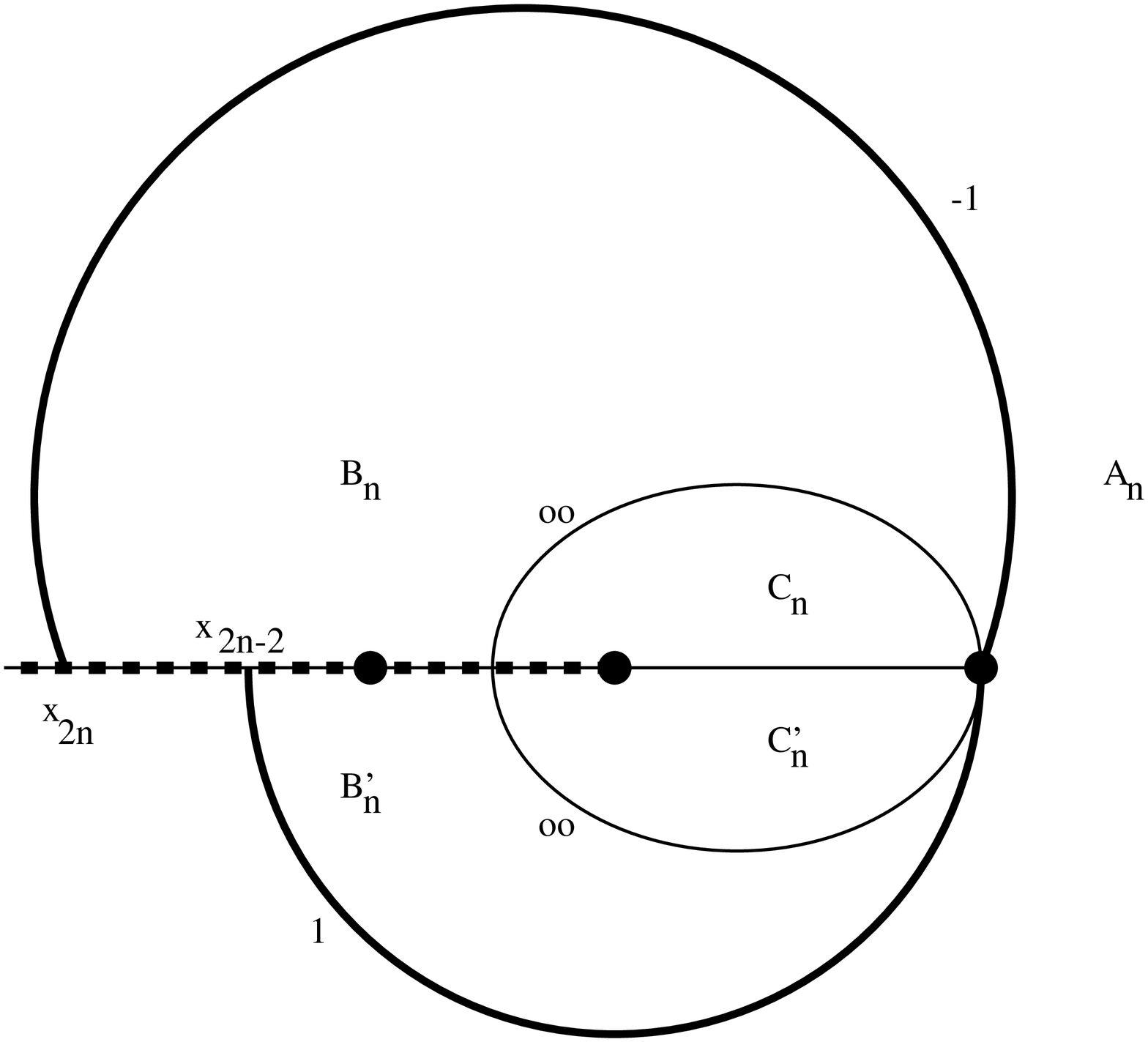}{10cm}
\figlabel\figxiv
%
Let first $2n \ge 4$.
For all these dyons the three curves $r_{\rm d}=\pm 1$ and $\infty$ are as sketched in Fig. \figxiv.

In each half plane there are only two curves. The $r=\infty$ curves are the same for
all $n$, while the $r_{\rm d}=1$ curve in the lower half plane starts at the point $x_{2n-2}$ on
the real axis where the $r_{\rm d}=-1$ curve in the upper half plane of the dyon $(2n-2,1)_0$ had
ended. In particular, $x_{2n}$ increases as $n (>0)$ is increased.
Since the regions $B_n,\, B_n',\, C_n$ and $C_n'$ all 
are bounded by a portion of the cut
$[\s_1,\s_2]$, by the  argument given at the end of the subsection on the W-boson, the  dyons $(2n,1)_0$ 
having $\vert n_e\vert \ge 2$,
cannot exist in either of these regions: they can only exist in regions $A_n$. 
The decay on
the $r_{\rm d}=-1$ curve (which is ${\cal C}^+_{2n}$) is $(2n,1)_0\to (1,0)_{-1} +(2n-1,1)_{1}$
while the decay on the $r_{\rm d}=1$ curve (which is ${\cal C}^-_{2n}$) is
$(2n,1)_0\to (1,0)_{1} +(2n-1,1)_{-1}$.
We will see below that  $(2n-1,1)_{1}$, resp. $(2n-1,1)_{-1}$ decay on curves that are
inside the curves ${\cal C}^+_{2n}$, resp. ${\cal C}^-_{2n}$, so that the final states of
the decays of $(2n,1)_0$ indeed exist. Note the following consistency check: if we
transport the dyon $(2n,1)_0$ from region $A_n$ through the interval $[x_{2n}, x_{2n-2}]$
into the upper half plane, it is described on the other side of the cut as $M_{1\cdot
2\cdot 3}^* (2n,1)_0=(2-2n,-1)_0=-(2n-2,1)_0$ which indeed does exist in the region above 
$[x_{2n}, x_{2n-2}]$.

Now consider the dyon $(2,1)_0$, i.e. $n=1$. Then the point $x_{2n-2}\equiv x_0$ coincides with 
the singularity $\s_1$, but this does not change the general argument just given. What is different however,
is the appearence of an additional curve in the lower half plane, namely $r_{\rm d}=-1$ which goes 
from $\s_2$ to $\s_3$ and separates region $C_1'$ into a region $C'$ and an innermost region $D$.
The above argument that the dyon cannot exist in region $C_n'$ now applies to $C'$. To show that 
$(2,1)_0$ cannot exist in $D$ either, it is enough to show
that it cannot decay across this $r_{\rm d}=-1$ curve separating $C'$ and $D$. The only kinematically
possible decay on this curve would be $(2,1)_0 \to (1,0)_{-1} +(1,1)_1$. But the quark $(1,0)_{-1}$ does not exist 
in the neighbourhood on either side of this curve as we have shown above. So $(2,1)_0$ 
cannot exist in region $D$ either. This also follows from the
RG-flow from $m=0$ or by applying $M_3^*$ when crossing the cut, which provides some 
consistency checks.

Contrary to the other states considered before,  the dyons $(2n,1)_0$ with $n\ge 1$  
have  existence domains that are asymmetric with respect to
reflection by the real axis, i.e. under complex conjugation. However this is not surprising:
reflection corresponds to $CP$ and for the states considered before, $CP$ only mapped them
to their antiparticles which must exist in the same domains. On the other hand,
$CP$ maps the dyon $(2n,1)_0$ 
to $(-2n,1)_0$ which is a different BPS state. $CP$ then tells us that
the dyons $(-2n,1)_0$   exist in domains
$\overline {A_n}$ that are the complex conjugate of the existence domains $A_n$ of the $(2n,1)_0$. 
The domains $\overline {A_n}$ are bounded by the curves ${\cal C}^-_{-2n}=\overline{ {\cal
C}^-_{2n} }$ in the upper half plane and by  ${\cal C}^+_{-2n}=\overline{ {\cal
C}^+_{2n} }$ in the lower half plane ($n>0$).

Next we will turn to the dyons with odd $n_e$ and $s=\pm 1$. Since the dyons
$(2n+1,1)_{-1}$ are  the $CP$ conjugates of the dyons $(-2n-1,1)_{+1}$ it is enough to
consider the dyons $(2n+1,1)_1$ for all $n\in {\bf Z}$. For each of these dyons of odd $n_e$, there is an
infinity of a priori possible decay curves, one for every odd $r$ and one for $r=\infty$.

\subsection{The dyon $(-1,1)_1$}

%
\fig{Sketch of the possible decay curves of the dyon $(-1,1)_1$ for 
$m<{\ld\over 2}$ labelled by the values of $r_{\rm d}$. The thick curve
(${\cal C}_0^+$)  is the one that turns out to be relevant.}{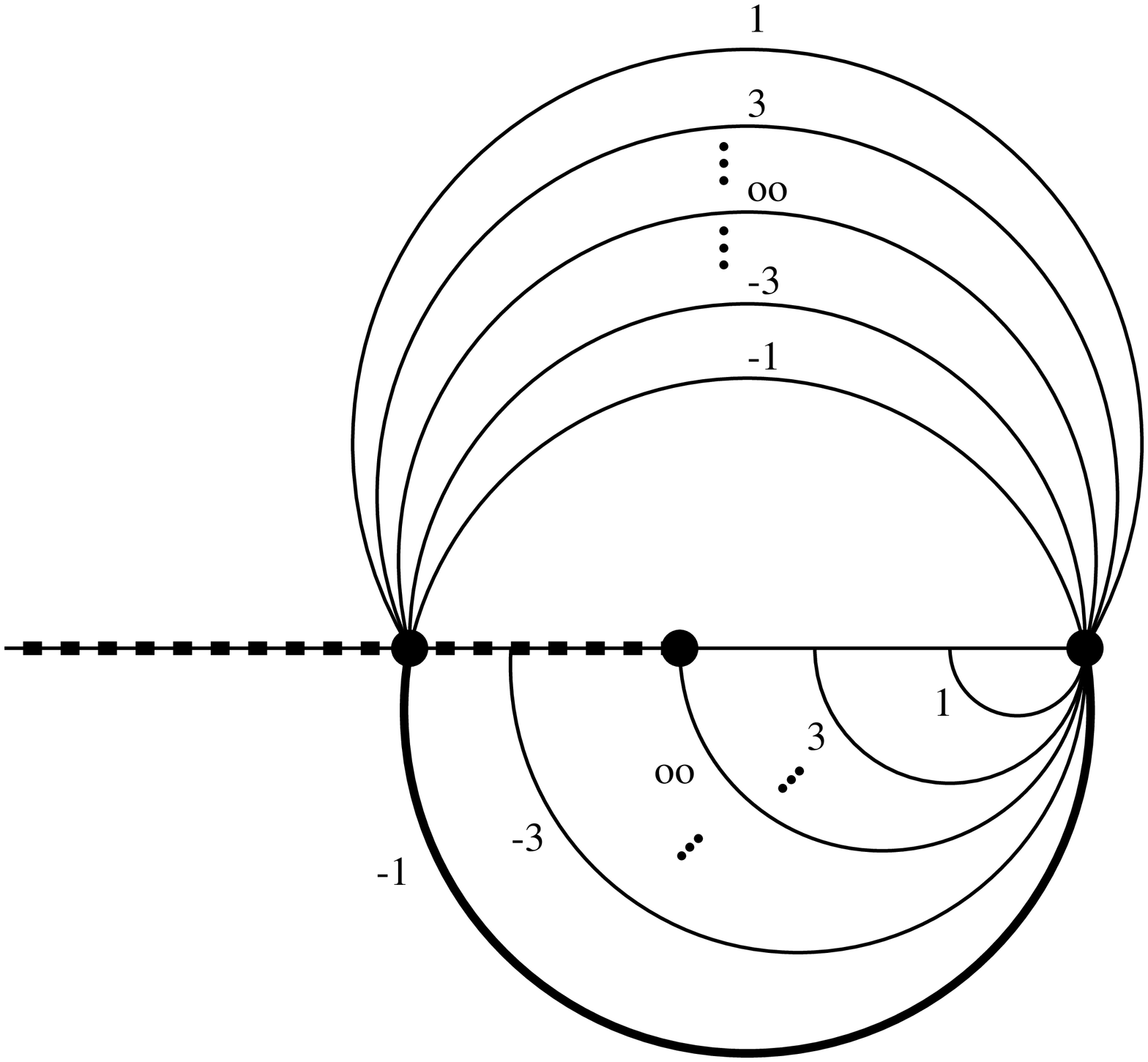}{8cm}
\figlabel\figxv
%
All curves go through $\s_3$, and the curves in the upper half plane all end at
$\s_1$ where this dyon is massless. The curves are shown in Fig. \figxv.

Since the dyon $(-1,1)_1$ exists at $\s_1$ when approached from the upper half plane, we
can always transport it to any point in this  half plane without crossing any curve,
and thus we know that it exists everywhere in the upper half plane. Let us now focus on
the lower half plane. The curves with $r_{\rm d}=-2p-1$ correspond to the decays
$(-1,1)_1\to 2\times (2p,1)_0+(-1)\times (4p+1,1)_{-1}$. Clearly, if $\vert p\vert$ is too
large, the final states will be too massive and the decay is impossible. We find that on
the $r_{\rm d}=-1$ curve ($p=0$) the decay $(-1,1)_1\to 2\times (0,1)_0 +(-1)\times
(1,1)_{-1}$ is indeed possible, i.e. $(-1,1)_1$ decays into the states that are massless at $\s_3$ and
$\s_1$ in the lower half plane. We also find that no decays are kinematically possible on
the $r_{\rm d}=-3,\, -5,\, \ldots$ curves. In any case, $(-1,1)_1$ cannot exist in any of the
regions in the lower half plane between two curves $r_{\rm d}=-2p-1$ and $r_{\rm d}=-2p-3$ ($p=0,\, 1,\,
2,\, \ldots$) because these regions touch part of the cut $[\s_1, \s_2]$ and hence the
existence of $(-1,1)_1$ would imply the existence of $M_{2\cdot 3}^* (-1,1)_1=(1,1)_3$
just above the cut. However, such a state has $s=3$ and cannot exist. Similarly,
$(-1,1)_1$ cannot exist between any of the curves $r_{\rm d}=2p+1$ and $r_{\rm d}=2p+3$ ($p=0,\, 1,\,
2,\, \ldots$) in the lower half plane because they touch part of the cut $[\s_2, \s_3]$,
and we would similarly conclude that a state $M_3^* (-1,1)_1=(-1,3)_1$ would exist just
above the cut. We conclude that the dyon $(-1,1)_1$ exists everywhere except in the region
of the lower half plane that is bounded by the cut $[\s_1, \s_3]$ and the $r_{\rm d}=-1$ curve
which is nothing else than ${\cal C}_0^+$ (or actually its part in the lower half plane).

As discussed above, it follows from $CP$ that the dyon $(1,1)_{-1}$ exists everywhere
outside the mirror image of the region just described, which is bounded by $[\s_1, \s_3]$
and the part in the upper half plane of the curve ${\cal C}_0^+$.

\subsection{The dyon $(1,1)_1$}

%
\fig{Sketch of the possible decay curves of the dyon $(1,1)_1$ for 
$m<{\ld\over 2}$ labelled by the values of $r_{\rm d}$. The thick curve
(${\cal C}_0^-$)
is the one that turns out to be relevant.}{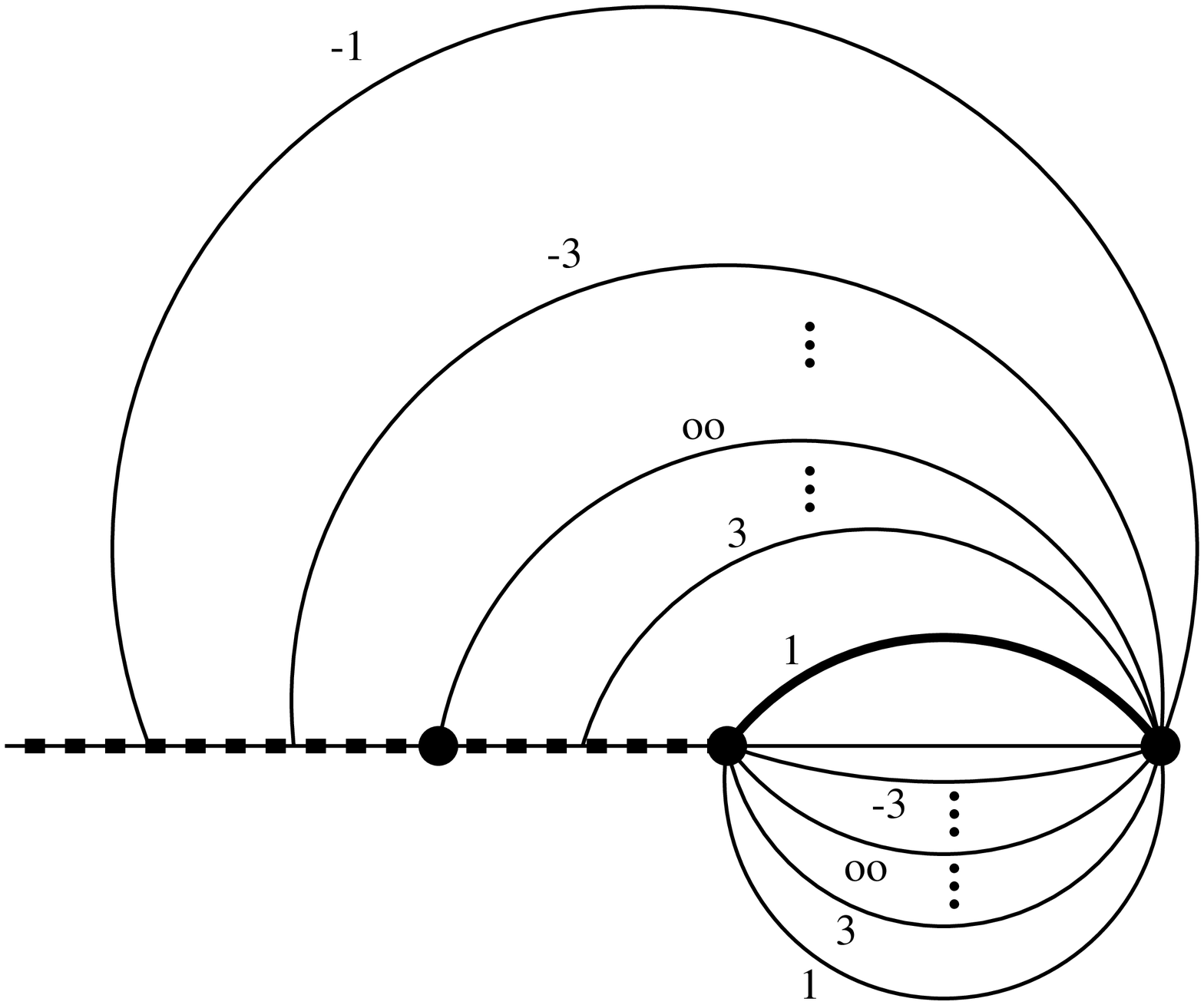}{8cm}
\figlabel\figxvi
%
All curves go through $\s_3$, and the curves in the lower half plane all start at $\s_2$
where this dyon is massless. They are shown in Fig. \figxvi.

Again, any point in the lower half plane can be reached from $\s_2$ without crossing any
curve, and hence $(1,1)_1$ exists everywhere in the lower half plane. What about the upper
half plane? It turns out that there the only curve on which a decay is kinematically
possible is the $r_{\rm d}=1$ curve with the decay $(1,1)_1\to 2\times (0,1)_0+(-1)\times
(-1,1)_{-1}$ with the final states being again those BPS states that are massless at
$\s_3$ and $\s_2$ in the upper half plane. It is then clear that this decay must happen as
$(1,1)_1$ cannot exist inside the $r_{\rm d}=1$ curve. This can be seen either from the RG-flow
argument from $m=0$, or else since the existence of $(1,1)_1$ in this region would imply
the existence of $(M_3^*)^{-1} (1,1)_1=(1,3)_1$ below the cut $[\s_2,\s_3]$ which is
excluded. We conclude that the dyon $(1,1)_1$ exists everywhere outside the region bounded
by  $[\s_2, \s_3]$ and the $r_{\rm d}=1$ curve in the upper half plane which is nothing else than
the corresponding part of ${\cal C}^-_0$. By $CP$ we see that the dyon $(-1,1)_{-1}$
exists everywhere outside a region bounded by $[\s_2, \s_3]$ and the part of 
${\cal C}^-_0$ in the lower half plane.

\subsection{The dyons $(2n+1,1)_1$ with $n>0$}

\vskip 2.mm
\fig{Sketch of the possible decay curves of the dyons $(2n+1,1)_1$ for 
$m<{\ld\over 2}$ labelled by the values of $r_{\rm d}$. The thick curves are
those that turn out to be relevant.
They are ${\cal C}^\infty$ in the upper half plane and ${\cal C}^-_{2n}$ 
in the lower half plane.}{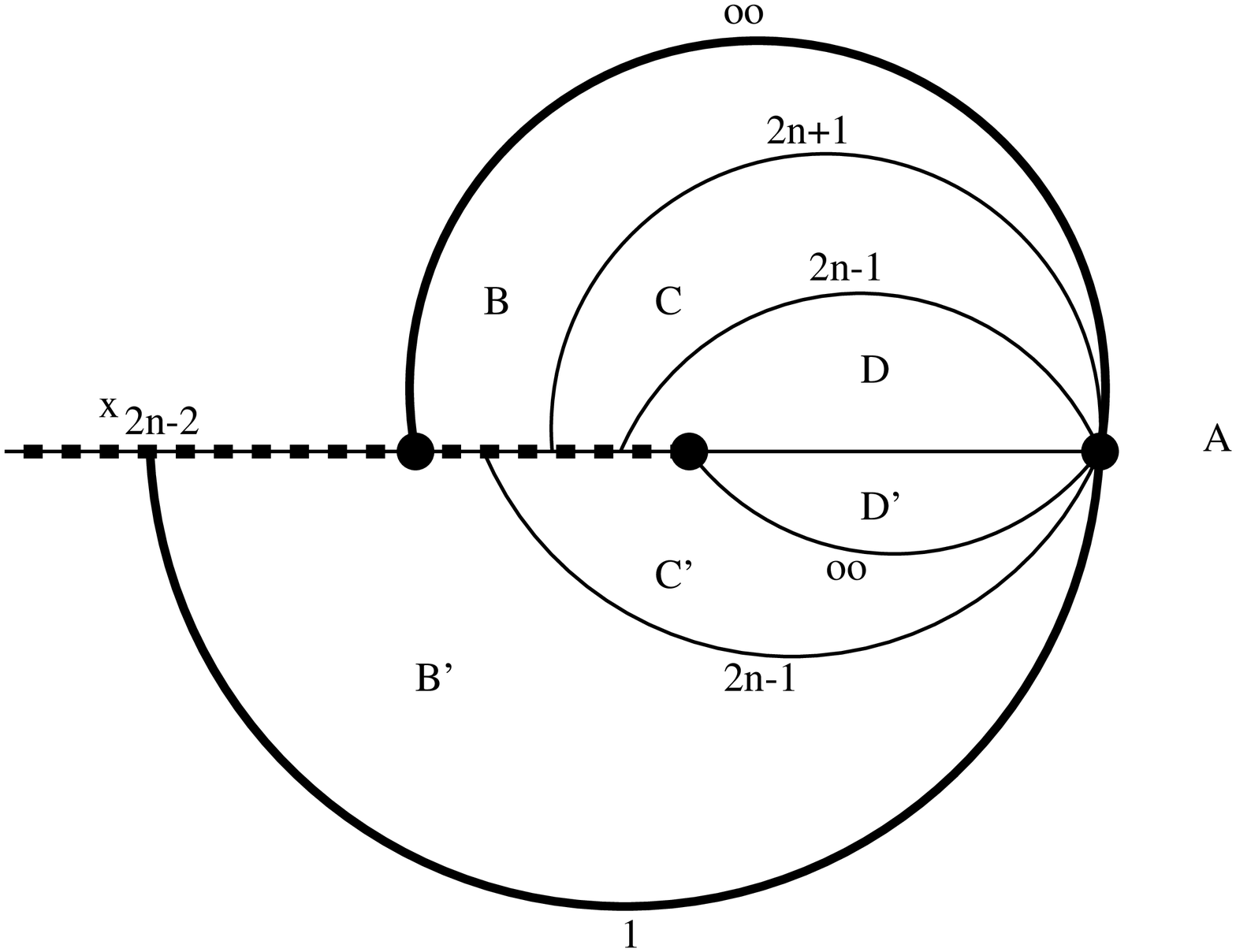}{10cm}
\figlabel\figxvii 
\vskip 2.mm
%
Among the infinity of a priori possible decay curves, we find that the decays are
kinematically possible only on the following ones: in the upper half 
plane on $r_{\rm d}=\infty$, $r_{\rm d}=2n+1$ and $r_{\rm d}=2n-1$ and 
in the lower half plane on $r_{\rm d}=\infty$,
$r_{\rm d}=2n-1$ and $r_{\rm d}=1$. These curves are shown in Fig. \figxvii.

The curves shown in Fig. \figxvii\ are generic, except for $2n+1=5$ where the $r_{\rm d}=2n-1=3$
curve in the upper half plane ends at $\s_2$, and for $2n+1=3$ where there is no
$r_{\rm d}=2n-1=1$ curve in the upper half plane and where the $r_{\rm d}=2n-1$ curve in the lower half
plane  coincides with the $r_{\rm d}=1$ curve and starts at $x_0\equiv \s_1$.

There are two kinematically possible decays on the $r_{\rm d}=1$ curve:
$(2n+1,1)_1\to (1,0)_1 + (2n,1)_0$ and 
$(2n+1,1)_1\to 2\times (1,0)_1 + (2n-1,1)_{-1}$. But this curve is ${\cal C}^-_{2n}$ and
hence also the decay curve of $(2n,1)_0$ (where the latter decays into $(1,0)_1 + (2n-1,1)_{-1}$), 
so that the first of the two decays cannot take
place (or actually is identical to the second). 
On the other hand, we will see below that $(2n-1,1)_{-1}$ exists in the lower half
plane everywhere outside the curve ${\cal C}^\infty$ which is well inside ${\cal
C}^-_{2n}$ ($n>0$), so that the second of the two decays can indeed take place. In the
upper half plane, on the $r_{\rm d}=\infty$ curve (which is ${\cal C}^\infty$) we can have the
decay $(2n+1,1)_1\to (n+1)\times (1,1)_1 + (-n)\times (-1,1)_1$ into final states that do
exist there. The existence of $(2n+1,1)_1$ in regions $B,\, C,\, D,\, B',\, C'$ 
and $D'$ is ruled out
by the by now familiar arguments of transporting the state through the cuts $[\s_1,\s_2]$
or $[\s_2,\s_3]$ and thus generating states $(M_{2\cdot 3}^*)^{\pm 1} (2n+1,1)_1$ or 
$(M_{3}^*)^{\pm 1} (2n+1,1)_1$ that would have $\vert n_m\vert \ge 2$.

\subsection{The dyons $(2n+1,1)_1$ with $n<-1$}

The decay curves on which the decays are kinematically possible look qualitatively very
much like those of Fig. \figxvii. The only curve that starts in the lower half plane to
the left of $\s_1$ is the curve ${\cal C}^+_{2n+2}$ (which is now $r_{\rm d}=-1$ and replaces the $r_{\rm d}=1$ curve
of Fig. \figxvii), while in the upper half plane we still have the same $r=\infty$ curve
${\cal C}^\infty$. By the same  arguments, using  $(M_{2\cdot 3}^*)^{\pm 1}$ and 
$(M_{3}^*)^{\pm 1}$, we can show that the dyons under consideration only exist outside
these two curves. The decay on ${\cal C}^\infty$ is e.g.
$(2n+1,1)_1\to (n+1)\times (1,1)_1 + (-n)\times (-1,1)_1$ as before, while on ${\cal
C}^+_{2n+2}$ it is $(2n+1,1)_1\to 2\times (-1,0)_1 + (2n+3,1)_{-1}$ which do exist there.

\subsection{The dyons $(2n+1,1)_{-1}$}

These dyons are the $CP$ conjugate states of $(-2n-1,1)_1$. Hence, in the lower half plane
they always exist everywhere outside the curve ${\cal C}^\infty$, while in the upper half
plane they only exist outside curves ${\cal C}^+_{2n}$ if $n>0$ and only outside curves
${\cal C}^-_{2n+2}$ if $n<-1$. As a consistency check, consider transporting a dyon
$(2n+1,1)_1$ with $n>0$ through  the cut $(-\infty, \s_1]$ (see Fig. \figxvii) into the
region below the cut. Then this state is described there by $(M^*_{1\cdot 2\cdot
3})^{-1}(2n_1,1)_1=-(2n+3,1)_{-1}$. This state indeed exists in this region because it is
outside ${\cal C}^\infty$ (in the lower half plane).

\section{\bf  Decay curves and BPS spectra for large mass ($m>{\ld\over 2}$)}

\subsection{The general picture}

%
\fig{Shown are a sketch of the relative positions of the relevant decay curves 
for $m>{\ld\over 2}$ as well as
the BPS states that decay across these curves (for not too 
large $\vert n_e\vert$). The BPS states that decay across ${\cal C}^\infty$ in the upper
half plane are the W-boson $(2,0)_0$ as well as all dyons $(2n+1,1)_1,\ n\ne -1,0$,
while the states decaying across ${\cal C}^\infty$ in the lower
half plane are again the W-boson $(2,0)_0$ as well as all 
dyons $(2n+1,1)_{-1},\ n\ne -1,0$.
The only states existing inside the ${\cal C}^+_0$ curve are the three dyons $(\e,1)_\e$,
$(-\e,1)_\e$, $(-\e,1)_{-\e}$ as well as the quark $(1,0)_1$ 
and the monopole $(0,1)_0$.}{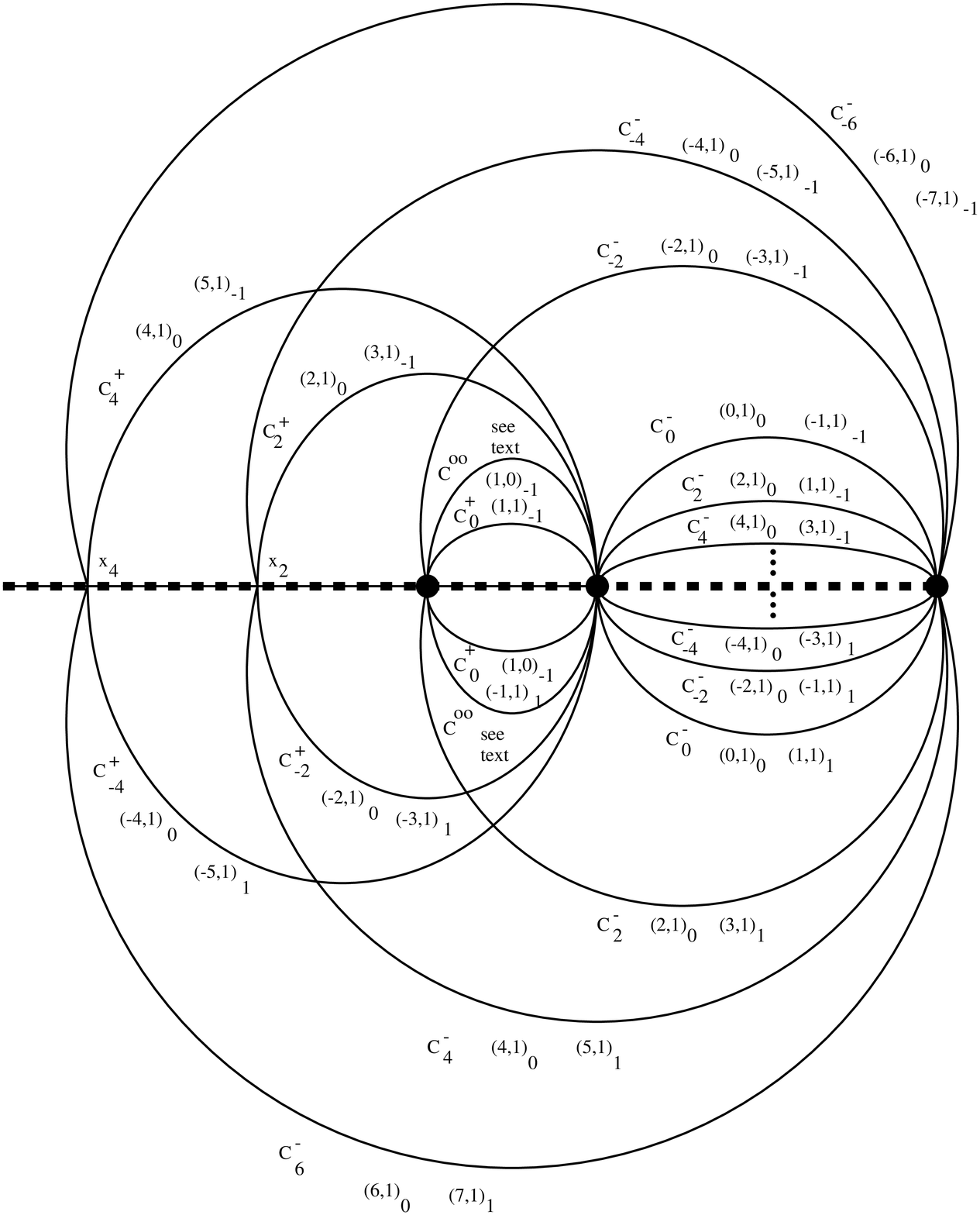}{15cm}
\figlabel\figxix 
As $m$ is increased from $m<{\ld\over 2}$ to $m>{\ld\over 2}$ one goes through  $m={\ld\over 2}$
where the singularities $\s_2$ and $\s_3$ coincide and where new states become massless.
At $m>{\ld\over 2}$, the quantum numbers of the massless states
at these two singularities have changed. One must however keep in mind that this is a somewhat ``local" effect,
where locality here refers to the distance on the Coulomb branch: these
rearrangements do not much affect $a(u)$ or $\ad(u)$ for $u$ far away, i.e. such that $\vert u-\s_2\vert
\gg \vert \s_2-\s_3\vert$. What does this mean for the relevant decay curves and existence domains of the
BPS states of Fig. \figxviii? We expect, and verify below, that all what can happen is the following: as
$m={\ld\over 2}$ and $\s_2=\s_3$, the curve ${\cal C}_0^-$ has shrunk to a point. Then as $m$ is
increased beyond ${\ld\over 2}$ and $\s_3$ moves off to the right, all curves (except  ${\cal C}_0^-$)
remain ``attached" to the real axis ``at the left" at points $\s_1,\ x_{2n}$ (that, of course, 
do move as $m$ is varied), while at their right some curves remain attached at $\s_2$
while others are attached at $\s_3$. It turns out that all curves ${\cal C}^+_{2n},\, n\in \Z$ are
attached to $\s_2$, and all curves ${\cal C}^-_{2n}$ with $n<0$ for $\e>0$ and $n>0$ for $\e<0$ are
attached to $\s_3$. But there appears also a new family of relevant curves, attached to $\s_2$ and
$\s_3$: these are the curves ${\cal C}^-_{2n}$ with $n\ge 0$ for $\e>0$ and $n\le 0$ 
for $\e<0$. All this is
shown in Fig. \figxix. 
The only states existing inside the ${\cal C}^+_0$ curve are the three dyons $(\e,1)_\e$,
$(-\e,1)_\e$, $(-\e,1)_{-\e}$ as well as the quark $(1,0)_1$ and the monopole $(0,1)_0$. 
The dyon $(-\e,1)_{-\e}$ and the monopole $(0,1)_0$
however decay on the curve ${\cal C}^-_0$ between $\s_2$ and $\s_3$. So, as always, the
only BPS states that exist everywhere are the three states that are responsible for the
singularities, namely the two dyons $(\pm \e,1)_{\e}$ and the quark $(1,0)_1$.

Note that, as $m\to \infty$, the special dyons $(-\e,1)_\e$ and $(\e,1)_\e$ have 
finite masses as do all dyons $(2n+1,1)_\e$ as well as the W-boson. These states thus
survive the RG flow to the pure gauge theory. These surviving states 
all decay on one
and the same curve, ${\cal C}^\infty$ which one may consider as fixed under the flow
(except for $(-\e,1)_\e$ and $(\e,1)_\e$, of course). After transforming to the natural
quantum numbers $(\net,n_m)_\st$ of the pure gauge theory, see eq. \nqn,
we see that these states all have $\st=0$ and
precisely constitute the spectrum of the pure gauge theory as established in [\FB].
On the other hand, the states that do not survive this flow either simply disappear from
the spectrum because their masses diverge, as is the case of the quarks or the 
special dyons
$(\pm 1,1)_{-\e}$ or $(0,1)_0$ inside ${\cal C}^\infty$, 
or because they are ``hit" by their decay curve that moves outwards as
$m\to\infty$. Indeed,  any point $u$ which for some $m$ still
is in the existence domain of a given BPS state will be hit by the corresponding curve
for large enough $m$. Note that again, as for $m<{\ld\over 2}$, all BPS states exist at
$\s_3$ and in a narraow wedge extending to the right of it, in particular on the real
interval $[\s_3,\infty)$. Of course, this does not prevent any real point $u$ to be hit
by the curves: As $m\to\infty$ we also have $\s_3\to\infty$ and any point $u$ will
eventually end up inside the curves.

Now let us briefly comment
on the discrepancy  with the published version of [\BRASTI]. The authors of [\BRASTI], 
for $N_f=2$ with equal bare masses, consider a  real 
point $u$ to the {\it right} of the singularity $\s_3$ 
and write that a certain dyon does no longer exist
there for a certain mass $m>{\ld\over 2}$. 
In the light of what we have said so far, it is clear that this cannot be true. 
This fact is even more obvious since, if it
were true, we would need to have decay curves crossing the real $u$ axis to the right of
$\s_3$ where there are no cuts - a situation that cannot occur as one can easily see,
also without our detailed analysis.
The authors of [\BRASTI] have checked this point again, and told us that they actually agree with
our result, the discrepancy being only due to some error when writing up their paper.

In the remainder of this subsection, we will again discuss the decay curves for
each type of BPS state, thus proving Fig. \figxix\ to be correct.

\subsection{The dyons $(2n,1)_0$}

Recall that decays can only happen on the curves \cxxi\ for $r=\pm 1$ and $\infty$. But numerically
we find that the $r_{\rm d}=\infty$ curve has disappeared. Actually, as $m$ approaches ${\ld\over 2}$ from
below, one can see how this curve (which goes through $\s_3$, see Figs. \figxii\ to \figxiv)
shrinks to a point, and its absence at $m>{\ld\over 2}$ is perfectly compatible with a smooth RG
flow as $m$ is increased.
\def\doublefig#1#2#3#4#5{
\par\begingroup\parindent=0pt\leftskip=1cm\rightskip=1cm\parindent=0pt
\baselineskip=11pt
\global\advance\figno by 1
\midinsert
\centerline{\hbox{\epsfxsize=#3\epsfbox{#2}
\hskip 1.cm 
\epsfxsize=#5\vbox{\epsfbox{#4}\vskip 1.5cm\null}}}
\vskip 12pt
{\bf Fig. \the\figno:} #1\par
\endinsert\endgroup\par
}
\def\figlabel#1{\xdef#1{\the\figno}}
\def\encadremath#1{\vbox{\hrule\hbox{\vrule\kern8pt\vbox{\kern8pt
\hbox{$\displaystyle #1$}\kern8pt}
\kern8pt\vrule}\hrule}}
\doublefig{In the left figure we show a sketch of the relative positions of the  decay curves of the dyons $(2n,1)_0$
($n>1$) for $m>{\ld\over 2}$. All curves shown are relevant. The figure on the right shows
a sketch of the relative positions of the  decay curves of the monopole $(0,1)_0$.
The thick curves are the relevant ones.}{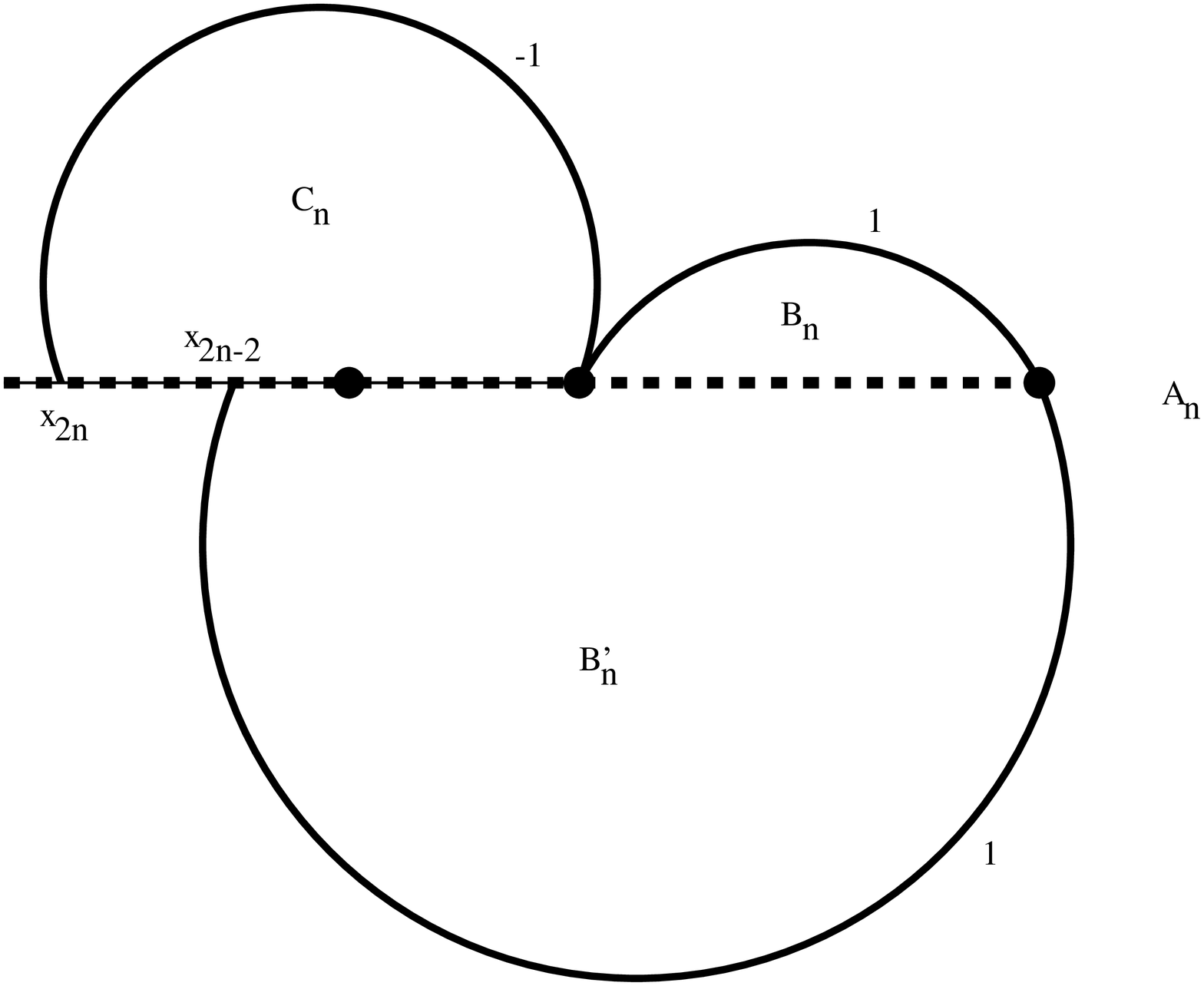}{7cm}{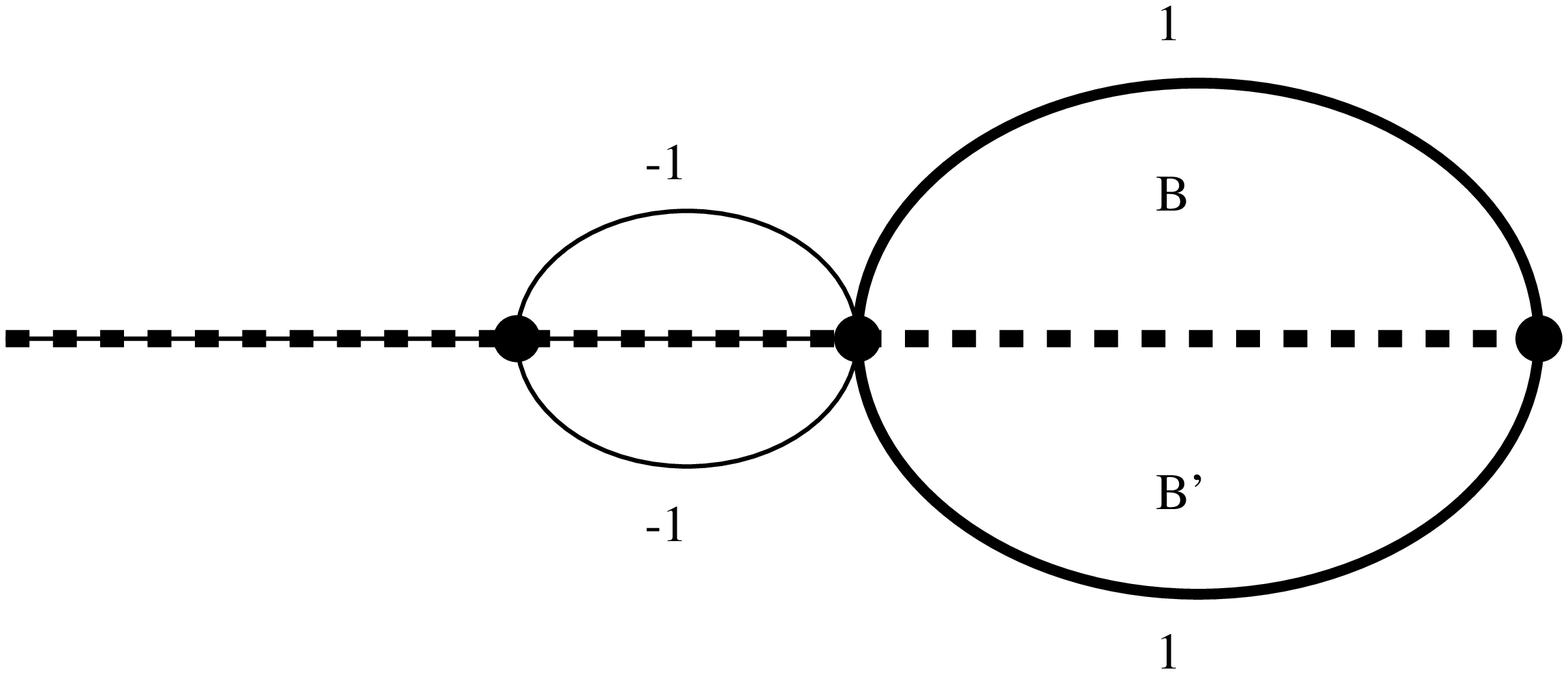}{7cm}
\figlabel\figxx 
For a generic dyon $(2n,1)_0$ with $n\ge 2$ one finds for the $r_{\rm d}=\pm 1$ curves the situation
depicted in Fig. \figxx\ on the left. The $r_{\rm d}=1$ curve (for $\e>0$ and $\e<0$) is the curve ${\cal C}^-_{2n}$ while
the $r_{\rm d}=-1$ curve ($\e>0$) is ${\cal C}^+_{2n}$.
All curves turn out to be relevant with the following kinematically possible decays: $(2n,1)_0\to
(1,0)_{-1}+ (2n-1,1)_1$ on $r_{\rm d}=-1$, $(2n,1)_0\to (-1)\times (1,0)_1 + (2n+1,1)_1$ 
on $r_{\rm d}=1$ for $\e>0$
and $(2n,1)_0\to (1,0)_1 +(2n-1,1)_{-1}$ on $r_{\rm d}=1$ for $\e<0$. To show that $(2n,1)_0$ cannot exist
in regions $C_n,\ B_n$ and $B_n'$ we use the same type of argument as before. The monodromy
$M_{2\cdot 3}$ around $\s_2$ and $\s_3$ is the same as for small mass, and just as before,
transporting the dyon $(2n,1)_0$ from $C_n$ to $B_n'$
through the cut $[\s_1,\s_2]$  or vice versa would result in states with
$\vert n_m\vert =2n$ that cannot exist.  To prove the non-existence of $(2n,1)_0$ in
$B_n$, transport it to $B_n'$ where it is described as $(M_3^*)^{-1}(2n,1)_0=(2n-2,1)_{-2}$ which
has $s=-2$ and cannot exist. We conclude that $(2n,1)_0$ only exists in regions $A_n$ outside these
curves.
The only difference for $n=1$, i.e. for the dyon $(2,1)_0$ is that the point $x_{2n-2}\equiv x_0$
coincides with $\s_1$, but this does not change the preceeding discussion.

The dyons $(2n,1)_0$ with $n<0$ are the $CP$ conjugates of the dyons $(-2n,1)_0$ with $-2n>0$, and
thus exist  in the complex conjugate domains $A_n=\overline {A_{-n}}$.

The situation is slightly different for the monopole $(0,1)_0$, as shown in Fig.  \figxx\ on the right.
No decays are possible on the $r=-1$ curve while on the $r=1$ curve one has $(0,1)_0\to (-1)\times
(1,0)_1 + (1,1)_1$ for $\e>0$ and $(0,1)_0\to (1,0)_1 +(-1,1)_{-1}$ for $\e<0$. The monopole cannot
exist in $B$ or $B'$ since this would lead to $(M_3^*)^{\pm 1} (0,1)_0=(\pm 2,1)_{\pm 2}$ on the
other side of $[\s_2,\s_3]$ which, having $\vert s\vert =2$, is excluded.

\subsection{The dyons $(2n+1,1)_1$}

\def\doublefig#1#2#3#4#5{
\par\begingroup\parindent=0pt\leftskip=1cm\rightskip=1cm\parindent=0pt
\baselineskip=11pt
\global\advance\figno by 1
\midinsert
\centerline{\hbox{\epsfxsize=#3\vbox{\null\vskip 2.cm\epsfbox{#2}}
\hskip 1.cm 
\epsfxsize=#5\epsfbox{#4}}}
\vskip 12pt
{\bf Fig. \the\figno:} #1\par
\endinsert\endgroup\par
}
\def\figlabel#1{\xdef#1{\the\figno}}
\def\encadremath#1{\vbox{\hrule\hbox{\vrule\kern8pt\vbox{\kern8pt
\hbox{$\displaystyle #1$}\kern8pt}
\kern8pt\vrule}\hrule}}
\doublefig{Shown is a sketch of the relative positions, for $m>{\ld\over 2}$,
of the  decay curves of the dyon $(1,1)_1$ (left) and of the dyon $(-1,1)_1$
(right). The thick curves are the relevant ones.}{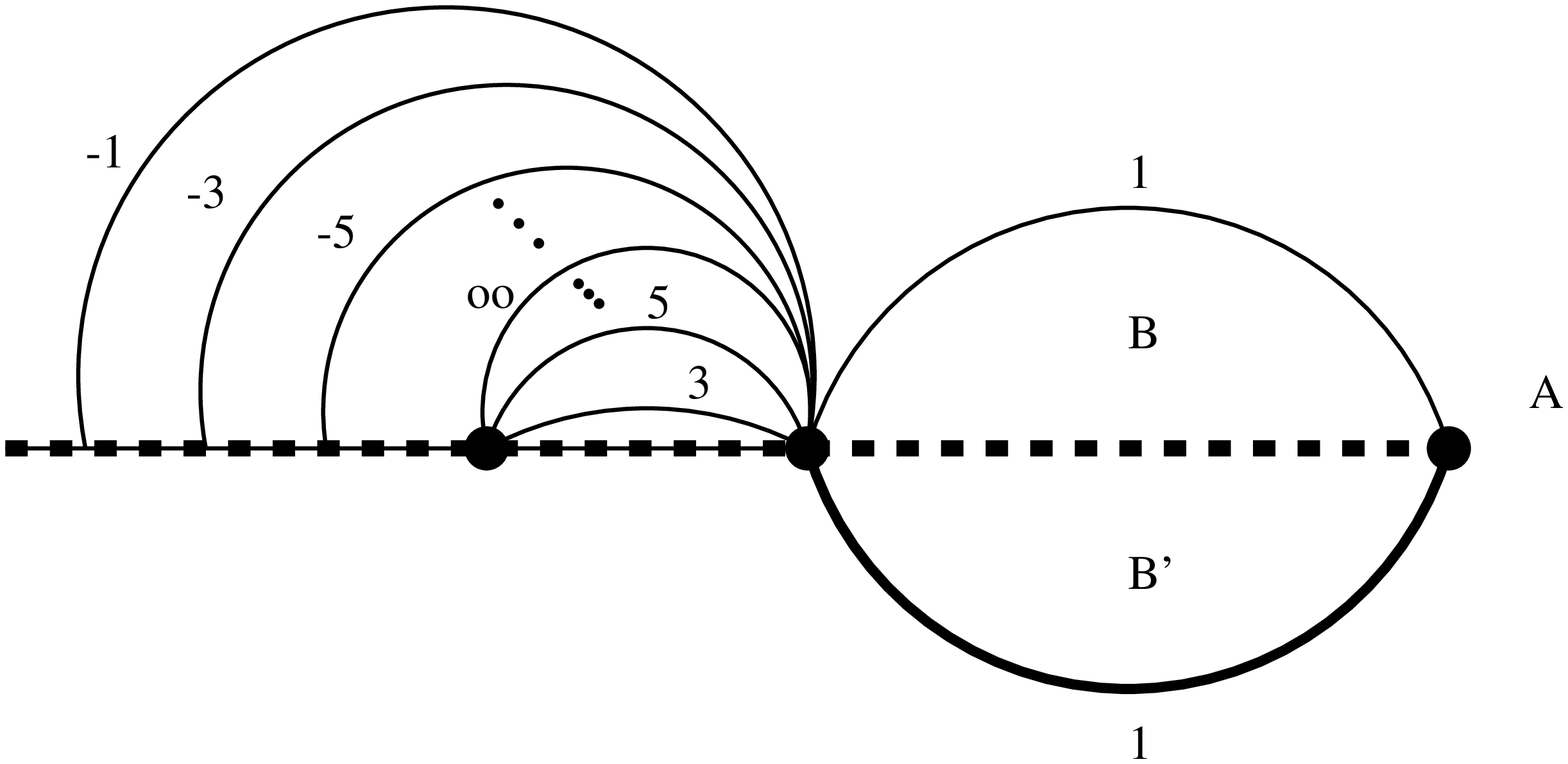}{7cm}{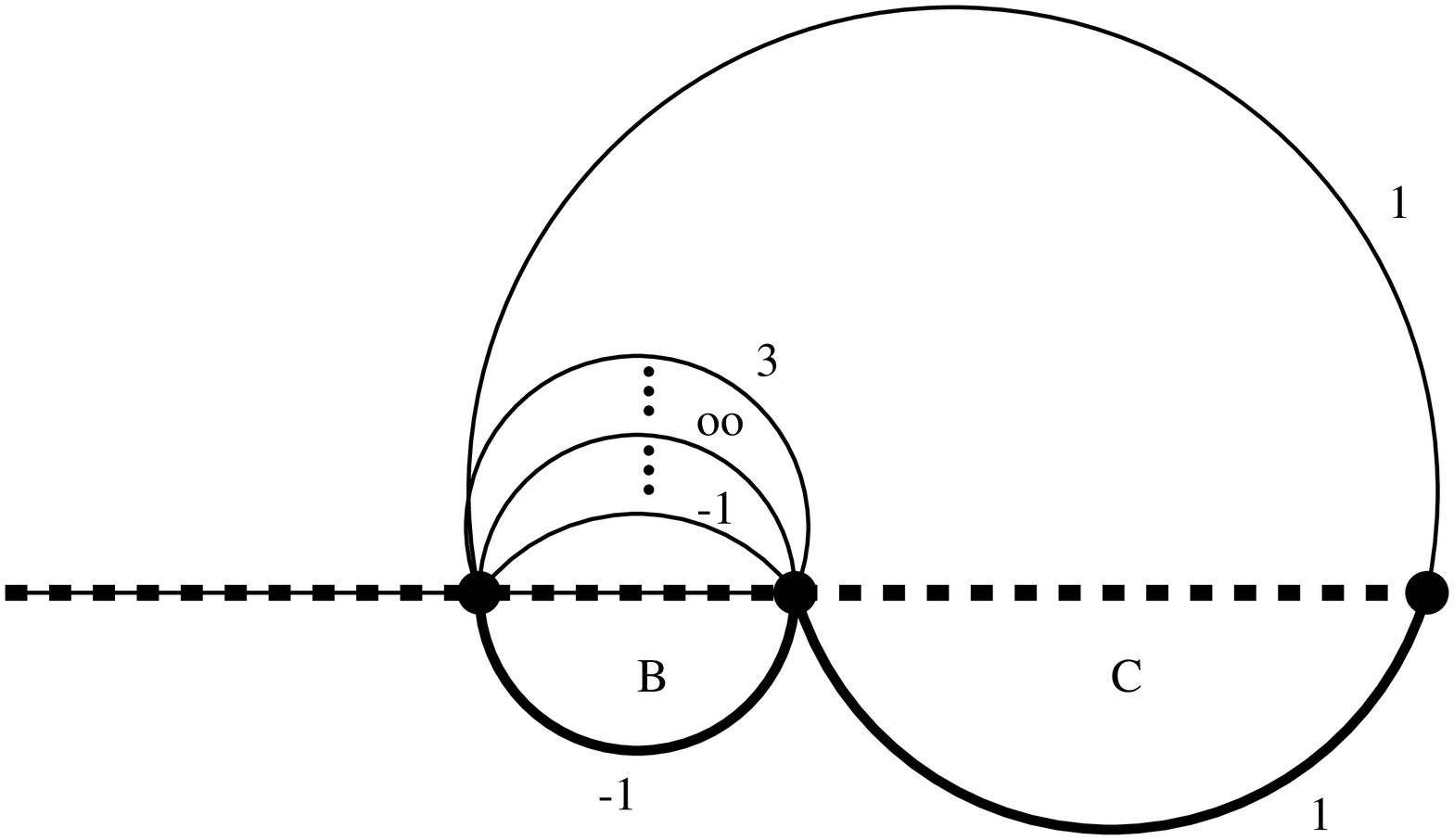}{7cm}
\figlabel\figxxii 
First, look at the dyon $(1,1)_1$ for which the curves are as shown in the left
Fig. \figxxii. All curves go
through $\s_2$ where this dyon is massless when approached from $\e>0$, and we see that it must
exist in the whole upper half plane. It cannot exist in region $B'$ since otherwise this would
imply the existence of $M_3^* (1,1)_1=(3,1)_3$ above the cut. Note that conversely, the existence
of $(1,1)_1$ in $B$ only implies the existence of $(M_3^*)^{-1} (1,1)_1=(-1,1)_{-1}$ below the cut
which is perfectly consistent (see Fig \figxix). 
The decay across the $r_{\rm d}=1$ curve for $\e<0$ (${\cal C}^-_0$) is
$(1,1)_1\to 2\times (1,0)_1 + (-1,1)_{-1}$ into final states that are massless in the lower half
plane at $\s_3$ and $\s_2$.

Next, consider the dyon $(-1,1)_1$ with curves shown in the right Fig. \figxxii. 
All curves in the upper
half plane go through $\s_1$ where this dyon is massless. Hence it exists everywhere in the upper
half plane. It cannot exist in $B$ or in $C$ since $M_{2\cdot 3}^* (-1,1)_1=(1,1)_3$ and $M_3^*
(-1,1)_1=(1,1)_3$. The decays in the lower half plane are $(-1,1)_1\to (-1)\times (1,0)_{-1} +
(0,1)_0$ on $r_{\rm d}=-1$ (${\cal C}_0^+$) and 
$(-1,1)_1\to 2\times (1,0)_1 +(-3,1)_{-1}$ on $r_{\rm d}=1$ (${\cal C}^-_{-2}$).

\def\doublefig#1#2#3#4#5{
\par\begingroup\parindent=0pt\leftskip=1cm\rightskip=1cm\parindent=0pt
\baselineskip=11pt
\global\advance\figno by 1
\midinsert
\centerline{\hbox{\epsfxsize=#3\epsfbox{#2}
\hskip 1.cm 
\epsfxsize=#5\epsfbox{#4}}}
\vskip 12pt
{\bf Fig. \the\figno:} #1\par
\endinsert\endgroup\par
}
\def\figlabel#1{\xdef#1{\the\figno}}
\def\encadremath#1{\vbox{\hrule\hbox{\vrule\kern8pt\vbox{\kern8pt
\hbox{$\displaystyle #1$}\kern8pt}
\kern8pt\vrule}\hrule}}
\doublefig{Shown is a sketch, for $m>{\ld\over 2}$, 
of the relative positions of the  decay curves of a generic dyon 
$(2n+1,1)_1$ with $n\ge 1$ in the left figure and $n\le -2$ in the right figure.
We have not shown the curves  inside the $r=\infty$
curves. The thick curves are the relevant ones.}{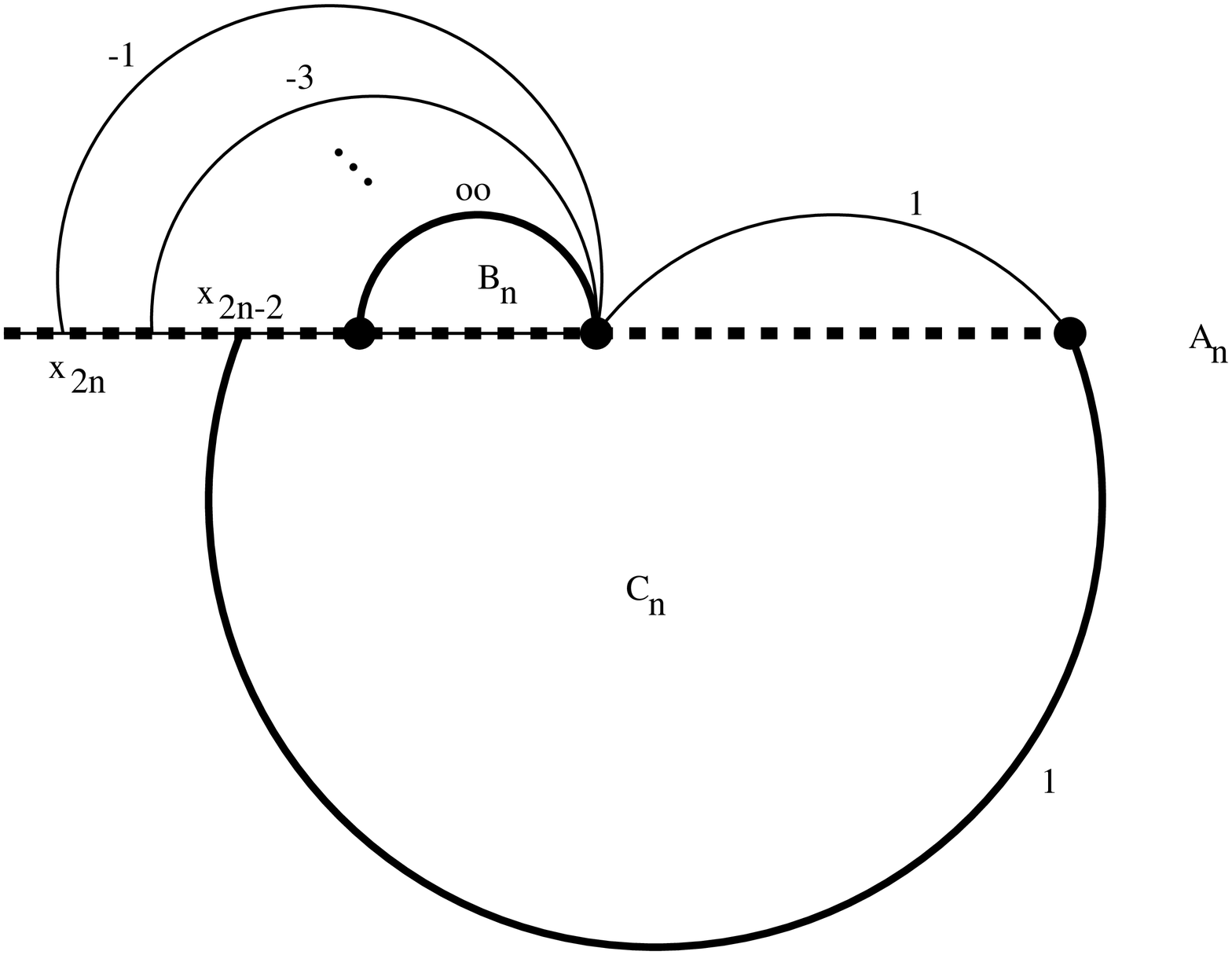}{7cm}{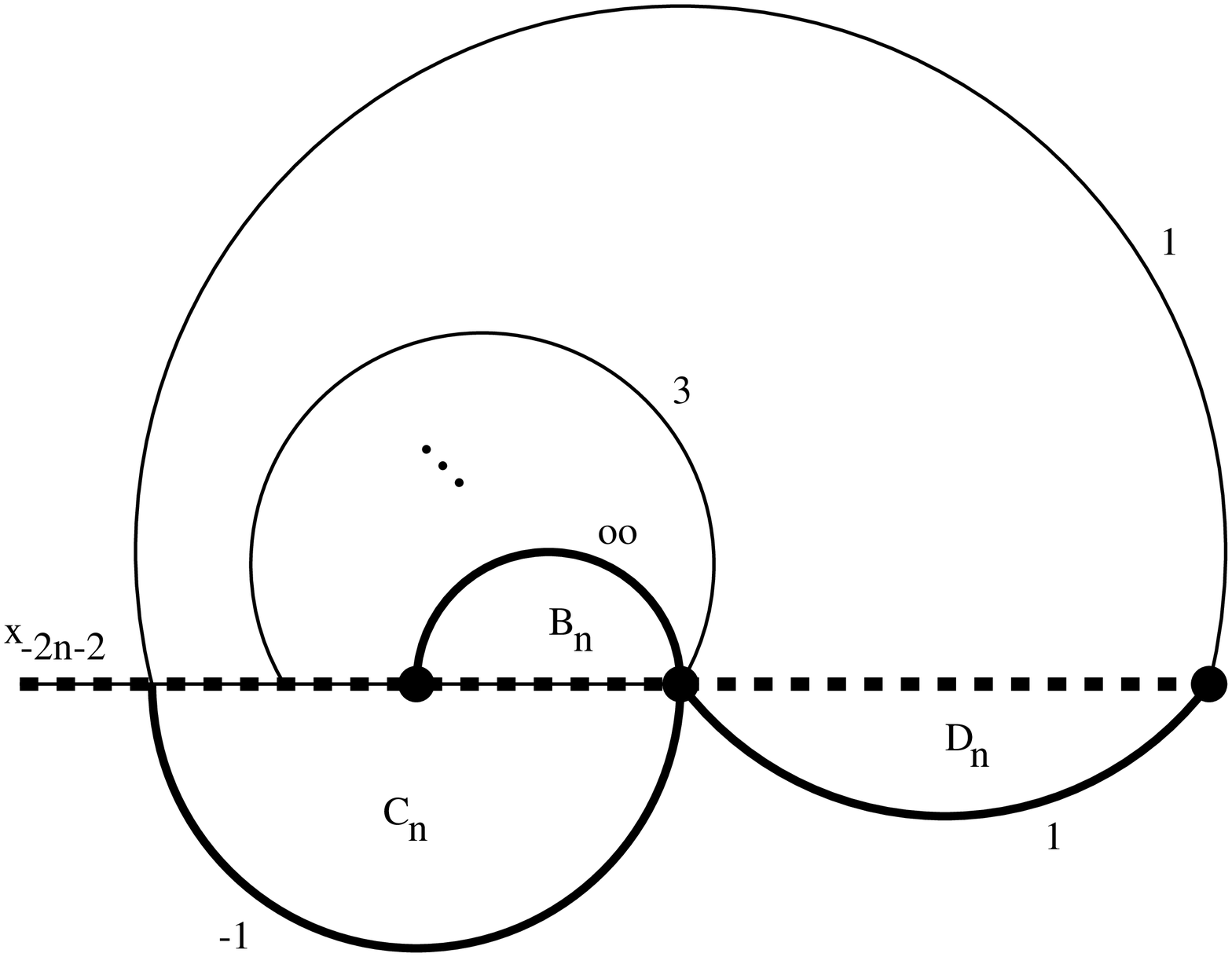}{7cm}
\figlabel\figxxiv 
Now, for a generic dyon $(2n+1,1)_1$ with $n\ge 1$ one has the situation of 
the left Fig. \figxxiv. Of
course, for $n=1$, i.e. for the dyon $(3,1)_1$, the $r=1$ curve for $\e<0$ starts at $x_0\equiv
\s_1$, but this makes not much difference. Decays are only possible on the $r=\infty$ curve and the
$r=1$ curve for $\e<0$. By the same arguments as above it is clear (for $n\ge 1$) that $(2n+1,1)_1$
cannot exist in regions $B_n$ or $C_n$. The decays are $(2n+1,1)_1\to (n+1)\times (1,1)_1 +
(-n)\times (-1,1)_1$ on the $r=\infty$ curve which is ${\cal C}^\infty$ in the upper half plane,
and $(2n+1,1)_1\to 2\times (1,0)_1+(2n-1,1)_{-1}$ on the $r=1$ curve which is ${\cal C}^-_{2n}$ in
the lower half plane.

Next, for the dyons $(2n+1,1)_1$ with $n\le -2$, we have the situation of 
the right Fig \figxxiv.
The kinematically possible decays are:
$(2n+1,1)_1\to 2\times (1,0)_1 +(2n-1,1)_{-1}$ on the $r_{\rm d}=1$ curve for $\e<0$ (${\cal C}^-_{2n}$), 
$(2n+1,1)_1\to (-2)\times (1,0)_{-1}+ (2n+3,1)_{-1}$ 
on the $r_{\rm d}=-1$ curve for $\e<0$ (${\cal C}^+_{2n+2}$), and 
$(2n+1,1)_1\to (-n)\times (-1,1)_1 +(n+1)\times (1,1)_1$ on the $r_{\rm d}=\infty$ curve for $\e>0$ (${\cal
C}^\infty$). By the $(M_{2\cdot 3}^*)^{\pm 1}$ and $M_3^*$ arguments one sees that $(2n+1,1)_1$ with $n\le
-2$ cannot exist in regions $B_n,\ C_n$ and $D_n$.

The dyons $(2n+1,1)_{-1}$ exist in the complex conjugate domains of where the $(-2n-1,1)_1$ exist.
All this is shown in Fig. \figxix.

\subsection{The quarks $(1,0)_{\pm 1}$}

\def\doublefig#1#2#3#4#5{
\par\begingroup\parindent=0pt\leftskip=1cm\rightskip=1cm\parindent=0pt
\baselineskip=11pt
\global\advance\figno by 1
\midinsert
\centerline{\hbox{\epsfxsize=#3\epsfbox{#2}
\hskip 1.cm 
\epsfxsize=#5\vbox{\epsfbox{#4}\vskip 1.0cm\null}}}
\vskip 12pt
{\bf Fig. \the\figno:} #1\par
\endinsert\endgroup\par
}
\def\figlabel#1{\xdef#1{\the\figno}}
\def\encadremath#1{\vbox{\hrule\hbox{\vrule\kern8pt\vbox{\kern8pt
\hbox{$\displaystyle #1$}\kern8pt}
\kern8pt\vrule}\hrule}}
\doublefig{Shown is a sketch, for $m>{\ld\over 2}$, of the relative positions of the  decay curves of the quark $(1,0)_1$
(left) and of the quark $(1,0)_{-1}$ (right). For the latter,
we have not shown the curves $r>0$ inside the $r=\infty$
curve. The thick curves are the relevant ones.}{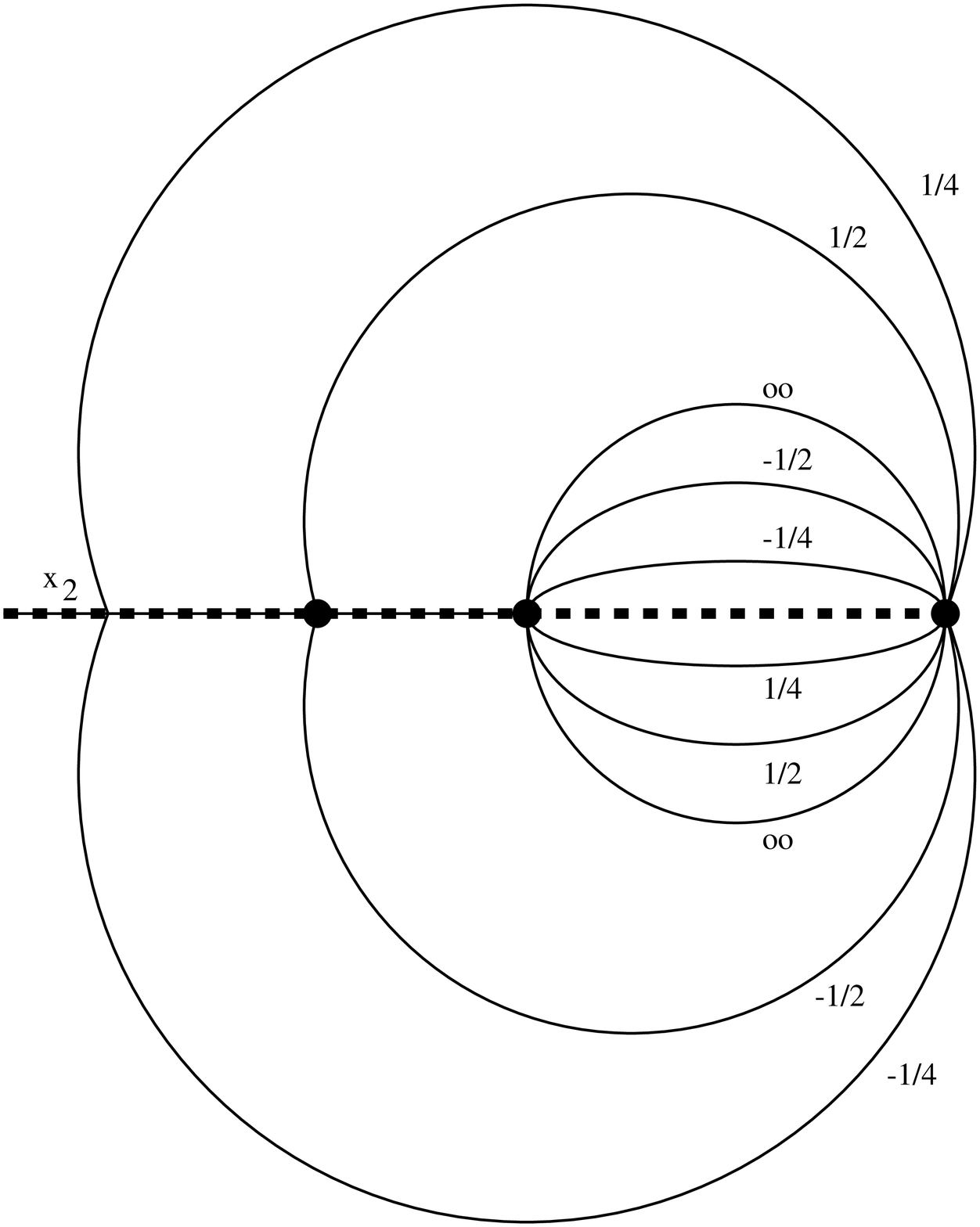}{6cm}{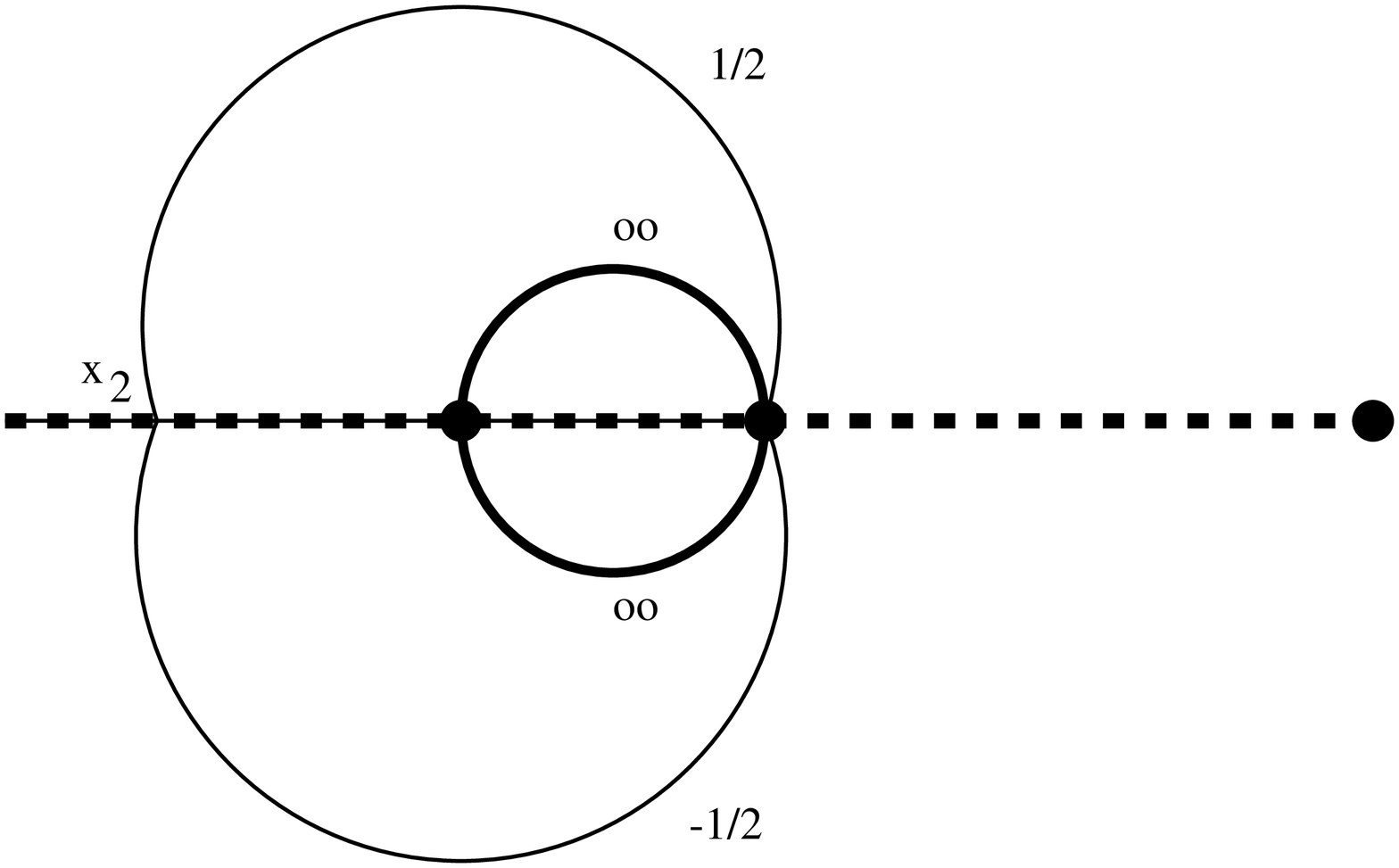}{7cm}
\figlabel\figxxvi 
First, the quark $(1,0)_1$ is massless at $\s_3$. All curves go through $\s_3$, see Fig. \figxxvi, and we
conclude that this quark exists everywhere!

For the other quark $(1,0)_{-1}$, the curves are shown in the right Fig. \figxxvi. Decays are only possible on the
$r=\infty$ curve (${\cal C}_0^+$) where they are $(1,0)_{-1}\to (0,1)_0+(-1) \times (-1,1)_1$ for $\e>0$
and $(1,0)_{-1}\to (-1)\times (0,1)_0 +(1,1)_{-1}$ for $\e<0$. The quark $(1,0)_{-1}$ cannot exist inside
the $r=\infty$ curve since again this would imply the existence of a state $(M_{2\cdot 3}^*)^{\pm 1} (1,0)_{-1} =
(0,\mp 1)_{-2}$ that does not exist.

\subsection{The W-boson $(2,0)_0$}

For the W-boson, there is no $r=\infty$ curve, and the $r=\pm 1$ curves exist only for 
$\e=\pm 1$, where
they actually are ${\cal C}^\infty$. By the $(M_{2\cdot 3}^*)^{\pm 1}$ argument, 
the W-boson cannot exist
inside ${\cal C}^\infty$. The decays are $(2,0)_0\to (1,1)_\e + (-1)\times (-1,1)_\e$.
{\bf \chapter{Physical discussion : the $N=2$ superconformal fixed points}}
\sectionnumber=0
In the light of our preceding results, we now discuss the physics of 
the superconformal points. 

At a point $u_{*}$ on the Coulomb branch where mutually non-local dyons 
become massless, the low energy theory is believed to be an 
interacting superconformal field theory [\AD]. In the theories we 
studied in this paper, such a point occurs when two singularities coincide 
[\APSW]. Near $u_{*}$ which 
always lies in a region where the low energy theory is 
strongly coupled (independently of the choice of variables), the masses 
of the particles becoming massless at $u_{*}$ are much smaller than 
the masses of all the other  particles (which are of order $\Lambda$) 
and set a new mass scale $M$. This implies that one can give an intrinsic 
meaning to the superconformal field theory (SCFT) independently of its embedding in 
the original non-abelian gauge theory, by letting 
$\Lambda\rightarrow\infty$. Actually, the {\it same} $N=2$ SCFT can be 
embedded in different $N=2$ non-abelian gauge theories.
Argyres and Douglas were able to show, 
in a particular but generic case that corresponds in our language of 
Section 2.1 to a $k=1$ SCFT, that the low energy coupling does 
not depend on the separation of scales $\Lambda /M$. This is a 
strong indication that the coupling does not run between the scales 
$M$ and $\Lambda$, and thus  very convincing evidence that
the theory indeed has conformal invariance. One of the most striking 
properties of these SCFTs is that they do not contain any gauge boson 
that could contribute to the $\beta$ function with a minus sign. We have seen
this in great detail in Section 4 for the $k=2$ superconformal point appearing in the 
$N_f=2$ theory, since the only 
spin one particle at this point has quantum numbers $(2,0)_{0}$ and a 
mass $\ld /\sqrt{2}\ne 0$. It is easy to realize that this
is also the case at any superconformal point appearing in the $1\leq N_f\leq
3$ theories. To explain that the $\beta $ function can nevertheless be zero,
Argyres and Douglas [\AD] suggested a simple ansatz which
consists in computing the perturbative
contribution of each hypermultiplet which occurs in the SCFT separately.
This is done 
by using duality to go to a formulation of the theory where the given hypermultiplet 
is described locally, and then to apply the inverse duality transformation
to the contribution to the beta function. 
The individual terms are then simply added 
to obtain the total $\beta$ function: 
$$ {\partial \tau\over\partial\log\mu } = -{i\over 2\pi}\
\sum \Bigl( n_m \tau -n_e \Bigr) ^2.\eqn\betaad $$
where each state $(n_e,n_m)_s$ in the sum should be counted with its correct
multiplicity $d$. Of course, it is by no means obvious why this should give the 
correct answer.\foot{
An early and somewhat related discussion can be found in ref. \CARDY .}
In this equation, $\tau =\theta /\pi + 8i\pi /g^2$ is the (generalized)
coupling of a theory containing the hypermultiplets $(n_e,n_m)$ over which
the sum is done. This is not the low energy coupling of the original gauge
theory which, to avoid confusion, we will  hereafter  denote
$\tau _{\rm eff}=da_D /da$ . 

To discuss the validity of this ansatz \betaad\ remark that it
has three immediate and important consequences noted in [\AD].
The first one is that the $\theta$ angle does run, due to the contribution of
magnetically charged states. This is quite surprising in a U(1) theory, and
this effect may well remain qualitatively
valid and have interesting
consequences in non-supersymmetric theories as well, perhaps by shading a new
light on the strong CP problem. However, we will have nothing new to say
about this in the following. The second consequence of \betaad\ is that 
we have an RG fixed point with a fixed point coupling $\tau _*$ satisfying
$$ \sum \Bigl( n_m \tau _* -n_e \Bigr) ^2 =0. \eqn\sumrule $$
The third consequence is that this fixed point is IR stable, the slope 
$\omega$ of the $\beta$ function being positive, with \betaad\ giving
$$ \omega = \RE\lim _{\tau\rightarrow\tau _*} {1\over \tau -\tau _*}
{\partial (\tau -\tau _*)\over\partial\log\mu} =
{\sum n_m^2\over\pi}\, \IM\tau _*>0 .\eqn\slopead$$
The positivity of $\omega$ is also required by unitarity [\MACK].
In the following, we will argue that \sumrule\ is correct, but that
\slopead\ is wrong. This was already suspected in [\AD], since \slopead\
gives irrational values for $\omega$. We will see that the correct value for 
$\omega$ is indeed a rational number.

Before presenting our discussion, it is necessary to stress the following: although
the effective coupling $\tau _{\rm eff}(u)$ and the running coupling $\tau _u (\mu
)$ of the microscopic theory at fixed $u$ and scale $\mu$
are different physical quantities,
they are nevertheless related. The easiest way to understand
this is to first choose $u$ near a singular point $u_0$ where only locally
related states become massless, a familiar case.
At fixed $u$, the particle spectrum of the
theory consists of the BPS states becoming massless at $u_0$
and having  masses $M\sim t(u-u_0)$ where $t$ is a constant, and of other
states of mass $\sim \Lambda$ or higher. The dependence of $M$ on $u$
simply means that $u$ is a good local coordinate near $u_0$, or, in terms
more appropriate for the discussion to come, that $u$ has dimension one. This
is equivalent to the fact that the low energy theory is {\it free} massless
super-QED. The effective coupling $\tau _{\rm eff}$ for a given $u$ then
corresponds to the microscopic running coupling $\tau _u(\mu )$ at a scale
$\mu \sim M$ or lower. Between the scales $M$ and $\Lambda $, $\tau _u(\mu)$
is given reliably, after a duality transformation which renders the
theory weakly coupled in the IR,
by a one loop calculation
in ordinary $N=2$ super-QED coupled only with the 
states of masses $\sim M$. One then identifies $\tau _{\rm eff}(u)$
with $\tau _u (\mu =M)$ to find the asymptotic behaviour of $\tau _{\rm eff}$
near the singularity [\SWI ,\SWII]. Similarly, using the same argument
backwards, and repeating this reasoning
near a superconformal point $u_*$ where $\tau _{\rm eff}$ is known via our
explicit formulas (e.g. \dxxv\ for the $k=2$ SCFT), one can deduce the form
of $\tau _u (\mu)$. Hence the effective coupling at $u_*$ must
coincide with the fixed point coupling $\tau _*$: 
$$\tau _{\rm eff}(u_*)=\tau_* \ .
\eqn\fpc$$
Furthermore, we will see below that $\tau _{\rm eff}$ has the following asymptotic
behaviour,
$$ \tau _{\rm eff}(u) = \tau _* + C (u-u_*) ^{\gamma} + o \bigl( 
(u-u_*)^{\gamma} \bigr).\eqn\tauasym$$
Moreover, it follows from eq. \dxii \ that at $u$
the masses of the particles becoming
massless at $u_*$ are $M\sim t (u-u_*)^{1/\alpha}$, and that
$\alpha $ is the anomalous dimension of $u$. We deduce that
$\tau - \tau _* \sim \mu ^{\alpha\gamma}$ and thus that the slope $\omega$ of the
$\beta$-function is given by
$$\omega = \alpha\gamma .\eqn\ome$$

Let us first exploit the fact that $\tau _{\rm eff}(u_*)=\tau _*$ to discuss
\sumrule . Naively, at a superconformal point where two singularities
coincide, one has precisely two massless states. For instance, at the $k=3$
superconformal point of the $N_f=3$ theory depicted in Fig. 2, 
one would expect to 
have the states $(1,1)_2$ and $(1,2)_3$. However, with these
states, \sumrule\ does not give the correct answer for the fixed point
coupling, which in this case is $\tau _*= e^{i\pi /3}=\tau _{\rm eff}
(u_*)$. The solution to this puzzle is not to give up \sumrule ,
but to realize that
there are other states which become massless at the superconformal points. 
One way to find them is to look for massless states in
the maximal set of BPS states of the $N_f=3$ theory. This is correct
because one can show, along the lines of Section 4, that
all the states belonging to the maximal set must exist
at the superconformal point. Thus we deduce that we have six massless states,
three triplets $(1,1)_2$, $(0,1)_1$ and $(1,0)_1$, and three singlets 
$(1,2)_3$, $(-1,1)_0$ and $(2,1)_3$. With these states \sumrule\ gives the
right answer. One interesting point to note is the following: the states being
massless at the potentially coinciding singularities 
for small masses ($m<\lt /8$), i.e.
$(1,1)_2$ and $(1,2)_3$, and the states becoming massless for large
masses ($m>\lt /8$), i.e. $(-1,1)_0$ and $(1,0)_1$, are
four {\it distinct} states and not simply analytic continuations of each other. 
The mechanism which allows to change the quantum numbers at the
singularities in the asymptotically free theories, a phenomena which 
was known to occur since the work of Seiberg and Witten [\SWI ,\SWII], 
is thus fully elucidated: at a point where two singularities collide,
new states become massless ``accidentally''
and, when the singularities split again, these new states
can take over the r\^oles of the original massless states which were
responsible for the singularities before the collision.

\def\doublefig#1#2#3#4#5{
\par\begingroup\parindent=0pt\leftskip=1cm\rightskip=1cm\parindent=0pt
\baselineskip=11pt
\global\advance\figno by 1
\midinsert
\centerline{\vbox{\epsfxsize=#3\epsfbox{#2}
\vskip 1.cm 
\epsfxsize=#5\vbox{\epsfbox{#4}\vskip 1.0cm\null}}}
\vskip 12pt
{\bf Fig. \the\figno:} #1\par
\endinsert\endgroup\par
}
\def\figlabel#1{\xdef#1{\the\figno}}
\def\encadremath#1{\vbox{\hrule\hbox{\vrule\kern8pt\vbox{\kern8pt
\hbox{$\displaystyle #1$}\kern8pt}
\kern8pt\vrule}\hrule}}
\doublefig{The curves of marginal stability and stable BPS states in the
upper half $u$-plane near the
singularities in the $N_f=2$ theory with $m_1=m_2=m$. The upper configuration 
corresponds to $m<\ld /2$ and the lower configuration
to $m>\ld /2$.}{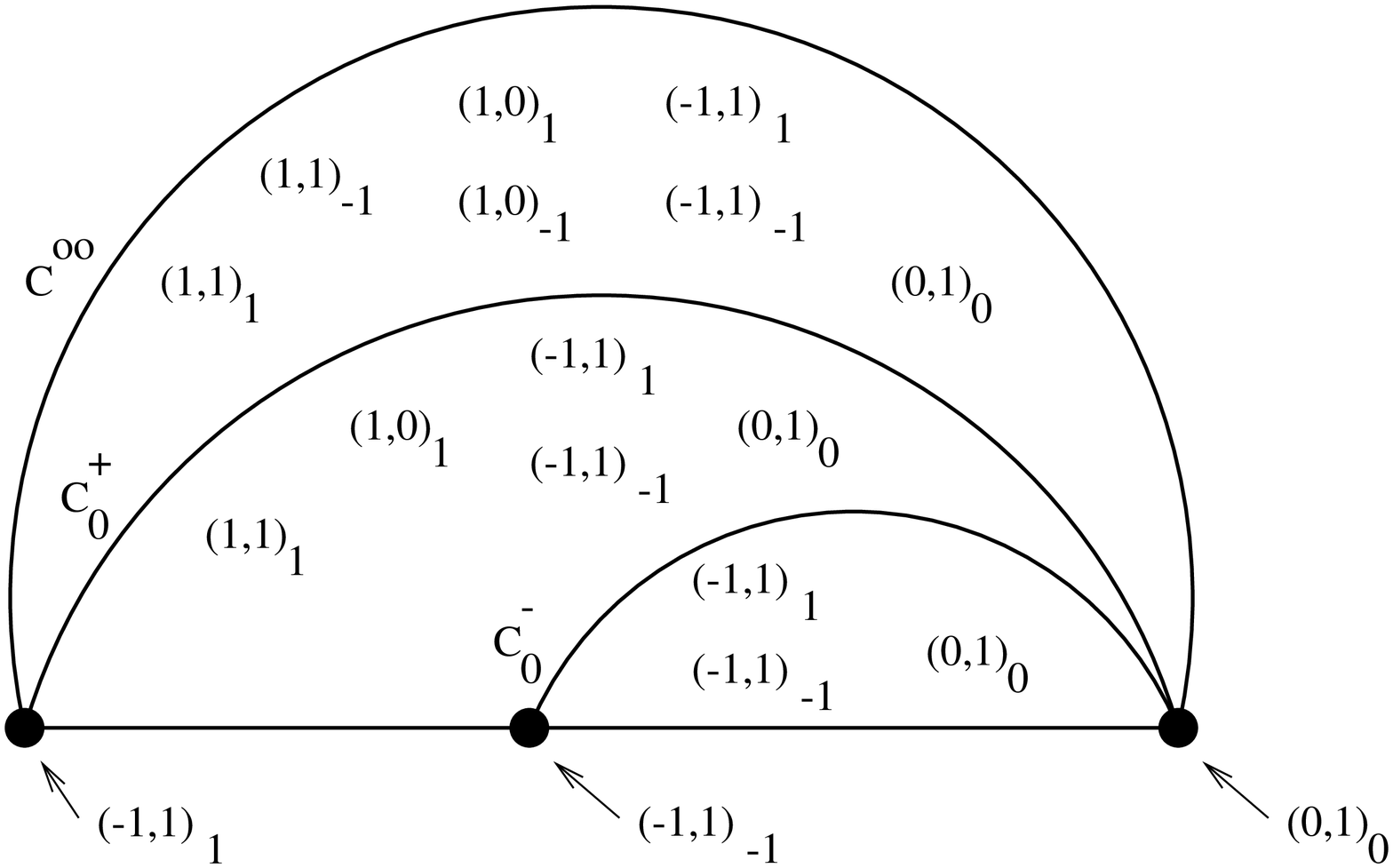}{13cm}{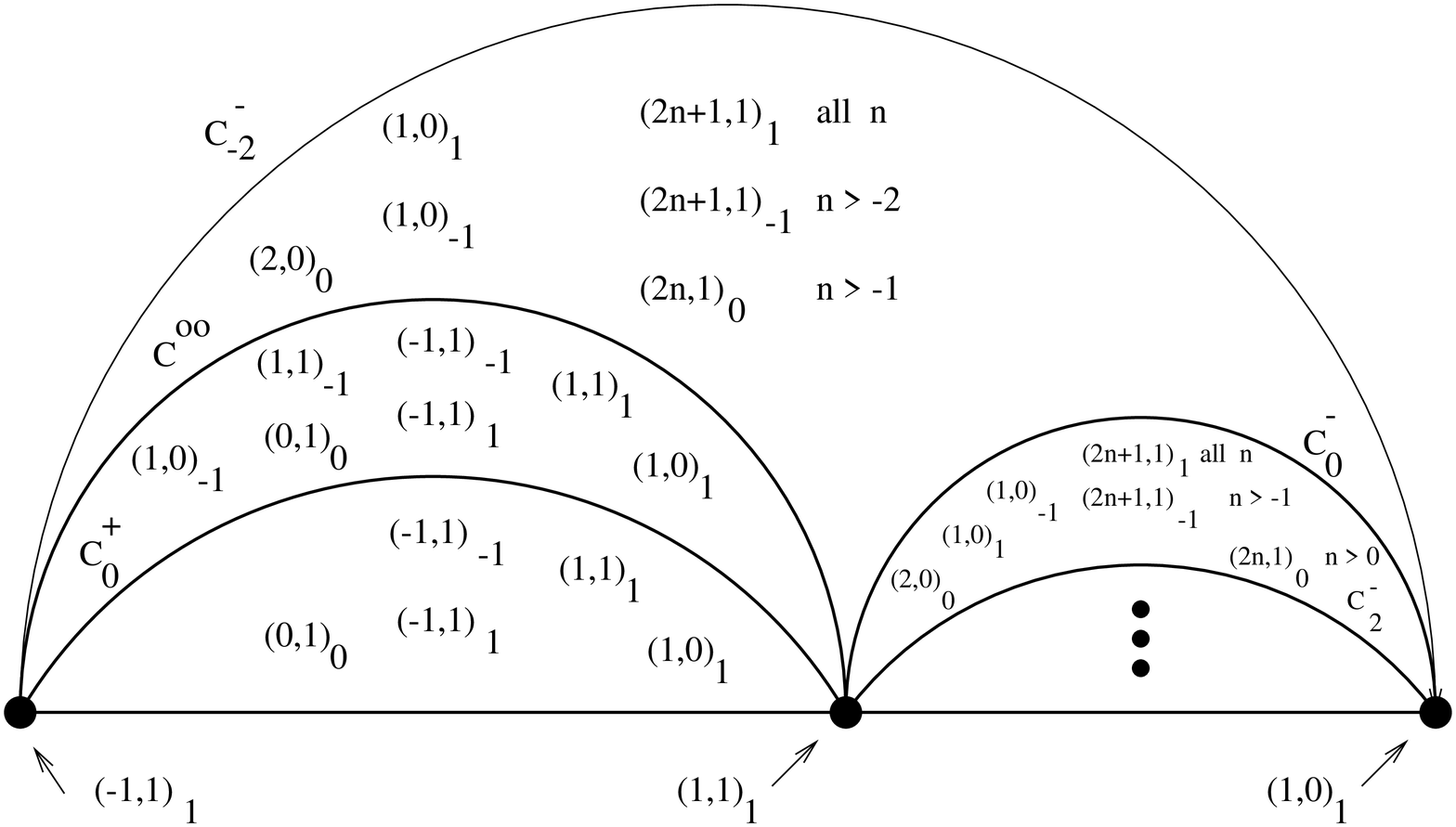}{13cm}
\figlabel\figdoublet
Let us illustrate this mechanism in more detail in the case of the $N_f=2$
theory studied in Section 4. Figure \figdoublet\  represents an enlargement of
Figs. \figxviii\ and \figxix\ near the singularities. 
We explicitly indicate the existing
states in the different regions surrounded by the curves of marginal
stability. Since $\ad=a-{m\over \rd}=0$ at the superconformal point, 
it is clear from Figure \figdoublet\ 
that the states becoming massless at this point
are the two singlets $(-1,1)_{-1}$ and $(1,1)_1$, and the two
doublets $(0,1)_0$ and $(1,0)_1$. Equation
\sumrule\ then  gives the correct fixed point coupling, $\tau _*=i$.
When $m<\ld /2$, the states responsible for
the singularities are $(-1,1)_{-1}$ and $(0,1)_0$, and exist everywhere on
the Coulomb branch. The states $(1,1)_1$ and $(1,0)_1$ are also very special,
since they exist everywhere except inside the innermost curve ${\cal C}_0^-$.
This is a nice test of the spectra derived in Section 4, since $(1,1)_1$ and
$(1,0)_1$, being massless at the superconformal point, must exist everywhere
except possibly inside curves of marginal stability shrinking to zero when
such a point is approached. This is indeed what happens for the curve ${\cal
C}_0^-$. When $m>\ld /2$, the r\^oles of the states $(-1,1)_{-1}$, $(0,1)_0$
and $(1,1)_1$, $(1,0)_1$ are exchanged but the picture  still is perfectly
coherent because $(-1,1)_{-1}$ and $(0,1)_0$ again exist everywhere outside ${\cal
C}_0^-$. 

Let us  mention that at the $k=1$ SCFT appearing in the
$N_f=1$ theory, we have three massless states which correspond exactly
to the three states which were identified in [\AD] at the $\Z _3$ point of
the SU(3) moduli space. This was expected since the two SCFTs are believed [\APSW] to
be the same.
It would be interesting to check whether such a correspondence between
states still holds for the $k=2$ and $k=3$ SCFT when these theories are
embedded in the SU(4) or SO(8) pure gauge theories [\ADEN].

There is still one unsatisfying point in the above discussion. Though we saw
that \sumrule\ gives the correct fixed point coupling, its application
requires the knowledge of the set of all the massless states, 
which from the above discussion seems to depend crucially on the theory in
which the SCFT is embedded. This is not natural since $\tau _*$ is, at least
modulo a duality transformation, a characteristic of the SCFT.
Actually, the massless spectra can be deduced from the local data only. Indeed,
as the original massless states exist everywhere on the Coulomb branch and
cannot decay, one can produce new massless states by encircling the
superconformal singular point. The massless spectra must then be invariant
under the monodromy $M_{\rm sc}$ at the superconformal point. This invariance
only generates a finite number of new states because $M_{\rm sc}$ is of
finite order. Using this invariance, one can deduce straightforwardly
the massless BPS spectra uniquely from the quantum numbers of the colliding
singularities. For instance, by applying \monodsc\ on the states
$(-1,1)_{-1}$ and $(0,1)_0$, one generates exactly the two missing states
$(1,1)_1$ and $(1,0)_1$. 

It is very tempting to conjecture that the matrix
$M_{\rm sc}$ always 
corresponds to an exact quantum duality symmetry of the SCFT. 
In addition to the invariance of the spectrum, this is supported by the 
(related) fact
that the fixed point couplings satisfy $M_{\rm sc}\cdot\tau _* =\tau _*$ and
by the fact that the symmetry $M_{\rm sc}$ can be realized as a global
symmetry e.g. as  the $\Z _3$ symmetry of the $k=1$ SCFT when 
embedded in the SU(3) pure gauge theory.  
Note that the surprising
idea that global symmetries can be related to non-trivial
duality transformations was first realized and used
in [\FB] in the case of the $\Z _2$ symmetry acting on 
the moduli space of the SU(2) pure gauge theory. 
We find a very explicit realization of this fact with the superconformal
points studied in this paper.
Very recently, this idea has also been used very nicely to 
study $S$ duality in some finite
gauge theories [\FGT]. We strongly feel that this is a very interesting point
of view, since it relates a priori purely quantum symmetries to more
conventional symmetries which have a classical origin. 

Let us now turn to the computation of $\omega$. 
We will also obtain $\alpha$, recovering the results of [\APSW]. 
To compute these two critical exponents, it is enough to find the asymptotic
expansions of $a_D$ and $a$ near $u_*$ up to the second order.
An easy way to deduce
the result is the following: diagonalizing the monodromy matrix $M_{\rm sc}$, its
eigenvalues are $x_{\pm} =\exp (\pm {i\pi\over 6}(1+k))$, which shows that
the leading power in the expansion,  corresponding to $1/\alpha$, see
\dxii, must be of the form 
$$ {1\over\alpha}=\pm {1+k\over 12} + n \ , \quad n\in\Z .$$
Imposing $\alpha\geq 1$ which is a necessary condition for a unitary SCFT
[\APSW], and $\alpha\leq 2$ which amounts to saying that the operator $u$ is
relevant, the only possibility is that $\alpha$ is related to the eigenvalue
$x_-$ with $n=1$,
$$\alpha = {12\over 11-k}\cdotp\eqn\exalpha$$
The other eigenvalue $x_+$ is related to the subleading power 
$1/\beta>1/\alpha$ in the expansion by
$${1\over\beta}= +{1+k\over 12} + 1={13+k\over 12}\cdotp\eqn\exbeta$$
The formulae \exalpha\ and \exbeta\ can be checked using the
explicit solutions given in \dxxv\ for $N_f=2$ or those in  
Appendix C for the other cases. 
By using 
$$\eqalign{
a_D(u) =& a_D(u_*) + c_D (u-u_*)^{1/\alpha} + c_D' (u-u_*)^{1/\beta}
+ o\bigl( (u-u_*)^{1/\beta} \bigr) \cr
a(u) = & a(u_*) + c (u-u_*)^{1/\alpha} + c' (u-u_*)^{1/\beta}
+ o\bigl( (u-u_*)^{1/\beta} \bigr)\cr}$$
and the relation $\tau _{\rm eff} = da_D/da$, we deduce that
the exponent $\gamma$ of \tauasym\ is given by $\gamma = 1/\beta -
1/\alpha$, and thus $\omega = \alpha\gamma$ is 
$$\omega = {2+2k\over 11-k}\cdotp\eqn\exomega$$
This is a rational number which differs from \slopead . This implies that
despite its appealing features, the Argyres-Douglas ansatz \betaad\
cannot be exact. The problem of finding the general form of the $\beta$
function for these non-local field theories thus remains open. Maybe one can
guess the result by using the exact solutions for $\tau _{\rm eff}$
presented in this paper around the superconformal points and use the
arguments above to relate $\tau _{\rm eff}$ and $\tau$.

A simple inspection of the formulae \exalpha\ and \exomega\ reveals that the
two exponents $\alpha$ and $\omega$ are related by
$$ \omega = 2(\alpha -1).\eqn\screl$$
Such a ``scaling" relation is possible because, due to $N=2$ supersymmetry, the only free
parameter is $k$ and thus one expects only one independent critical exponent.
It is actually possible to understand \screl\ on  general grounds, 
following the line of reasoning of [\APSW]: if we denote by $U$ the U(1)
$N=2$ vector superfield contained in the low energy effective action at a
generic point on the Coulomb branch, a deviation of the coupling constant
from its fixed point value corresponds to the irrelevant (since the fixed
point is IR stable) operator
$$ \delta L _{\rm eff} = \Lambda ^{-\omega} \int d^4 \theta\,  U^2,$$
where the fermionic integration is performed over half of $N=2$ superspace.
As the dimension of $L _{\rm eff}$ must be 4 since the action is
dimensionless, and the dimension of $d^4 \theta $ is two, we must have
$-\omega + d_{U^2} =2$ where $d_{U^2}$ is the dimension of the operator
$U^2$. Because of {\it super}conformal invariance, we must have
$d_{U^2}=2d_U=2\alpha $ because these scaling
dimensions are directly proportional to the R charges of $U$ and $U^2$ 
under the exact {\it quantum} U(1)$_R$ symmetry of the superconformal 
algebra. We thus recover the relation \screl . Conversely, one can consider
\screl\ as a non-trivial test that the theory has indeed $N=2$ superconformal
invariance.

\endpage

{\bf \Appendix{A}}

For reference, in this appendix we briefly discuss the 
positions of the singularities and their flows for the  massive $N_f=1,2,3$
theories. For simplicity, all non-zero bare masses are 
taken to be equal.

\subsection{$N_f=3$ with $m_1=m_2=m_3\equiv m$}

For $m\ne 0$ the flavour symmetry is ${\rm SU}(3)$, while for $m=0$ it is 
${\rm spin}(6)\simeq {\rm SU}(4)$. 
For the present case, the discriminant of the cubic polynomial \di, \dii\ in $x$ is
$$\eqalign{\Delta={\lt^2\over 16} & \left(-u+m^2+{\lt\over 8} m\right)^3 \cr
& \times \left[ u^2 +\left( {3\over 8}\lt m -{\lt^2\over 256}\right) u
-\lt m^3 -{3\over 256}\lt^2 m^2 -{3\over 2048}\lt^3 m \right] }
\eqn\dvi$$
showing that there is a triple singularity $\s_t$, 
transforming as the {\bf 3} of ${\rm SU}(3)$,
and two singlet singularities $\s_\pm$
at
$$\s_t = m^2+{\lt\over 8} m\quad , \quad 
\s_\pm = -{3\over 16} \lt m + {\lt^2\over 512} 
\pm \lt^{1/2} \left( m+ {\lt\over 64}\right)^{3/2} \ .
\eqn\dvii$$
For small $m$ one has $\s_+ \simeq {\lt^2\over 256}$ and $\s_- \simeq -{3\over 8}
\lt m$, and at  $m=0$, $\s_-$ and $\s_t$ coincide and one has a quadruple singularity at $u=0$
corresponding to a massless dyon of magnetic charge one 
which transforms as the {\bf 4} of ${\rm SU}(4)$.
The singularity at $\s_+$ is due to a massless dyon of magnetic charge two
[\SWII] (see Fig. \figiia). As $m$ is increased, the singlet $\s_-$ moves to the left and the triplet
moves to the right.
The singularities $\s_t$ and $\s_+$  meet at
$$m={\lt\over 8}\quad , \quad \s_t=\s_+={\lt^2\over 32}=2m^2 \ .
\eqn\dixbis$$
which is the superconformal point. As $m$ is increased further, as discussed in Section
2,
the quantum numbers at the singularities are changed and now are
$(1,1)_0$ and $(-1,1)_0$ at $\s_-$ and $\s_+$ and
a quark triplet at $\s_t$.
The latter disappears to infinity as $m\to\infty$, $\lt\to 0$,\ $m^3\lt=\lz^4$
fixed, while
$\s_\pm\to\pm \lz^2$, and indeed, (after rotating the quark cut to the right and
shifting $\ad\to\ad+a$) we are left with the pure gauge theory
$N_f=0$.

\subsection{$N_f=3$ with $m_1=m_2=0$ and  $m_3\equiv m$}

For this case, the roots of the cubic  are at $x=u$ and $x={\lt^2\over 128} 
\pm \left[ {\lt^2\over 256}\left( {\lt^2\over 64}
-4u+4m^2\right)\right]^{1/2}$.
Thus there is a singlet singularity at $\s_s$ and two doublet singularities at
$\s_d^\pm$:
$$\s_s=m^2+{\lt^2\over 256} \quad , \quad \s_d^\pm=\pm {\lt\over 8} m \ .
\eqn\dxiii$$
As $m$ is increased from 0, the
quadruple singularity at the origin splits into two doublets of massless
monopoles (having different values of $s$) while the singlet at $\s_s$ starts
to move to the right. But $\s_d^+$ moves faster to the right and meets $\s_s$
at the superconformal point
$$m={\lt\over 16}\quad , \quad \s_d^+=\s_s={\lt^2\over 128}=2 m^2 \ .
\eqn\dxv$$
As $m$ is increased beyond ${\lt\over 16}$, $\s_s$
remains larger than $\s_d^+$ and the singularity at $\s_s$ is now due to a
(singlet) massless quark, while the other two 
doublet states at $\s_d^+$ and $\s_d^-$ now are a magnetic monopole and the dyon
with $n_e=\mp 1$ both having vanishing $s$. As $m\to\infty$, $m\lt=\ld^2$ fixed, the
massless quark at $\s_s$ dissappears to infinity, while $\s_d^\pm\to \pm
{\ld^2\over 8}$, and one is left with the massless $N_f=2$ theory.

\subsection{$N_f=2$ with $m_1=m_2\equiv m$}

This case was discussed in great detail in the main
body of this paper, see Section 2.

%
%
%
%

\subsection{$N_f=2$ with $m_1=0$ and $m_2\equiv m$}

In this case the flavour symmetry is only ${\rm U}(1)\times {\rm U}(1)$ for $m\ne 0$. 
The discriminant of the cubic  is
$$\Delta={\ld^4\over 16} \left[ u^4 -m^2 u^3 -{\ld^4\over 32} u^2
+{9\over 64} \ld^4 m^2 u +{\ld^8\over 4096} -{27\over 256} \ld^4 m^4 \right] \ .
\eqn\dxxi$$
For $m=0$ we have two doublets of massless monopoles and dyons in one and the
other spinor representation of ${\rm spin}(4)$ at $u=\pm {\ld^2\over 8}$. For small
non-zero mass $m$ each of these doublets splits into two singlets at
$$\s_-^\pm = -{\ld^2\over 8} \pm {\ld\over 2}m + {\cal O} (m^2) \quad , \quad
\s_+^\pm = {\ld^2\over 8} \pm i {\ld\over 2}m + {\cal O} (m^2) \ .
\eqn\dxxii$$
\vskip 2.mm
\fig{The flow of the singularities on the complex $u$-plane
for $N_f=2$, $m_1=0$ and small $m_2=m$}{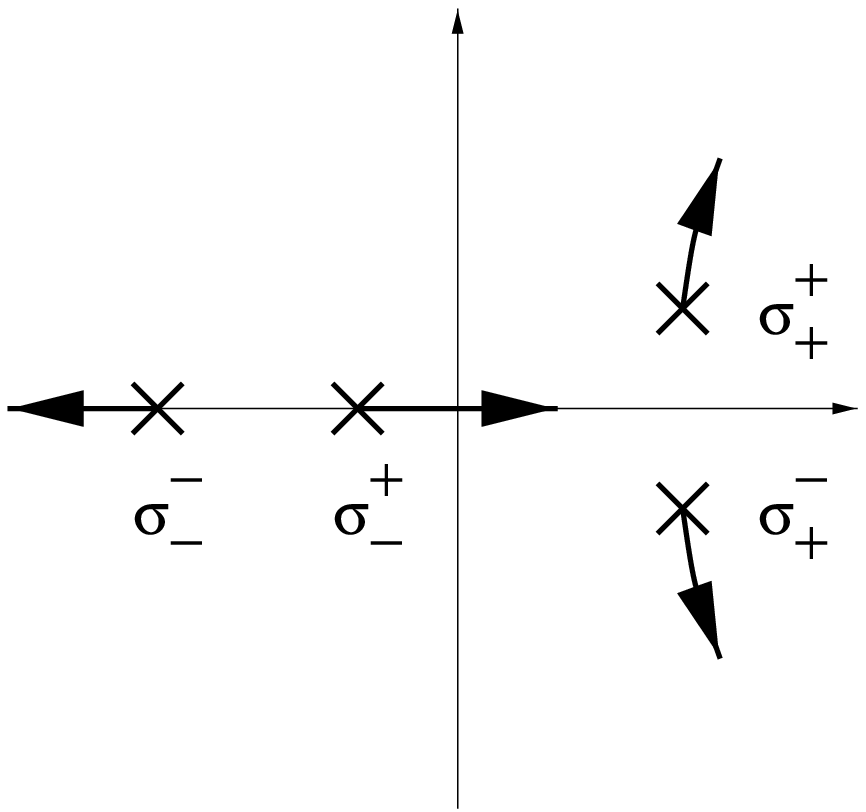}{4cm}
\figlabel\figi
\vskip 2.mm
Two singularities move off the real axis into the complex plane, see Fig. \figi. As
$m$ is increased $\s_-^+$ moves to the right on the real axis, much faster than
the other singularities. It passes between $\s_+^+$ and $\s_+^- = (\s_+^+)^*$. As
$m\to\infty$, $\s_-^+\to\infty$, while $\s_-^-$ and $\s_+^\pm$ arrange themselves
on a circle at the roots of $u^3+{27\over 256} \ld^4 m^2 =0$. Since
$m\ld^2=\lu^3$ these are precisely the singularities of the massless $N_f=1$
theory. In the present case, while flowing from $m=0$ to $m=\infty$, we never encounter a
superconformal point where singularities coincide. Nevertheless, the quantum
numbers of the massless states change by the monodromy matrix of another singularity as
explained in Section 2.

\subsection{The $N_f=1$ theory}

The discriminant is
$$\Delta={\lu^6\over 16}\left[ -u^3 +m^2 u^2 +{9\over 8} \lu^3 m u -\lu^3 m^3
-{27\over 256}\lu^6\right] \ .
\eqn\dxxiiibis$$
The positions of the singularities are given by the three roots of this cubic
polynomial in $u$:
$$\s_0=\a_++\a_-  +{m^2\over 3}\quad , \quad 
\s_\pm = {\rm e}^{\pm 2i\pi/3} \a_+ +{\rm e}^{\mp 2i\pi/3} \a_- +{m^2\over 3} \ .
\eqn\dxxiv$$
The massless states are all singlets. The $\a_\pm$ are given by
$$\a_\pm = {1\over 3} \left\{ m^6 -{135\over 16} \lu^3 m^3 -{729\over 512}\lu^6
\pm \left[ 27 \lu^3 \left( {27\over 64} \lu^3 - m^3\right)^3 \right]^{1/2} 
\right\}^{1/3} \ .
\eqn\dxxvbis$$
\vskip 2.mm
\fig{The flow of the singularities for $N_f=1$. The left figure shows
the positions of the singularities for small mass $m$, the middle
figure shows them at the superconformal point  where $\s_+$ and
$\s_-$ coincide, while the right figure shows them for large
$m$.}{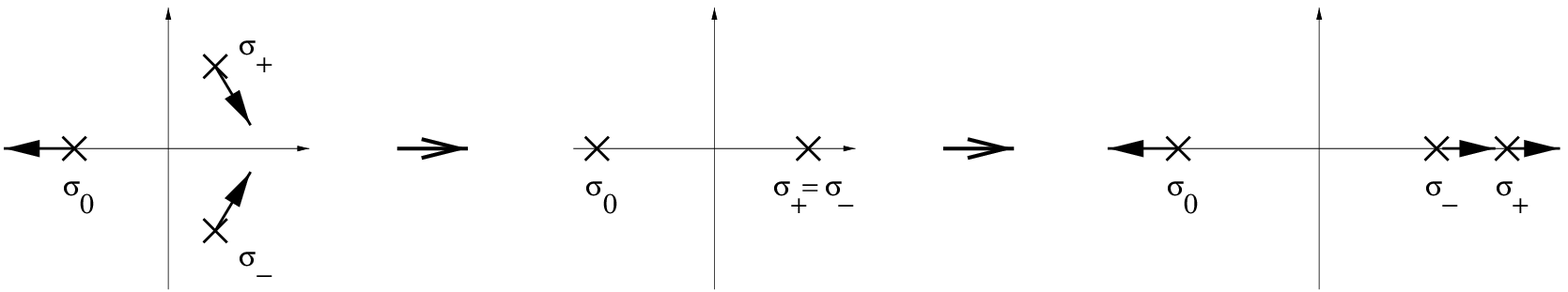}{16cm}
\figlabel\figii
\vskip 2.mm
\noindent
While the choice of phase of the square root does not really matter since one can
always exchange the r\^oles of $\a_+$ and $\a_-$, the phase of the cubic root has
to be determined carefully, so that the three singularities $\s$ are either all
real or one real and the other two complex conjugate, and such that the
singularities vary continuously as $m$ is varied from 0 to $\infty$. In
particular, for $0<m^3<{27\over 64}\lu^3$ the curly bracket in \dxxvbis\ is real and
negative, and we choose $\a_\pm = -{1\over 3} \vert \{ \ldots \}\vert^{1/3}$.
This ensures that $\s_0$ is real and negative, while $\s_+$ and $\s_-$ are
complex conjugate. For $m=0$ one simply has 
$\s_0=-u_0\, , \ \s_\pm ={\rm e}^{\pm i\pi/3} u_0$ with 
$u_0=3 \lu^2/( 4\cdot 2^{2/3} )$. As $m$ is increased, $\s_0$
moves to the left and $\s_\pm$ approach the real axis, see Fig. \figii. 
For small $m$
one has
$$\s_0\simeq -u_0(1+2\mu)\quad , \quad
\s_\pm \simeq u_0 \left( {\rm e}^{\pm i\pi/3} + 2\mu {\rm e}^{\mp i\pi/3} \right) 
\eqn\dxxvi$$
where $\mu= 2\cdot 2^{1/3} m/(3\lu) \ll 1$. When the square root in \dxxvbis\
vanishes, $\a_+=\a_-$, so that  $\s_+$ and $\s_-$ meet on  the real axis. This is a
superconformal point:
$$m={3\lu\over 4} \quad , \quad \s_+=\s_-={4\over 3} m^2={3\lu^2\over 4} \ .
\eqn\dxxvii$$
As $m$ is increased further, the curly bracket in \dxxvbis\ is now complex. We
choose $[\ldots ]^{1/2}=-i \vert [ \ldots ] \vert^{1/2}$. (The opposite choice
only exchanges $\s_+$ and $\s_-$.) The phase of the cubic root is chosen so that
$\a_+=\a_-^*$ with ${2\pi\over 3}<\arg \a_+ < \pi$ and 
$-\pi <\arg\a_-< - {2\pi\over 3}$. With this choice, the singularities will flow
continuously: $\s_0$ keeps moving to the left, while $\s_\pm$, now both real,
move to the right, but $\s_+$ moves faster, see Fig. \figii. As $m\to\infty$, one has
$$\s_0\simeq -(\lu^3 m)^{1/2} \quad , \quad \s_-\simeq (\lu^3 m)^{1/2} 
\quad ,\quad \s_+\simeq m^2
\eqn\dxxviii$$
up to terms ${\cal O}(\lu^3/m)$. Since $\lu^3 m=\ld^4$, $\s_0$ and $\s_-$ flow to
the singularities of the pure gauge theory, while $\s_+$ disappears to infinity.

{\bf \Appendix{B}}

In this appendix we will show how to express the integrals 
$I_i^{(j)}$ in terms of the standard elliptic
integrals $K(k),\ E(k), \ \Pi_1(\nu,k)$, and also give their 
expressions in terms of the Weierstra\ss\ $\wp$ function,
as well as
give certain relations between elliptic
integrals.
Basic references are [\ERD, \WW, \ELL]. 
Our conventions are those of [\ERD]. Unless
otherwise stated, we will assume that the roots $e_i$ of the cubic
$\eta^2=4\xi^3 - g_2\xi - g_3$ are all different. They obey $\sum_i e_i=0$,
and $\eta^2=4\prod_i (\xi-e_i)$.

The $K, E$ and $\Pi_1$ only depend on the square of
$k$ which was defined in terms of the $e_i$ as $k^2={e_2-e_3\over e_1-e_3}$. 
We use the notation\foot{
Note that in Mathematica which we used extensively for the numerical
determination of the curves of marginal stability, one denotes 
$K(k)={\tt EllipticK[}k^2{\tt ]}$,
$E(k)={\tt EllipticE[}k^2{\tt ]}$ and
$\Pi_1(\nu,k)={\tt EllipticPi[}-\nu,k^2{\tt ]}$.}
of [\ERD]:
$$\eqalign{
K(k) &= \int_0^1 {\d x\over \left[ (1-x^2) (1-k^2 x^2) \right]^{1/2} }
\cr
E(k) &= \int_0^1 \d x \left( {1-k^2 x^2 \over  1-x^2 }\right)^{1/2} 
\cr
\Pi_1(\nu,k) &= \int_0^1 
{\d x\over \left[ (1-x^2) (1-k^2 x^2) \right]^{1/2} (1 +\nu x^2)}
\ . }
\eqn\txxiii$$
We will now show 
that the integrals over the cycle $\g_1$ are
$$\eqalign{
I_1^{(1)} = 2 \int_{e_3}^{e_2} {\d\xi\over \eta} 
&= {2\over (e_1-e_3)^{1/2}} \, K(k)  \cr
I_2^{(1)} = 2 \int_{e_3}^{e_2} {\xi \d\xi\over \eta} 
&= {2\over (e_1-e_3)^{1/2}} \, \left[ e_1 K(k) +(e_3-e_1) E(k) \right] \cr
I_3^{(1)} = 2 \int_{e_3}^{e_2} {\d\xi\over \eta (\xi-c)} 
&= {2\over (e_1-e_3)^{3/2}} \, \Bigg[ {1\over 1-\tilde c +k'}\, K(k)\cr
& \phantom{XXX}+ {4 k'\over 1+k'}\, {1\over (1-\tilde c)^2- k'^2}\, 
\Pi_1\left( \nu(c), {1-k'\over 1+k'}\right) \Bigg]  }
\eqn\txxiv$$
where
$$k'^2=1-k^2 \quad , \quad
\tilde c = {c-e_3\over e_1-e_3} \quad , \quad
\nu(c) = - \left( {1-\tilde c+k'\over 1-\tilde c -k'} \right)^2
\left({1-k'\over 1+k'}\right)^2 \ .
\eqn\txxv$$
The corresponding integrals $I_i^{(2)}$ over the cycles $\g_2$ are
obtained from equations \txxiv\ and \txxv\ by exchanging everywhere in
these equations the roots $e_1$ and $e_3$. In particular, this exchanges
$k$ and $k'$, and results in the exchange of $\tilde c$ and $1-\tilde
c$, so that $\nu(c)$ gets replaced by
$- \left( {\tilde c+k\over \tilde c -k} \right)^2
\left({1-k\over 1+k}\right)^2$.

To convert the $I_j$ into the $K,E,\Pi_1$ one needs to transform the cubic
curve into a quartic curve. Let's demonstrate this for
$I_1^{(1)}$:
$$I_1^{(1)} = 2 \int_{e_3}^{e_2} {\d\xi\over \eta} = 
\int_{e_3}^{e_2}  {\d\xi\over [ (\xi-e_1)(\xi-e_2)(\xi-e_3) ]^{1/2}}
= {1\over (e_1-e_3)^{1/2}} \int_0^{k^2} 
{\d\tilde\xi\over [ (\tilde\xi-1)(\tilde\xi-k^2)\tilde\xi ]^{1/2}}
\eqn\ai$$
where we changed variables , $\tilde\xi=(\xi-e_3)/(e_1-e_3)$, and introduced
$k^2$ as above. The transformation to a quartic curve is achieved by the
further change of variables
$$\tilde\xi=1 + k' +{1\over \zeta - {1\over 2 k'} }
\eqn\aii$$
where the complementary modulus $k'$ is given by \txxv. (The choice of sign
for $k'$ does not matter.) A final rescaling $x=2k'\, {1+k'\over 1-k'}\, \zeta$
yields
$$ I_1^{(1)} = {1\over (e_1-e_3)^{1/2}}\, {2\over 1+k'} \int_{-1}^1 
{\d x\over \left[ \left( 1- \left( {1-k'\over 1+k'} \right)^2 x^2 \right)
(1-x^2) \right]^{1/2} }
= {2\over (e_1-e_3)^{1/2}}\, {2\over 1+k'}\, K\left({1-k'\over 1+k'} \right) \ .
\eqn\aiii$$
Using one of the standard relations between elliptic integrals with
different moduli [\ERD]
$$ {2\over 1+k'}\, K\left({1-k'\over 1+k'} \right) = K(k) \quad , \quad
(1+k')\, E\left({1-k'\over 1+k'} \right) = E(k) + k' K(k) \ ,
\eqn\aiv$$
one obtains the first equation \txxiv.\foot{
Note that it is quite non-trivial to keep track of the correct overall sign. For example, to
get the last equality of eq. \aiii\ one needs to define carefully where the cuts of the
differents square roots lie. For real masses $m_i$ and real $\Lambda$, all one can get is a
sign ambiguity that may depend on ${\rm sign}(\IM u)$. This ambiguity is the same for all
three integrals $I_1,\, I_2$ and $I_3$, so that it is most easily fixed by comparing the
resulting $a$ and $\ad$ with the required asymptotics \dxxiii.}

Going through exactly the same steps for the integral $I_2^{(1)}$ leads to
$$\eqalign{ I_2^{(1)} &= {1\over (e_1-e_3)^{1/2}}\, {2\over 1+k'} \int_{-1}^1 
\d x { 
(e_1-e_3) \left[ 1+k' -2k'  \left( 1- \left( {1-k'\over 1+k'} \right)^2 x^2
\right)^{-1} \right] + e_3
\over
 \left[ \left( 1- \left( {1-k'\over 1+k'} \right)^2 x^2 \right)
(1-x^2) \right]^{1/2} } \cr
&= 
{2\over (e_1-e_3)^{1/2}}\, {2\over 1+k'} 
\Bigg[ \left( (e_1-e_3) (1+k') + e_3 \right) K\left({1-k'\over 1+k'}\right) \cr
& \phantom{= {2\over (e_1-e_3)^{1/2}} {2\over 1+k'} X}
-2 k' (e_1-e_3) \
\Pi_1\left( -\left( {1-k'\over 1+k'} \right)^2,\, {1-k'\over 1+k'}\right)\Bigg]
\ .}
\eqn\avi$$
Using the relation $(1-\bar k^2)\Pi_1(-\bar k^2,\bar k) = E(\bar k)$ with $\bar
k=(1-k')/(1+k')$, as well as the relations \aiv, one arrives at the second equation
\txxiv.

Finally for $I_3^{(1)}(c)$, going through the same steps leads to
$$\eqalign{  I_3^{(1)}(c) = {1\over (e_1-e_3)^{3/2}}\, {2\over 1+k'} \int_{-1}^1 
&{\d x \over  \left[ \left( 1- \left( {1-k'\over 1+k'} \right)^2 x^2 \right)
(1-x^2) \right]^{1/2} }  \cr
&\times {1\over 1+k'-\tilde c} 
\left[ 1 - {2k'\over 1+k'-\tilde c}\, {1\over {1-k'\over 1+k'} x - 
{1-k'-\tilde c\over 1+k'-\tilde c}} \right] \cr }
\eqn\aviii$$
where $\tilde c=(c-e_3)/(e_1-e_3)$. The last term in this expression is of the
type ${1\over ax-b}={ax+b\over a^2x^2-b^2}$ and can be replaced by 
${b\over a^2x^2-b^2}$. It is then clear, using again \aiv\ that one gets the
third relation \txxiv.

The integrals $I_3^{(j)}(c)$ can be simplified if $c$ is one of the roots $e_i$.
We will always consider $I_3^{(1)}(c)$. Everything can be translated for 
$I_3^{(2)}(c)$ if we permute everywhere $k$ and $k'$ as well as $e_3$ and $e_1$.
Let first $c=e_1$. Then $\tilde c =1$ and 
$\nu(c)=-\left( {1-k'\over 1+k'} \right)^2 \equiv - \tilde k^2$ (see \txxv). One
has [\ERD]
$$\Pi_1(-\tilde k^2,\tilde k) ={1\over 1-\tilde k^2}\, E(\tilde k) = {1+k'\over 4
k'} \left( E(k)+ k' K(k)\right) \ .
\eqn\aix$$
Hence
$$I_3^{(1)}(e_1)= {2\over (e_1-e_3)^{3/2}}\, {E(k)\over k^2-1} \ .
\eqn\ax$$
This can also be directly obtained since $I_3^{(1)}(e_1)= 2 {\d\over \d e_1}
I_1^{(1)}$. Using $k^2 {\d\over \d k^2} K(k) = - {1\over 2} K(k) + {1\over 2}
{E(k)\over 1-k^2}$ this gives again \ax. Obviously, this constitutes a
consistency check for the third equation \txxiv. Let now $c=e_2$. Then $\tilde
c=k^2$ and $\nu(c) =-1$. 
$\Pi_1$ is singular for $\nu=-1$, but $I_3^{(1)}(e_2)$ can be obtained in exactly the
same way since $I_3^{(1)}(e_2)=2 {\d\over \d e_2} I_1^{(1)}$ provided we keep the
integration cycle $\g_1$ fixed and away from $e_2$ (and $e_3$). We then immediately
get
$$I_3^{(1)}(e_2) = {2\over (e_1-e_3)^{3/2}}\, {1\over k^2} \left( {E(k)\over
1-k^2}- K(k)\right) \ .
\eqn\axii$$
Finally, if $c=e_3$, one has $\tilde c=0$ and also $\nu(c)=-1$, and again 
$\Pi_1$ is singular, but we can still use $I_3^{(1)}(e_3)=2 {\d\over \d e_3} I_1^{(1)}$ 
which readily gives
$$I_3^{(1)}(e_3) = {2\over (e_1-e_3)^{3/2}}\, {1\over k^2} \left( K(k)-E(k)
\right) \ .
\eqn\axiii$$
Of course, using the first two equations \txxiv,
one can reexpress $K(k)$ and $E(k)$ in terms of
$I_1^{(1)}$ and $I_2^{(1)}$ and hence we have shown how to express 
$I_3^{(1)}(e_j)$ in terms of $I_1^{(1)}$ and $I_2^{(1)}$.

We also need to Taylor expand  $I_3^{(j)}(c+\delta c)$, hence we want to
compute the derivative ${\d\over \d c} I_3^{(j)}(c)$. For any cycle $\g_j$ we
have
$$ {\d\over \d c} I_3^{(j)}(c) =\oint_{\g_j} {\d\xi\over \eta (\xi-c)^2} \ .
\eqn\axiv$$
Now observe that
$$ 0= \oint \d\xi\, {\d\over \d\xi} {\eta\over \xi-c} =
 \oint \d\xi\,  { 12\xi^2-g_2\over 2 \eta(\xi-c)} - 
 \oint \d\xi\,  {\eta^2 \over \eta (\xi-c)^2} \ .
\eqn\axv$$
This is a sum of integrals containing $I_1, I_2$ and $I_3$, as well as the
integral \axiv. Solving for the latter gives
$$(4c^3-c g_2 -g_3) \oint {\d\xi\over \eta (\xi-c)^2} 
= -2c\, I_1 +2\, I_2 -{1\over 2}(12 c^2-g_2)\, I_3(c) \ .
\eqn\axvi$$
Inserting now the appropriate values of $g_2$ and $g_3$ for the $N_f=1$ curve
and taking $c=-u/3$ yields the desired relation for the derivative $I_3'(c)\equiv 
{\d\over \d c} I_3(c)$:
$$-{\lu^6\over 32}  I_3' \left(- {u\over 3}\right) = {u\over 3}
I_1 +I_2 -{m\lu^3\over 4} I_3\left(- {u\over 3}\right)\ .
\eqn\axvii$$

A word of caution is in order: in eq. \txxiv\ we have replaced the 
integrals over the cycle $\g_1$ by twice the integrals from $e_3$ to
$e_2$. Depending on the detailed form of the cycle, these two
definitions may differ for $I_3$, which has a pole at $\xi=c$, by terms
$2\pi i\, {\rm res} {\d\xi\over \eta}$. So we allow the freedom to add
such terms ``by hand", this being equivalent to changing the definition
of the cycle $\g_1$ with respect 
to the position of the pole. A related point is that
for $c=e_2$ or $c=e_3$, one has $\nu(c)=-1$ and
$\Pi_1(-1,\tilde k)$ diverges. Nevertheless, we can remove the divergence by an
appropriate choice of integration contour.\foot{
This is so because $\nu\to 1$ typically occurs under the RG flow as some bare mass
$m_j\to\infty$ and the divergent part of $\Pi_1$ precisely is some integer multiple
of ${m_j\over 2\rd}$.}
With this choice $\Pi_1(-1,\tilde k)$
should be understood as $K(\tilde k) -E(\tilde
k)/(1-\tilde k^2)$. Then, eqs. \axii\ and \axiii\ also follow in a straightforward
manner, using \aiv, from the third equation \txxiv.
Similar remarks apply to the integrals
$I_i^{(2)}$ over the cycles $\g_2$.

\vskip 2.mm
\fig{The elliptic curve in the $z$-plane where it is simply a
parallelogram with sides $\o_1$ and $\o_2$. The cycles $\g_1$ and
$\g_2$ are straight lines parallel to the sides of the
parallelogram.}{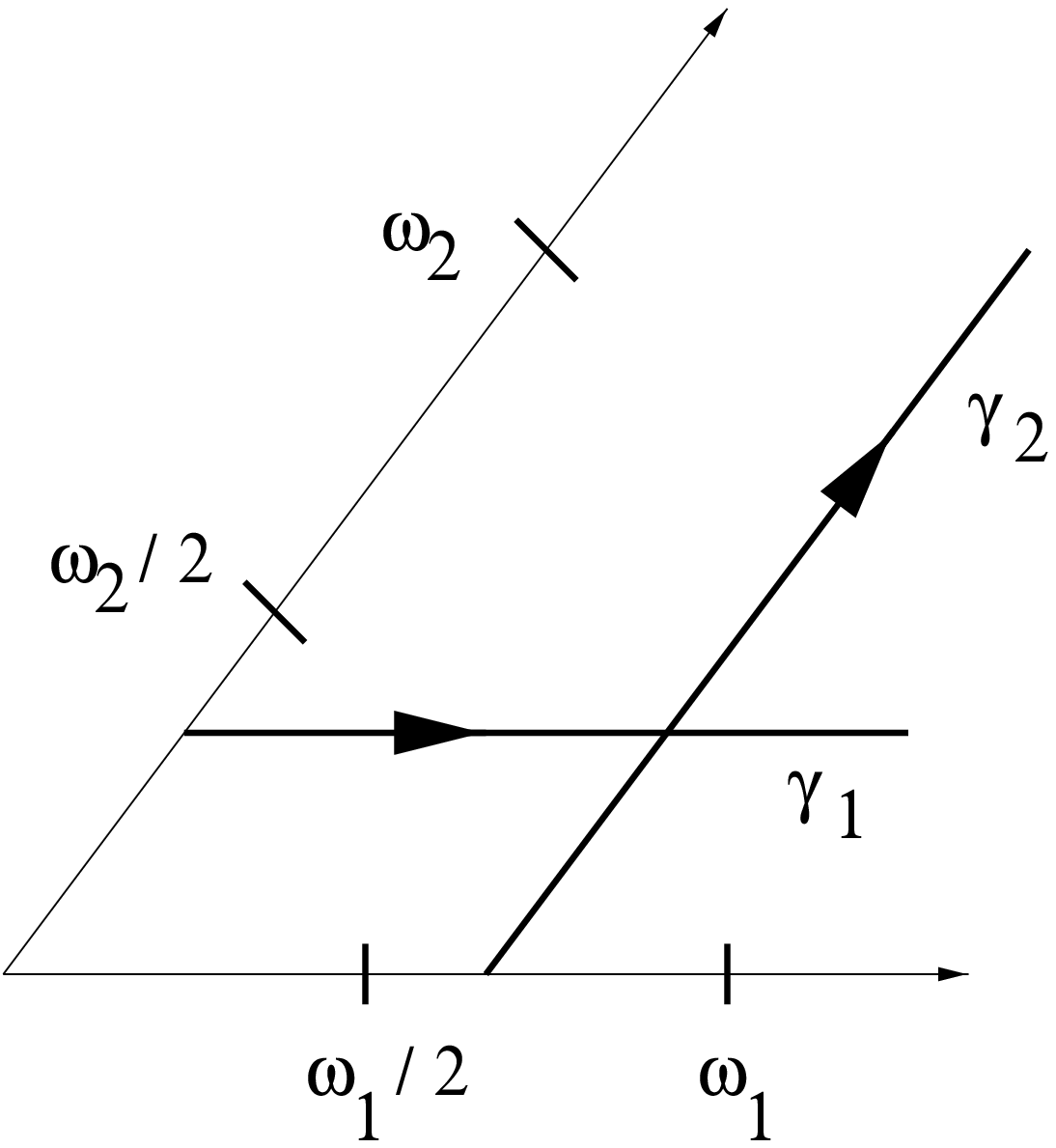}{5cm}
\figlabel\figiv
\vskip 2.mm
To gain a somewhat better control over the integration cycles $\g_1$ and
$\g_2$, in particular with respect to the positions of the poles, 
it is sometimes advantageous to introduce the uniformizing
variable $z$ via the doubly periodic Weierstra\ss\ $\wp$ function [\WW]
as $\xi=\wp(z)$, $\eta=\wp'(z)$. Then the integrals $I_1^{(i)}$ are
simply
$$I_1^{(i)}=\oint_{\g_i} {\d\xi\over \eta}
= \int_{\g_i} {\wp'(z) \d z\over \wp'(z)} = \int_0^{\o_i} \d z =\o_i
\eqn\txxvi$$
where $\o_1$ and $\o_2$ are the two periods of $\wp$ and are such that
$e_1=\wp\left({\o_1\over 2}\right)$,
$e_2=\wp\left({\o_1+\o_2\over 2}\right)$,
$e_3=\wp\left({\o_2\over 2}\right)$. The cycle $\g_1$ is mapped in the
$z$-plane to a straight line from $0$ to $\o_1$ (or any shifted copy of
it) while $\g_2$ is mapped to the straight line from $0$ to $\o_2$, see Fig. \figiv.

Obviously one has $\t={\o_2\over \o_1}$. Given the $e_1, e_2, e_3$, one
can obtain $\o_1$ and $\o_2$, hence $I_1^{(1)}$ and $I_1^{(2)}$ using
the inverse of the $\wp$ function. The latter is conveniently
expressed through the inverse of the Jacobi elliptic function ${\rm sn}$:
$$z_\xi \equiv \wp^{-1}(\xi)={1\over (e_1-e_3)^{1/2} } \  {\rm sn}^{-1}
\left[ \left({e_1-e_3\over \xi-e_3}\right)^{1/2}, k\right] \ .
\eqn\txxvii$$
For a numerical computation of the $\o_i=I_1^{(i)}$, eq. \txxiv\ is of
course simplest. Next one has (do not confuse the $\eta_i$ with $\eta$)
$$I_2^{(i)}= \oint_{\g_i} {\xi \d\xi\over \eta} = -2 \eta_i
\eqn\txxviii$$
where again numerical computation of the $\eta_i$ is easiest done via
\txxiv. Finally, one has
$$I_3^{(i)}= \oint_{\g_i} {\d\xi\over \eta (\xi-c)} =
{1\over \wp'(z_c)} \left[ 2 \o_i \zeta(z_c)-4 \eta_i z_c\right]
\eqn\txxix$$
where $z_c$ is given by \txxvii\ with $\xi=c$, and 
$$\zeta(z)={2\eta_1\over \o_1} z +{\pi\over \o_1} {\theta_1'\over
\theta_1}\left( {\pi z\over \o_1} \Big\vert \t\right) \ .
\eqn\txxx$$

{\bf \Appendix{C}}

In  this appendix we give the differentials $\l$ for the different massive theories as
well as the decomposition of the period integrals in terms of the three basic elliptic
integrals $I_1,\, I_2$ and $I_3$. We also check some RG flows at the level of these
integrals.

\subsection{One flavour: $N_f=1$}

A one-form $\l$ satisfying \div\ obviously is
$$\l=-{\rd\over 4\pi} {y\, \d x\over x^2} = 
{\rd\over 4\pi} \left[ \d x {\d\over \d x} \left({x\over y}\right)
-\left( 3x -2u+{\lu^3\over 4}{m\over x}\right) {\d x\over 2 y}\right] \ .
\eqn\tvii$$
The first term in the bracket is an exact form and vanishes upon
integration over a cycle. Converting to Weierstra\ss\ normal form \dxx\
by $\eta=2y,\ \xi=x-{u\over 3}$ we get
$$\int\l = {\rd\over 4\pi} 
\left[ u\, I_1 -3\, I_2 - {\lu^3\over 4}m\, I_3\left(-{u\over 3}\right) \right]
\ .
\eqn\tviii$$
We see from \tvii\ that $\l$ has  poles at $(x=0, y=\mp i \lu^3/8)$ with residues $\pm
{1\over 2\pi i} {m\over \rd}$ in agreement with eq. \dix.

\subsection{Two flavours: $N_f=2$}

We have [\SWII]
$$\eqalign{
\l&=-{\rd\over 4\pi} {y\, \d x\over x^2-{\ld^4\over 64}} 
=-{\rd\over 4\pi} {\d x\over y} {4 y^2\over \ld^2} 
\left( {1\over x-{\ld^2\over 8}} -  {1\over x+{\ld^2\over 8}} \right) \cr
&=-{\rd\over 4\pi}  {\d x\over y}
\left[ x-u-{\ld^2\over 16}  {(m_1-m_2)^2\over x-{\ld^2\over 8}} 
+ {\ld^2\over 16}  {(m_1+m_2)^2\over x+{\ld^2\over 8}} \right] \ .}
\eqn\tix$$
Converting to Weierstra\ss\ normal form of the cubic, again by 
$\eta=2y,\ \xi=x-{u\over 3}$ we arrive at
$$\int\l = {\rd\over 4\pi} 
\left[ {4\over 3} u\, I_1 -2\, I_2  
+ {\ld^2\over 8}  (m_1-m_2)^2\, I_3\left({\ld^2\over 8}-{u\over 3}\right) 
- {\ld^2\over 8}  (m_1+m_2)^2\, I_3\left(-{\ld^2\over 8}-{u\over 3}\right) 
\right]
\ .
\eqn\tx$$
One sees from \tix\ that $\l$ has  poles at $(x={\ld^2\over 8}, y=\pm i
\ld^2(m_1-m_2)/8)$ with residues
$\pm {1\over 2\pi i} {m_1-m_2\over 2\rd}$ and  at $(x=-{\ld^2\over 8}, y=\mp i
\ld^2(m_1+m_2)/8)$
with residues $\pm {1\over 2\pi i} {m_1+m_2\over 2\rd}$ [\SWII]. 
In particular, for $m_1=m_2=m$, there are only the poles 
at $(x=-{\ld^2\over 8}, y=\mp i \ld^2 m/4)$ with
residues $\pm {1\over 2\pi i} {m\over \rd}$.

\subsection{Three flavours: $N_f=3$}

This case is more complicated since the $y^2$ is no longer linear in
$u$, see eq. \di, \dii. There exist expressions for $\l$ in
the literature using a  quartic curve [\OHTA] instead.
Proceeding along the same lines as in [\OHTA] we can obtain $\l$ also for the
cubic curve \di, \dii: we write
$$\eqalign{
y^2 = G(x)-F^2(x) \quad , \quad 
F(x) &=\sqrt{a} \left( x-u-{x^2\over 2a} +{b^2\over 2a}\right) \cr
G(x) &= {x^4\over 4 a} - {b^2 x^2\over 2a} + {b^4\over 4a} +cx -d }
\eqn\txi$$
where we have set $a={\lt^2\over 64}$, $b^2=a\sum m_i^2$, $c={1\over 4}
\lt \prod m_i$ and $ d=a \sum_{i<j} m_i^2 m_j^2$. One can then check that
the following differential indeed satisfies \div
$$\l= - {\rd\over 8\pi} {x\, \d x\over \sqrt{a}\, y}
\left( {F(x) G'(x)\over 2 G(x)} -F'(x) \right) \ .
\eqn\txii$$

In general, to express $\int\l$ in terms of the three elliptic integrals
$I_i$ one needs to decompose the quartic polynomial $G(x)$
into linear factors. While this can be done in general, it is cumbersome and the
result not very illuminating. Here we will restrict ourselves to the
simpler case $m_1=m_2=0$, $m_3\equiv m$. Then, since $c=d=0$:
$G(x)={1\over 4a} (x^2-b^2)^2$ and ${G'\over 2G}={1\over x+b}+{1\over
x-b}$ where now $b=\lt m /8$. It is then easy to see that
$$\l= - {\rd\over 8\pi} {\d x\over y} \left[ x-2u +{b^2+bu\over x+b}
+{b^2-bu\over x-b} \right] 
\eqn\txiii$$
so that after introducing $\eta=2 y$ and $\xi=x-{u\over 3}-{\lt^2\over
192}$ we get
$$\eqalign{
\int\l = {\rd\over 4\pi} 
\Bigg[ \left({5\over 3} u -{\lt^2\over 192}\right) I_1 - I_2 
&- {\lt m\over 64}  (8u+\lt m) 
I_3\left(-{\lt m\over 8}-{\lt^2\over 192}-{u\over 3}\right) \cr
&+ {\lt m\over 64}  (8u-\lt m) 
I_3\left({\lt m\over 8}-{\lt^2\over 192}-{u\over 3}\right) 
\Bigg]
\ .}
\eqn\txiv$$
We see from \txiii\ that $\l$ has  poles at 
$(x= \lt m/8\, ,\ y=\mp i (\lt m/8 -u)\lt/8)$  and  at 
$(x= -\lt m/8\, ,\ y=\pm i (-\lt m/8 -u)\lt/8)$ with residues  
$\pm {1\over 2\pi i} {m\over 2\rd}$. Note that the integrals $I_3$ cancel in
the $m\to 0$ limit.

\subsection{The pure gauge theory: $N_f=0$}

Finally let us note that for the pure gauge theory in the conventions of
[\SWII] we have
$$\int\l = {\rd\over 4\pi}  
\left[ u I_1 -3 I_2 -{\lz^4\over 4} I_3\left(-{u\over 3}\right)  \right]
\ .
\eqn\txv$$
Since $-{u\over 3}$ is one of the roots $e_i$ one can use relations 
\ax\ to \axiii\
to reexpress $ I_3\left(-{u\over 3}\right)$ as a combination of $I_1$ and $I_2$. This
must be so for $N_f=0$ since $\l$ has no poles and thus  its integral must be
expressible through $I_1$ and $I_2$ only.

\subsection{RG flows of the integrals}

One can check the different RG flows at the level of the period
integrals. For example, starting with the $N_f=1$ periods \tviii\ and
letting $m\to\infty$, $\lu\to 0$ while keeping $m\lu^3=\lz^4$ fixed, we
immediately find that \tviii\ flows to \txv. The flow from $N_f=2$, eq.
\tx, to $N_f=1$, eq. \tviii, as $m_2\to\infty$, $\ld\to 0$, $m_2\ld^2=\lu^3$
fixed, is less trivial: starting with \tx\ we have
$$\eqalign{
\int\l \Big\vert_{N_f=2}\ \to\  {\rd\over 4\pi} 
\Bigg[ {4\over 3} u I_1 -2 I_2  
&+ {\lu^3\over 8}  (m_2-2m_1) I_3\left(-{u\over 3}+{\lu^3\over 8
m_2}\right) \cr
&- {\lu^3\over 8}  (m_2+2m_1) I_3\left(-{u\over 3}-{\lu^3\over 8
m_2}\right) 
\Bigg] \ .}
\eqn\txvi$$
Here, the integrals $I_i$ on the r.h.s. are meant to be those of
$N_f=1$. Taylor expanding $I_3(c+\delta c)=I_3(c) +I_3'(c)\delta
c+\ldots $, and using the relation \axvii\ for the derivative of
$I_3$ (valid for $N_f=1$), the r.h.s of eq. \txvi\
becomes exactly the r.h.s. of eq. \tviii, up to terms that vanish as $m_2\to\infty$.
Similarly one can check that as $m\to\infty$, the $N_f=3$ periods \txiv\
flow to the $N_f=2$ periods \tx\ with $m_1=m_2=0$. This requires to
reexpress $I_3(c)$ for $c$ a root $e_i$ in terms of $I_1$ and $I_2$
through the formulae  given in appendix B.

{\bf \Appendix{D}}

In this appendix we perform some checks on our equations \dxxv\ that express $a$ and
$\ad$ for $N_f=2$, $m_1=m_2=m$ in terms of the $I_i^{(j)}$ and hence of the complete
elliptic integrals $K$, $E$ and
$\Pi_1$. In particular, we will show that one indeed recovers the correct expressions of
the massless $N_f=2$ theory [\BF] as $m\to 0$, and that one gets the appropriate
expressions of $N_f=0$ in the $m\to\infty$ limit. 

First we examine the limit $m\to 0$. Then the extra term ${m\over \rd}$ in  \dxxv\
disappears and one has
$$\eqalign{
\oint_{\g_i} \l \to
& {\rd\over 4\pi} \left[ {4\over 3} u\, I_1^{(i)} - 2 I_2^{(i)} \right]
={\rd\over 4\pi} \oint_{\g_i} 
{\d\xi \left( {4\over 3} u - 2\xi\right)\over \eta} \cr
= & {\rd\over 4\pi} \oint_{\g_i} {\d x (u-x)\over y} = 
-  {\rd\over 4\pi} \oint_{\g_i}  
{\d x \sqrt{x-u}\over \sqrt{ x^2-{\ld^4\over 64} } }
\ . }
\eqn\txxxiii$$
Since $e_1\to {2u\over 3}$, $e_2\to -{u\over 3}+{\ld^2\over 8}$,
$e_3\to -{u\over 3}-{\ld^2\over 8}$
(or in terms of the variable $x$ the roots are $x_1=u$, $x_2={\ld^2\over 8}$),
these are precisely the integral expressions for $\ad$ and $a$ 
of the massless $N_f=2$
theory (which equal ${1\over 2}$ times the $\ad$ and $a$ of the $N_f=0$ theory)
[\SWI, \BF].

Now let $m\ne 0$. It is then straightforward to explicitly check that at the rightmost singularity,
$u=m^2+{\ld^2\over 8}$, we have the following: if $m<{\ld\over 2}$ one has $e_1=e_2$,
$k'=0$ and thus it follows that $\ad(u)=0\, $; if $m>{\ld\over 2}$, however, one has
$e_2=e_3$ so that $k=0$ and it then follows that $a(u)= {m\over \rd}$. Hence we see that
the massless BPS state at this singularity $u=m^2+{\ld^2\over 8}$ is a magnetic monopole
$(n_e,n_m)_s=(0,1)_0$ if $m<{\ld\over 2}$, while for $m>{\ld\over 2}$
it is a quark $(n_e,n_m)_s=(1,0)_1$. This perfectly agrees with the discussion of Section
2 and thus justifies the choice of adding the $ {m\over \rd}$ term in the first equation 
\dxxv.

Let us now examine the RG flow to the $N_f=0$ theory as $m\to\infty$, 
$\ld\to 0$, $m\ld =\lz^2$ fixed. The flow of the roots $e_i$ of eq. \dxxiv\ is
$$\eqalign{
e_1 & \to {u\over 6} + {1\over 2} \sqrt{u+\lz^2}\sqrt{u-\lz^2} \equiv e_1^{(0)} \cr
e_2 & \to -{u\over 3} \equiv e_3^{(0)} \cr
e_3 & \to {u\over 6} - {1\over 2} \sqrt{u+\lz^2}\sqrt{u-\lz^2} \equiv e_2^{(0)}  
\ . }
\eqn\txxxvi$$
where $e_i^{(0)}$ is the standard labeling of the roots for $N_f=0$.
We see that the flow exchanges the labelling of $e_2$ and $e_3$. Let us
first consider the flow of $a(u)$ as given by the first eq. \dxxv. Since the $N_f=2$ curve
flows to the $N_f=0$ curve, the integrands flow appropriately. Clearly, the
cycle $\g_1^{(N_f=2)}$ encircling $e_2$ and $e_3$  flows to the cycle $\g_1^{(N_f=0)}$.
Hence $I_1^{(1)} \to I_1^{(1)}\vert_{N_f=0}$ and $I_2^{(1)} \to I_2^{(1)}\vert_{N_f=0}$. The
integral $I_3^{(1)}$ is more subtle since it is the one that involves the pole at 
$c=-{u\over 3}-{\ld^2\over 8}=e_3^{(0)}-{\ld^2\over 8}$. 
As $\ld\to 0$, this pole approaches $e_3^{(0)}$. Hence,
for non-zero ${\ld^2\over 8}$ the pole at $\xi=c$ is outside the integration contour. As
$\ld\to 0$, the pole crosses the contour and $I_3^{(1)}$ picks up a contribution from the
residue which is $\delta I_3^{(1)}={4\pi\over \ld^2 m}$ which precisely cancels the additional
term ${m\over\rd}$ in eq. \dxxv. Hence
$$a(u) \to  {\rd\over 4\pi} 
\left[ {4\over 3}\, u I_1^{(1)} -2\,  I_2^{(1)}  
- {\lz^4\over 2}\,  I_3^{(1)}(e_3^{(0)}) \right] \Bigg\vert_{N_f=0} \ .
\eqn\txxxva$$
Using the relation 
$I_3^{(1)}(e_3^{(0)})\Big\vert_{N_f=0} = 
{4\over \lz^4} \left( I_2^{(1)} +{u\over 3}\, I_1^{(1)} \right)\Big\vert_{N_f=0} $ 
from appendix B,
we see that the r.h.s. of eq. \txxxva\ indeed coincides with the 
corresponding integral for $N_f=0$, and hence $a(u)\to
a(u)\Big\vert_{N_f=0} \equiv a^{(0)}(u)$. Next, let us discuss the flow of $\ad(u)$. The cycle
$\g_2$ encircles $e_1$ and $e_2$ which flow to $e_1^{(0)}$ and $e_3^{(0)}$. The corresponding
integral thus is the sum of the integral over a cycle $\g_2^{(0)}$ around $e_1^{(0)}$ and
$e_2^{(0)}$ and one over a cycle $\g_1^{(0)}$ around $e_2^{(0)}$ and $e_3^{(0)}$: $\g_2 \to
\g_2^{(0)} + \e \g_1^{(0)}$ where $\e={\rm sign}(\IM u)$. Again, the pole crossing the cycle
$\g_1^{(0)}$ gives a term $-\e {m\over \rd}$. As a result we have
$$\eqalign{
a(u) &\to  a^{(0)}(u) \cr
\ad(u) + \e\, {m\over \rd}  &\to  \ad^{(0)}(u)  + \e\,  a^{(0)}(u) 
\ . }
\eqn\txxxviia$$
This motivates us to define 
$\adt(u) = \ad(u) - \e \left( a(u)  -  {m\over \rd} \right)$
which is such that under the RG flow as $m\to\infty$ one has
$\at(u) \to a^{(0)}(u)$ and $\adt(u) \to \ad^{(0)}(u)$.

\ack

We are grateful to E. Br\'ezin for an interesting discussion, as well as to I. Bakas
for pointing out to us some properties of the elliptic integral of the third kind.

\refout
\end